\documentclass{aa}  

\usepackage{graphicx}
\usepackage[colorlinks=true, linkcolor=blue, urlcolor=blue, citecolor=blue]{hyperref}
\usepackage{subcaption}
\usepackage{multirow}
\usepackage[percent]{overpic}
\usepackage{txfonts}
\usepackage[export]{adjustbox}
\bibpunct{(}{)}{;}{a}{}{,} % to follow the A&A style

\usepackage{placeins}

\newcommand{\msun}{\mbox{$\rm M_{\sun}$}}

\newcommand{\kms}{\mbox{km~s$^{-1}$}}
\newcommand{\mloss}{\mbox{$\dot{M}$}}
\newcommand{\my}{\mbox{$\rm M_{\odot}$~yr$^{-1}$}}
\newcommand{\vexp}{\mbox{$v_{\rm exp}$}}

\newcommand{\ju}{$J_{\rm u}$}
\newcommand{\jl}{$J_{\rm l}$}

\newcommand{\rco}{\mbox{$r_{\rm CO}$}}

\newcommand{\nco}{\mbox{$N_{\rm CO}$}}
\newcommand{\rs}{\mbox{$\rm R_{\star}$}}
\newcommand{\trot}{\mbox{$T_{\rm rot}$}}

\newcommand{\eu}{$E_{\rm u}$}
\newcommand{\ctau}{$C_\tau$} 
\newcommand{\micron}{$\mu$m} 
\newcommand{\nc}{\mbox{$n_{\rm crit}$}} 
\newcommand{\mtot}{\mbox{$M_{\rm H_2}$}} 
\newcommand{\mass}{\mbox{$M_{\rm H_2}$}} 
\newcommand{\ncol}{\mbox{$N^{\rm col}_{\rm CO}$}} 
\newcommand{\teff}{\mbox{$T_{\rm eff}$}} 
\newcommand{\water}{\mbox{H$_2$O}} 
\newcommand{\hso}{{\it Herschel}}
\newcommand{\docem}{$^{12}$CO}
\newcommand{\trecem}{$^{13}$CO}

\def\cm#1{\ifmmode {\,{\rm cm^{-#1}}}                  % cm-1, cm-2, cm-3, ...
        \else \hbox{$\,${\rm cm$^{\rm -#1}$}}\fi}
\def\raw {\ifmmode\rightarrow\else$\rightarrow$\fi}
\def\ex#1{\ifmmode {\times 10^{#1}}         % x10$^{-1}$, x10$^{-2}$, etc
        \else \hbox{{$\times 10^{\rm #1}$}}\fi}
\begin{document} 

   \title{Warm CO in evolved stars from the THROES catalogue} %
    
   \subtitle{II. Herschel\thanks{{\it Herschel} is an ESA space observatory with science instruments provided by European-led Principal Investigator consortia and with important participation from NASA.}/PACS spectroscopy of C-rich envelopes}

   \author{J. M. da Silva Santos
          \inst{1} \inst{2}
          \and J. Ramos-Medina\inst{1} \and C. Sánchez Contreras\inst{1} \and P. García-Lario\inst{3}  
          }
   \institute{Centro de Astrobiología (CSIC-INTA), ESAC, Camino Bajo del Castillo s/n, 28691 Villanueva de la Cañada, Madrid, Spain
             \and 
	Institute for Solar Physics, Department of Astronomy, Stockholm University, AlbaNova University Centre, SE-106 91 Stockholm, Sweden. \email{joao.dasilva@astro.su.se}
             \and 
             European Space Astronomy Center, PO Box 78, 28691, Villanueva de la Cañada, Madrid, Spain
              }
   \date{\today}
              %\and
% \abstract{}{}{}{}{} 
% 5 {} token are mandatory
 
  \abstract
   {This is the second paper of a series making use of \textit{Herschel}/PACS spectroscopy of evolved stars in the THROES catalogue to study the inner warm regions of their circumstellar envelopes (CSEs).}
   {We analyze the CO emission spectra, including a large number of high-$J$ CO lines (from $J$=14-13 to $J$=45-44, $\nu$=0), as a proxy for the warm molecular gas in the CSEs of a sample of bright carbon-rich stars spanning different evolutionary stages from the Asymptotic Giant Branch (AGB) to the young planetary nebulae (PNe) phase.} %Our analogue study of O-rich and S-type objects is presented in paper I (Ramos-Medina et al. 2018, submitted to A\&A).
   {We use the rotational diagram (RD) technique to derive rotational temperatures (\trot) and masses (\mass) of the envelope layers where the CO transitions observed with PACS arise. Additionally, we obtain a first order estimate of the mass-loss rates and assess the impact of the opacity correction for a range of envelope characteristic radii. We use multi-epoch spectra for the well studied C-rich envelope IRC+10216 to investigate the impact of CO flux variability on the values of \trot ~and \mass.} %(around $\sim10^{15}$ and $\sim10^{16}$ cm for AGBs and pre-PNe/PNe, respectively) % (covering a wide range of excitation energies, $E_{\rm u}\sim580-5000$ K)
   {PACS sensitivity allowed the study of higher rotational numbers than before indicating the presence of a significant amount of warmer gas ($\sim$200-900\,K) not traceable with lower-$J$ CO observations at sub-mm/mm wavelengths. The masses are in the range \mass $\sim10^{-2}-10^{-5}\,\rm M_{\sun}$, anti-correlated with temperature. For some strong CO emitters we infer a double temperature (warm $\overline{T}_{\rm rot}\sim$400\,K and hot $\overline{T}_{\rm rot}\sim$820\,K) component. From the analysis of IRC+10216, we corroborate that the effect of line variability is perceptible on the \trot~of the hot component only, and certainly insignificant on \mass~and, hence, the mass-loss rate. The agreement between our mass-loss rates and the literature across the sample is good. Therefore, the parameters derived from the RD are robust even when strong line flux variability occurs, with the major source of uncertainty in the estimate of the mass-loss rate being the size of the CO-emitting volume.
   } %(with small variations within $\pm$15\% about the average at different epochs), (by a factor 3-4)
   {} 

   \maketitle
% ======================================================= %
% ======================================================= %
\section{Introduction}

The asymptotic giant branch (AGB) is a late evolutionary stage of
low-to-intermediate mass stars
(1\,\msun$\la$M$_\star$$\la$8\msun) which is largely dominated
by mass-loss processes. AGB stars can shed significant portions of
their outer atmospheric layers in a dust-driven wind, with mass-loss
rates of up to {\mloss$\sim$10$^{-4}\rm\,M_{\sun}\,yr^{-1}$}
\citep[e.g.][]{Habing1996,Hofner2018}. The material expelled by the
central star (with very low effective temperatures of
\teff$\sim$2000-3000\,K) forms a cool, dense circumstellar envelope
(CSE) that is rich in dust grains and a large variety of molecules.
After a significant decrease of the mass-loss
rate, the AGB phase ends and the `star+CSE' system begins to evolve to the Planetary Nebula
(PN) phase, at which the CSE is fully or almost fully ionized due to
the much higher central star temperatures
(\teff$\approx$10$^4$-10$^5$\,K) and more diluted envelopes. During the AGB-to-PN transition, shocks -- resulting from the
interaction between slow and fast winds at the end of the AGB phase
(or early post-AGB) -- also play an important role changing the
morphology, dynamics and chemistry of the CSEs \citep[e.g.][]{2000oepn.book.....K,2003ARA&A..41..391V,2006IAUS..234..193B}.
%% ------------------------

The CO\footnote{We abbreviate $\rm ^{12}C^{16}O$ as simply CO
throughout the paper.} molecule is an excellent tracer of the CSEs of AGB stars, post-AGB objects and PNe \cite[e.g][]{1996A&A...306..241G,2001A&A...368..969S,2002A&A...391..577S,2006A&A...450..167T}. The rotational transitions of the ground vibrational level over a wide range of excitation energies sample from cold ($\sim$10 K) to hot gas ($\sim$1000 K). Literature on observations of the cold, extended CO $J = 1-0$ to $J=6-5$ emission around many evolved stars is abundant at mm/sub-mm wavelengths
\citep[e.g.][]{1982ApJ...252..616K,1985ApJ...292..640K,1985ApJ...293..281K,1993ApJS...87..267O,1989A&A...219..256B,1998A&AS..129..363J,2010A&A...523A..59C,2012ApJS..203...16S,2014A&A...566A.145R}. Studies looking at far-infrared (FIR) observations of even higher $J$ CO transitions probing the warmest gas ($\sim$200-1000 K) in deeper layers of CSEs are much more scarce. Pioneering works based on observations with the Infrared Space Observatory ($ISO$) \citep[e.g.][]{2000A&A...360.1117J,2002A&A...391..577S} have continued more recently with \textit{Herschel} \citep[e.g.][]{2011A&A...526A.162G,2012A&A.537A.8B,2014A&A...561A...5K,2015A&A...581A..60D,2018arXiv180803467N}.

This is the second of a series of papers \citep[Paper I]{PaperI} where we analyze in an uniform and systematic way \textit{Herschel}/PACS spectra of a large sample of evolved stars from the THROES catalogue \citep{2017arXiv171105992R} to study their warm inner envelope regions
using high-$J$ CO transitions at FIR wavelengths. As in other previous studies, we divide our sample in O-rich and C-rich targets (papers I and II, respectively) since these two major chemistry classes correspond to progenitor stars with different masses, which follow somewhat different evolutionary paths, and also have a dissimilar dust composition, both facts potentially affecting the mass-loss process. We include targets with different evolutionary
stages: AGB, post-AGB (or pre-PNe) and young planetary nebulae (yPNe).

The goal of our study (papers I and II) is to obtain a first
  estimate of the average excitation temperature (\trot) and mass
  (\mass) of the warm envelope layers traced by the PACS CO lines in a
  uniform way using a simple analysis technique, the well-known
  rotational diagram (RD) method. The RD technique is useful to
  rapidly analyze large data sets (large number of lines and/or large
  samples) and to provide some constraints on these fundamental
  parameters. With the aim of benchmarking the results of the RD
  approximation, we obtain rough estimates of mass-loss rates
  (\mloss) and compare them to values in the literature, paying particular attention to studies including at least a few high-$J$ CO (FIR) transitions.

The impact of possible non-LTE effects on the results from the simple RD analysis was investigated in paper\,I. It was concluded that, though they are expected to be minor in general and probably only affecting the highest-$J$ CO transitions studied here ($J$$\ga$27) at most, their existence cannot be ruled out in the lowest mass-loss rate stars and/or the outermost layers of the PACS CO-emitting volume. We also showed that even under non-LTE conditions, the masses derived from the RDs are approximately correct (or, at the very least, not affected by unusually large uncertainties) since the average excitation temperature describes rather precisely the molecular excitation (i.e., the real level population). This is also corroborated by the good agreement found between our estimates of \mloss\ and those from detailed non-LTE excitation and radiative transfer (nLTEexRT) studies that exist for a number of targets. We stress that the RD method enables a characterization of the warm CSEs of evolved stars in a first approximation and that for a more robust and detailed study of the radial structure of the density, temperature and velocity in the CSEs, as well as for establishing potential mass-loss rate variations with time, more sophisticated analysis is needed \citep[e.g.,][]{1999A&A...345..841R,2002A&A...391..577S,2012A&A...539A.108D}.

Upon submission of this manuscript (and after our paper I was accepted for publication) we became aware of a recent work by \citet{2018arXiv180803467N} who independently presented \textit{Herchel}/PACS (and SPIRE) range spectroscopy of a sample of 37 AGB stars. These authors perform a similar RD analysis of the CO spectra and focus on deriving excitation temperatures (in contrast to our study, estimates of the envelope mass or mass-loss rates are not reported). Other differences with respect to the work by \citet{2018arXiv180803467N} is that we introduce a canonical opacity correction in the RDs and that we include post-AGBs and PNe.
 % 1. Introduction
\section{Observations}
\label{section:Observations}

\subsection{Observations and data reduction}
PACS is a photometer and medium resolution grating spectrometer \citep{2010A&A...518L...2P} onboard the Herschel Space Telescope \citep{2010A&A...518L...1P} probing the FIR wavelength range. The PACS spectrometer covers the wavelength range from 51 to 210 $\mu$m in two different channels that operate simultaneously in the blue (51-105 $\mu$m) and red (102-220 $\mu$m) bands. The Field of View (FoV) covers a 47\arcsec$\times$47\arcsec\ region in the sky, structured in an array of 5$\times$5 spatial pixels ("spaxels") with 9.4\arcsec$\times$9.4\arcsec. PACS provides a resolving power between 5500 and 940, i.e. a spectral resolution of approximately 55-320 $\rm km~s^{-1}$, at short and long wavelengths, respectively, and the PSF of the PACS spectrometer ranges from $\sim$\,9\arcsec\ in the blue band to $\sim$13\arcsec\ in the red band. The PSF is described in \cite{2016MsT..........1D} and \cite{2016A&A...591A.117B}. The technical details of the instrument can be found in the \textit{PACS Observer's Manual}\footnote{\url{herschel.esac.esa.int/Docs/PACS/html/pacs_om.html}}.

The PACS (1D) spectral data were taken from the THROES (caTalogue of HeRschel Observations of Evolved Stars) website\footnote{\url{https://throes.cab.inta-csic.es}} which contains fully reduced PACS spectra of a collection of 114 stars, mostly low-to-intermediate mass AGB stars, post-AGB and PNe. The data reduction is explained in \citet{2017arXiv171105992R} in full detail. The catalogue also consists of a compilation of previous photometry measurements at 12, 25, 60, 100 $\rm \mu m$ with the Infrared Astronomical Satellite (IRAS).
%The spectra were first obtained from the Herschel Science Archive (HSA) and interactively reprocessed with HIPE \citep{2010ASPC..434..139O} from raw data to science-ready products as described in \citet{2017arXiv171105992R} in full detail.

Table \ref{tab:1} offers a description of the observations used here where we
provide the target name as listed in the header of the PACS FITS files
and an alternative name for which some of the stars are more well
known in the literature. We only analyzed the spectra between [55-95]
and [101-190] $\rm \mu m$ because the flux densities are unreliable
above 190 $\rm\mu m$, below 55 $\rm \mu m$ and between 95-101 $\rm\mu
m$ due to spectral leakage. Two observation identifiers (OBSIDs) per
target, corresponding to the bands B2A, B2B and R1, are necessary to
cover the full PACS wavelength range. In the case of IRC+10216, 
  band B2A data exists \citep[see][]{2010A&A...518L.143D}, but these
  were acquired in a non-standard spectroscopy mode and have
  restricted access in the \hso\ Science Archive and, thus, are not
  included in the THROES catalogue. Instead, we used 4+3 OBSIDs
corresponding to seven different operational days: OD = [745, 1087,
  1257 and 1296] covering the spectral range [69-95, 140-190] $\rm \mu
m$ and OD = [894, 1113 and 1288] covering a narrower interval [77-95,
  155-190] $\rm \mu m$.

\begin{figure*}[t]
\centering
\begin{minipage}{.5\textwidth}
  \centering
  \includegraphics[width=.9\linewidth]{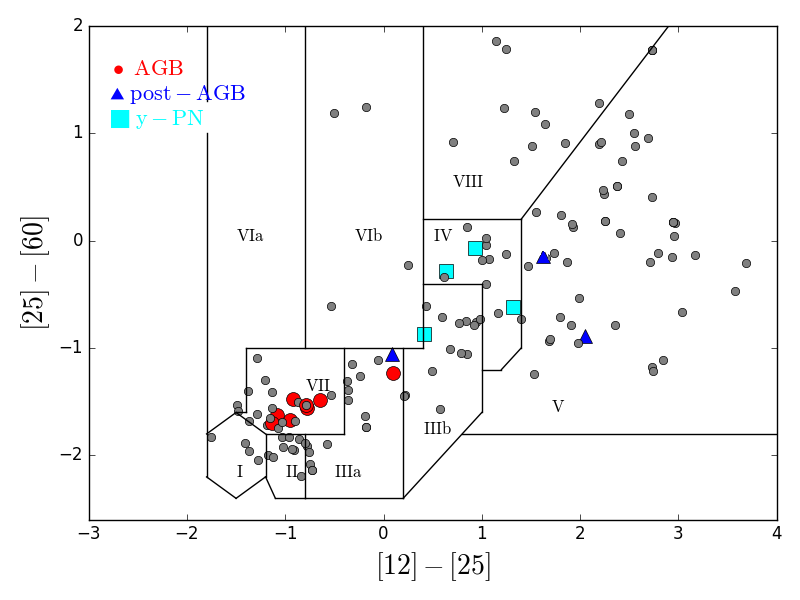}
\end{minipage}%
\begin{minipage}{.5\textwidth}
  \centering
  \includegraphics[width=.9\linewidth]{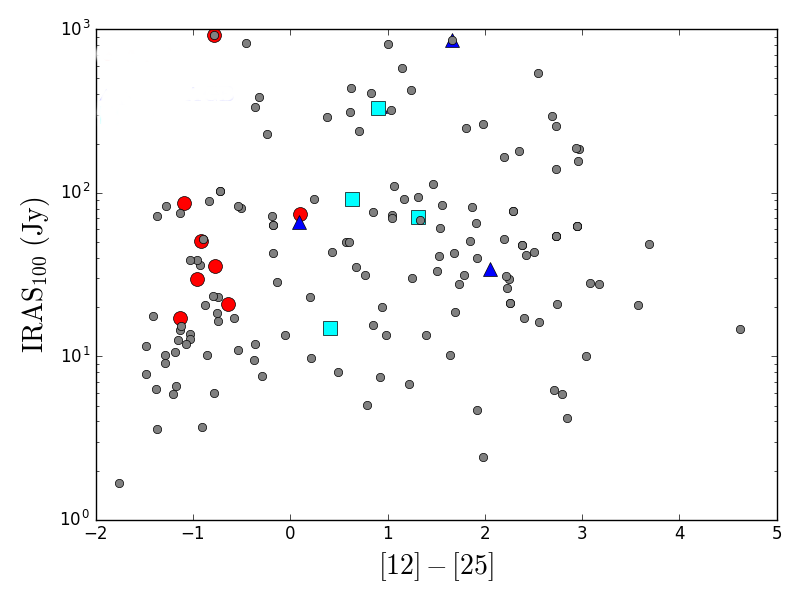}
\end{minipage}
\caption{IRAS color diagrams for the stars in the THROES catalogue. The colors are defined from the infrared fluxes at 12, 25 and 60  $\rm \mu m$. The boxes on the left panel are the ones defined in \citet{1988A&A...194..125V}, and the highlighted points correspond to the stars studied in this paper.} \label{fig:IRAS}
\end{figure*}

\subsection{Sample overview}

We searched for CO emission amongst the entire THROES catalog, but in
this paper we focus on C-rich CSEs (29\% of the entries) observed in
PACS range mode.  We found 15 evolved stars with at
least three CO emission lines with signal-to-noise ratio above 3.

This sample contains bright infrared targets spanning a range of
evolutionary stages from the AGB to the PN phase, but sharing similar
carbon chemistry with strong CO emission at high excitation
temperatures (up to $E_{\rm u}/k\approx5688$ K).  The AGBs, which are
mostly Mira variables, are known for their high mass-loss rates
compared to the mean value of
$\dot{M}\sim1.5\times10^{-7}\rm~M_{\sun}~yr^{-1}$ derived from studies
with large samples of carbon stars \citep{1993ApJS...87..267O}.  We
also include two mixed-chemistry post-AGBs \citep[Red Rectangle and
  IRAS\,16594-4656, ][]{1998Natur.391..868W,2005A&A...429..977W} and
two mixed-chemistry yPNes \citep[Hen 2-113 and CPD-56\degr8032,][and
  references therein]{1998MNRAS.296..419D,2015MNRAS.449L..56D}, that
is, they show both C-rich and O-rich dust grains.  In the Appendix we
provide Table \ref{tab:2} with a summary of some relevant properties
such as distance, effective temperature and gas mass-loss rate, along
with additional references.

Figure \ref{fig:IRAS} shows the classic IRAS color-color diagram
\citep{1988A&A...194..125V} featuring the colors $[25] - [60] =
-2.5\log\frac{IRAS_{25}}{IRAS_{60}}$ and $[12] - [25] =
-2.5\log\frac{IRAS_{12}}{IRAS_{25}}$ of all the stars in THROES. The
stars here studied are highlighted with colored filled symbols which
we will use consistently throughout the paper. This diagram is known
to be a good indicator of the evolutionary stage of
low-to-intermediate evolved stars, with AGBs populating the lower left
corner and more advanced stages being located on the diametrically
opposed, so-called cold, side of the diagram. It also shows an
evolution in terms of the mass-loss rate and/or progressively
increasing optical depths \citep{1987A&A...186..136B}. C-rich AGBs
clearly constellate in a different box compared to the O-rich stars in
Paper I, which has been interpreted as a consequence of different
grains' emissivities \citep{1986ApJ...311..345Z}. The AGB star that
falls outside the expected box with a clear 25 $\rm \mu m$ excess
($[12]-[25]=0.1$) is AFGL 3068 (LL Peg), which is an "extreme carbon
star": very dust-obscured by optically thick shells due to high
mass-loss rate \citep{1992ApJ...391..285V,1997A&A...326..305W}. The
post-AGB object HD\,44179, best known as The Red Rectangle, is also
outlying in this same box with respect to objects in a similar
evolutionary status beyond the AGB, which typically show much larger
25 $\rm \mu m$ excess indicative of detached cold dust (and gas)
envelopes.
 % 2. Observations
% ======================================================= %
\section{Observational results}
\label{section:obs_results}

\subsection{Features in PACS spectra}

\begin{figure*}
\centering
\includegraphics[width=0.98\linewidth]{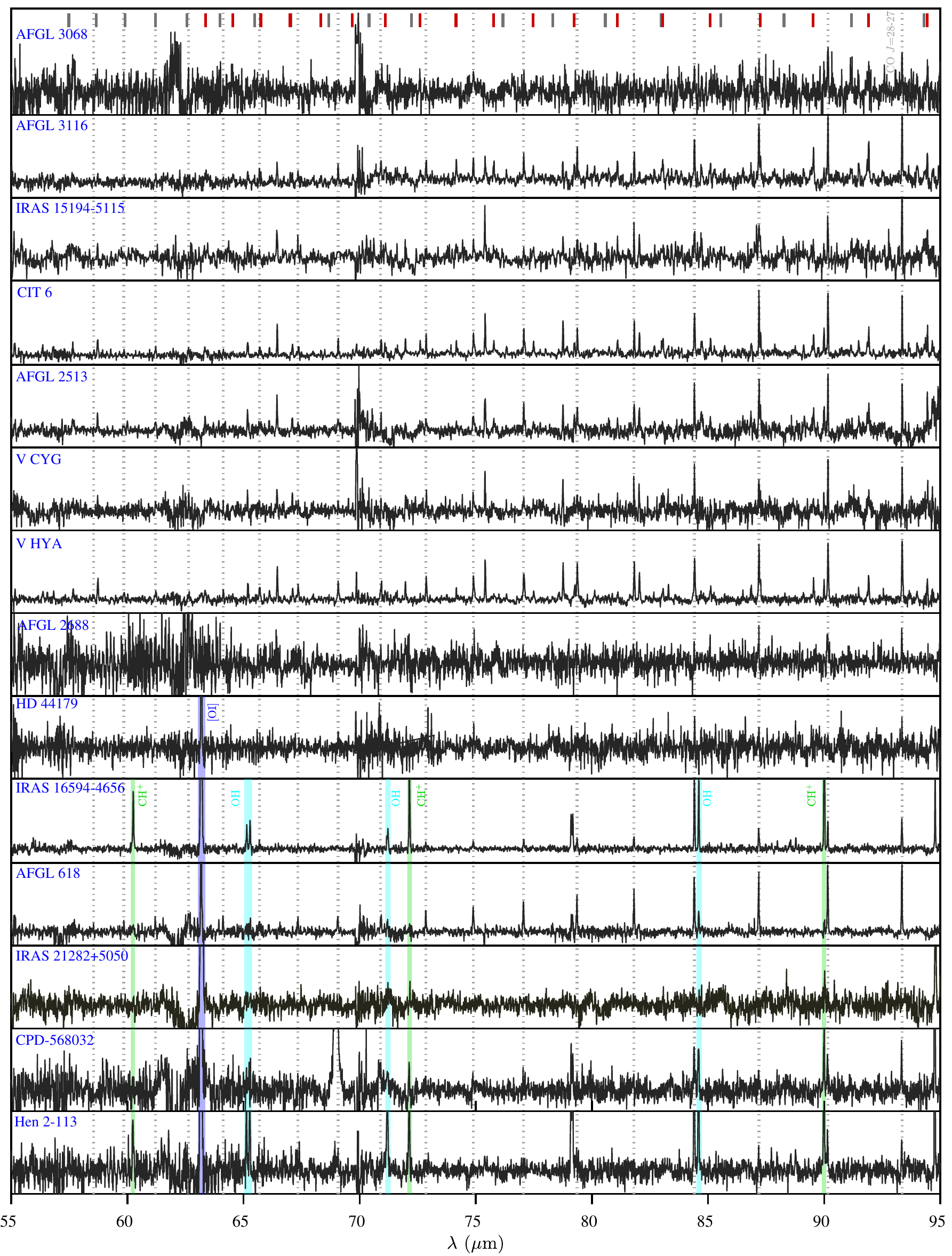}
\caption{Continuum-subtracted PACS spectra of our sample. The vertical segments (at the top) and bars indicate the rotational transitions of $\rm{}^{12}CO$ (light gray, dotted lines), $\rm{}^{13}CO$ (dark gray), HCN (red), CS (yellow), OH (cyan), $\rm CH^{+}$ (green) and forbidden lines (blue) of [\ion{C}{II}] $157.7\,\rm \mu m$, [\ion{O}{I}] $63.2, 145.5\,\rm \mu m$, [\ion{N}{II}] $121.9\,\rm \mu m$. } \label{fig:specs}
\end{figure*}

\begin{figure*}
\centering
\ContinuedFloat
\includegraphics[width=0.98\linewidth]{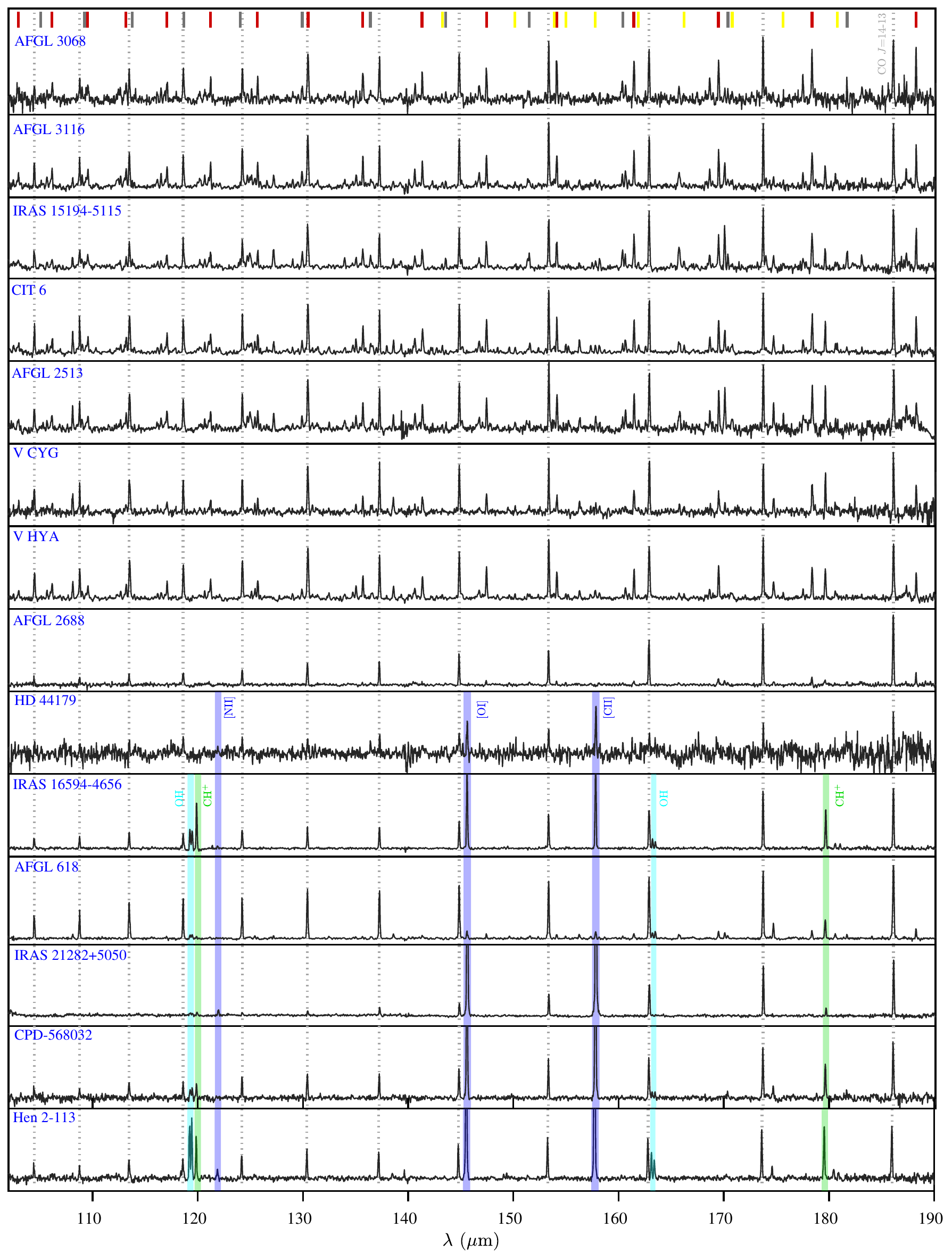}
\caption{Continued. } \label{}
\end{figure*}

\begin{figure*}
\centering
\includegraphics[width=\linewidth]{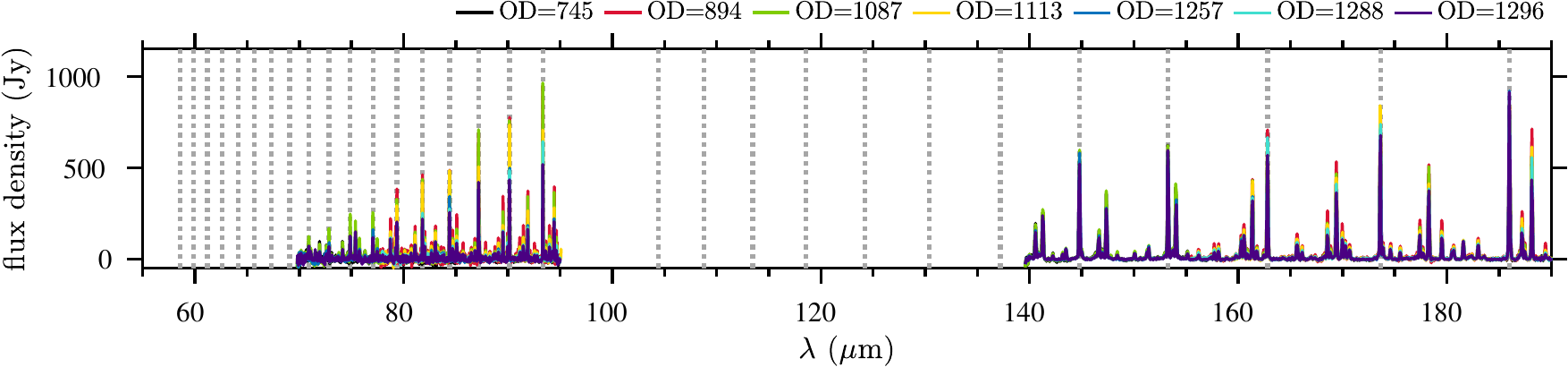} \\
\includegraphics[width=\linewidth]{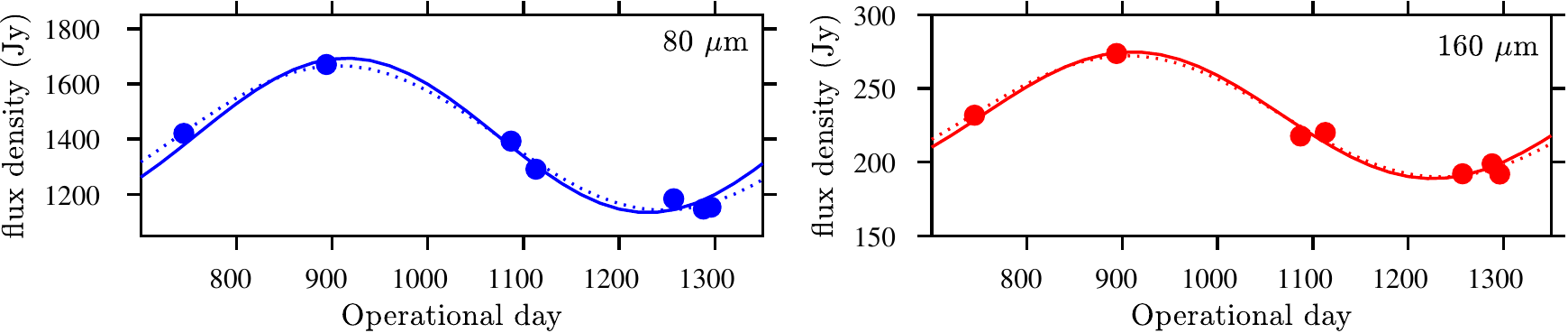} 
\caption{Line and continuum variability of IRC+10216 in the FIR. Top: Continuum subtracted spectra at different epochs (see also Fig.\,\ref{fig:irc10216_split}); the dotted lines sign the CO transitions. Bottom: Sine-wave fit to the continuum variability at two selected wavelengths for a fixed pulsation period of 630 days (solid lines) and free-period fit corresponding to 680 and 660 days for the blue and red dashed curves, respectively.} \label{fig:irc10216specs}
\end{figure*}

The PACS continuum-subtracted spectra are plotted in Fig.\,\ref{fig:specs} sorted in terms of stellar effective temperature (\teff\ increases from top to bottom) and arbitrarily scaled for better comparison of the CO spectra. The continuum was fitted using a non-parametric method after identifying the line-free regions of the spectrum for each target. A brief line-detection statistics summary is given in Table\,\ref{tab:lines}. Broad emission or absorption features due to instrumental artifacts are occasionally visible in some of the targets (e.g. near 62\,$\rm \mu m$). The spectra of IRC+10216 taken at seven different epochs are shown in the top panel of Fig.\,\ref{fig:irc10216specs}, and in more detail in Fig.\,\ref{fig:irc10216_split} in the Appendix.

Straightaway we see significant differences between the spectra of these targets such as the larger line density, resulting from a richer molecular content, in AGB stars compared to post-AGBs and yPNe as expected. In AGB stars the strongest spectral lines are due to CO emission, 
while post-AGBs and yPNe show much more prominent forbidden lines, except for AFGL\,2688 and AFGL\,618. Additional molecular features with intense emission are attributed to rotational transitions of common C-bearing species such as \trecem, HCN, CS, etc.

Not so common in C-rich stars is the presence of OH, yet we identify several OH doublet lines (at 79, 84, 119 and 163\,$\mu$m) in all yPNe and in the post-AGB star IRAS\,16594-4656, with the highest \teff$\sim$10\,000K in its class and where the presence of shocks is plausible. Two additional OH doublets at 65 and 71\,$\mu$m, with high-excitation energy of about 500 and 600\,K, respectively, are also observed in IRAS\,16594-4656 and in the yPNe Hen\,2-113.
%% A reference for the wavelengths and detection of these OH doublets in PACS spectra
%% towards YSO is: Wampfler et al. 2012, Goicoechea et al. 2015 
%% -----------------------------------------------------------
In those five (\teff$>$10\,000\,K) targets, we also identify pure
rotational transitions of CH$^+$, with the $J$=3-2 line at 119.86\,$\mu$m being one of the most clearly detected and best isolated lines\footnote{The lines of CH$^+$ covered by PACS are $J$=2-1, 3-2, 4-3, 5-4 and 6-5 at 179.61, 119.86, 90.02, 72.15 and 60.26\,$\rm \mu m$, respectively.}. The 179.61\,$\mu$m CH$^+$ line could be blended with the water 2$_{12}$-1$_{01}$
transition. The highest-$J$ lines of CH$^+$ are also present in IRAS\,16594-4656 and Hen\,2-113.
%%% J=1-0 ~357microns --- SPIRE
%% J=2-1 179.605
%% J=3-2 119.858
%% J=4-3  90.017
%% J=5-4  72.141
%% J=6-5  60.248
Interestingly, we do not see CH$^+$ in the spectrum of the Red Rectangle, where sharp emission features near 4225\,\AA\ were discovered and assigned to this ion by \cite{1992A&A...261L..25B}.
The detection of CH$^+$ in AFGL\,618 and the more evolved PNe NGC\,7027 had been previously reported by \citet[][\hso/SPIRE]{2010A&A...518L.144W} and \citet[][{\sl ISO} data]{1538-4357-483-1-L65}. As we see below in this section, the targets with CH$^+$ and OH emission also show the strongest atomic/ionic fine-structure lines.

We identify lines due to $\rm H_{2}O$ (orto- and para-) transitions in all AGBs, in the post-AGB AFGL\,2688 and in the yPNe AFGL\,618. The study of water lines in the C-rich AGB stars of our sample (except for AFGL\,2513) was already conducted by \citet{2016A&A...588A.124L} who suggested that both shocks and UV photodissociation may play a role in warm $\rm H_{2}O$ formation. Water lines had also been previously identified in the SPIRE spectrum of AFGL\,618 and AFGL\,2688 \citet{2010A&A...518L.144W}.

The fine-structure lines [\ion{O}{I}] at 63.18\,$\rm \mu m$ and 145.53\,$\rm \mu m$ and [\ion{C}{II}] at 157.74\,$\rm \mu m$ are very prominent in the spectra of all post-AGBs and yPNe in our sample with the exception of AFGL\,2688. The post-AGB object IRAS\,16594-4656 demonstrates clear signs of ionization since it shows a [\ion{C}{II}] line even stronger than that of AFGL\,618, which has a hotter central star (Table B.1). Weak [\ion{N}{II}] 121.89\,$\rm \mu m$ emission is also detected in Hen\,2-113 and IRAS\,21282+5050, and, tentatively in IRAS\,16594-4656 and HD\,44179.

The 69\,$\rm \mu m$ band of forsterite (Mg$_2$SiO$_4$) is visible in \,CPD-$56^{\circ}8032$, Hen\,2-113, and HD\,44179,
as already reported by \citet{2002MNRAS.332..879C} and \citet{2014A&A...565A.109B}. The presence of O-rich dust in such C-rich objects corroborates a mixed chemistry nature.

\subsection{Line flux variability in IRC+10216}

Variability of the continuum at optical and IR wavelengths due to stellar pulsations is a common property of AGB stars. We present PACS spectra at multiple epochs of the Mira-type C-rich star IRC+10216 displaying periodic variations both in the continuum and in the lines (see Fig.\,\ref{fig:irc10216specs} and Fig.\,\ref{fig:irc10216_split}).

\citet{2041-8205-796-1-L21} first reported on the discovery of strong intensity variations in high-excitation lines of abundant molecular species towards IRC+10216 using \hso/HIFI and IRAM 30m data. Line variability was attributed to periodic changes in the IR pumping rates and also possibly in the dust and gas temperatures in the innermost layers of the CSE. From the analysis of a 3 yr-long monitoring of the
molecular emission of IRC+10216 with \hso~(including HIFI, SPIRE and PACS data), \citet{2015ASPC..497...43T} concluded that intensity changes of CO lines with rotational numbers up to $J$=18 are within the typical instrument calibration uncertainties, but in the higher PACS frequency range ($J\geq28$), line strength variations of a factor
$\ga$1.6, and scaling with $J$, were found. More recently, \cite{2017ApJ...845...38H} reported 5\%-30\% intensity variability of additional mm lines with periodicities in the range 450-1180 days.

In section \ref{section:irc10216_var} we study in greater detail what is the impact of CO line variability on the estimate of \trot\ and \mass\ from the rotational diagram analysis using PACS data for different epochs. 
% --------------------------------------------------------------%
\subsection{CO line fluxes}
\label{section:fluxes}

We now focus on the purely rotational spectrum of CO in the ground vibrational state ($v$=0), which we use to study the physical properties of the warm regions of the molecular CSEs of our targets. In the PACS range one can potentially find very high CO rotational transitions, from $J=14-13$ (\eu$\sim$581\,K) to $J=45-44$ (\eu$\sim$5688\,K). The lack of data between 95 and 101 $\rm\mu m$ means we cannot detect the transitions $J$=26-25 and $J$=27-26. In the case of IRC\,+10216, the much larger gap in the PACS coverage ($\sim$95-140\,$\mu$m, Fig.\,\ref{fig:irc10216specs}) prevents detection of CO transitions with upper-level rotational number between \ju=19 and \ju=27. %The yPNe AFGL\,618 is the object with the highest-$J$ CO transition detected.
 
The line fluxes are given in Table\,\ref{tab:fluxes}. The quoted uncertainties correspond to the propagated statistical errors and do not contain absolute flux calibration errors (typically of $\sim$15\%-20\%) or underlying continuum subtraction uncertainties.

As expected, the resolving power of PACS ($\sim$80-300\,\kms) does not allow to spectrally resolve the CO profiles. This is true not only for AGB CSEs with full-widths-at-half-maximum (FWHM) similar to or smaller than the terminal expansion velocity of the envelopes (FWHM$\sim$10-25\,\kms, Table \ref{tab:2}), but also for post-AGB objects and yPNe, even in targets that are known to have fast ($\approx$100\,\kms) molecular outflows like AFGL\,618 \citep[e.g.][]{2010A&A...521L...3B}. For this reason we measured CO fluxes by simply fitting a Gaussian function to the PACS lines.

Due to insufficient spectral resolution some reported line fluxes are affected by line blend. These are identified with asterisks in the tables and figures. Some of the well-known line blends are CO\,$J$=30-29 with HCN\,$J$=39-38 and CO\,$J$=20-19 with HCN\,$J$=26-25 at 87.2 and 130.4\,$\mu$m, respectively. Also, the CO transitions $J$=21-20 and $J$=22-21 are blended with \trecem\ $J$=22-21 and $J$=23-22 at $124.2~\rm \mu m$ and $118.6~\rm \mu m$, respectively.

\begin{figure}[t]
\centering
\includegraphics[width=0.8\linewidth]{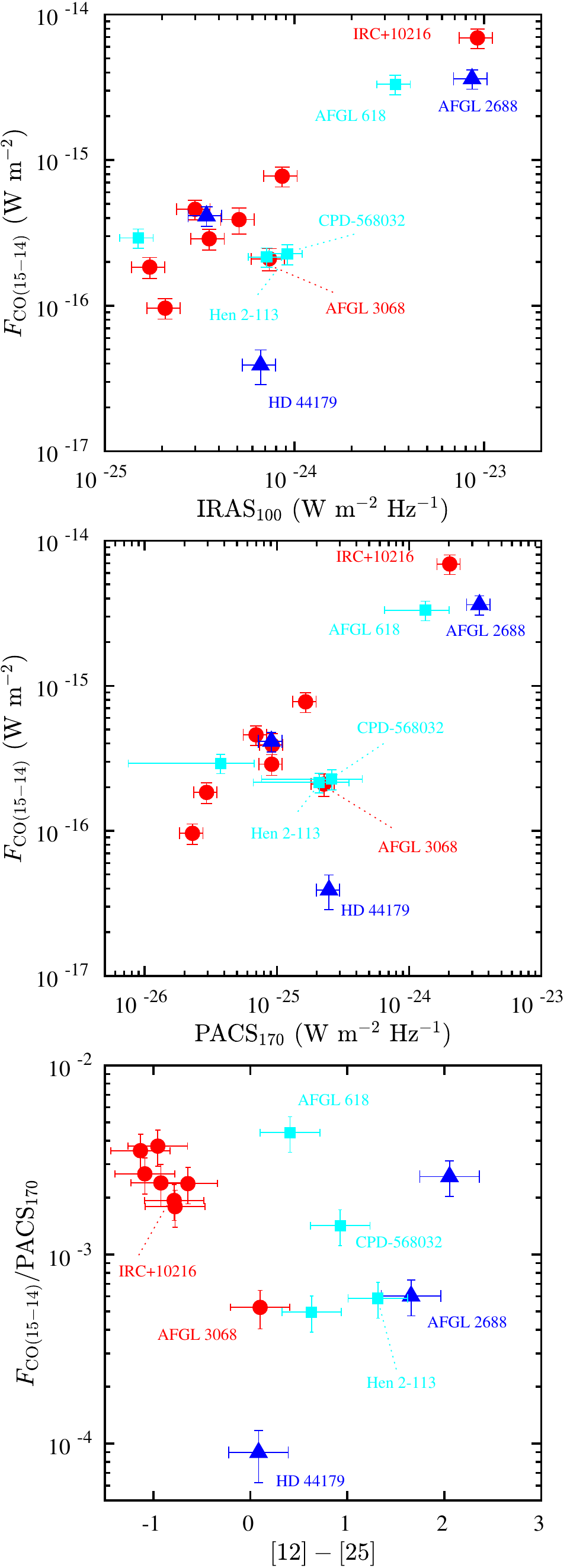}
\caption{Correlation between CO line flux and continuum intensities. Top: CO $J=15-14$ vs IRAS 100\,$\rm \mu m$ flux. Middle: CO $J=15-14$ vs PACS 170\,$\rm \mu m$ continuum flux. Bottom: continuum normalized CO $J=15-14$ flux vs IRAS [12]-[25] color. The symbols and colors are the same as in Fig.\,\ref{fig:IRAS}.} \label{fig:line_cont} 
\end{figure}

Figure\,\ref{fig:line_cont} compares the integrated flux of the CO\,($J$=15-14) line ($F_{\rm CO\,(15-14)}$) with the IRAS 100\,\micron\ flux (IRAS$_{\rm 100}$), the PACS continuum at 170\,\micron\ (PACS$_{\rm 170}$), i.e.\, near the CO\,($J$=15-14) line, and the [12]-[25] IRAS color. The $J$=15-14 transition is a strong, non-blended line, detected towards all of our targets, and it was also used by \citet{2016A&A...588A.124L} who analyzed PACS data of most of the AGBs here studied, which facilitates comparison.

As for the O-rich sample (paper I), there is a positive correlation between the CO line strength and the IRAS$_{100}$ and PACS$_{170}$ continuum fluxes. The correlation between the CO\,$J$=1-0 line and the IRAS fluxes of evolved stars of various chemical types had been previously reported \citep{1987A&A...183L..13O,1988A&A...196L...1O,1992A&A...257..701B}.
% We found a clear correlation with Pearson's correlation
%coefficient equal to $r=0.985$ (two-tailed $p<1\times10^{-4}$).
%Of course, in addition to an intrinsic relation between these two observational parameters (CO-line and IR-continuum flux), there is also a dependence with the distance to the sources, which is partially responsible of the observed relation. 
%The scatter of the points is significantly larger in the $F_{\rm CO\,15-14}$ vs. IRAS$_{100}$ distribution, which partially reflects the variability of the pulsating AGB stars given that PACS and IRAS observations were obtained at different epochs, as well as different instrument calibration uncertainties, background contamination sources, beam size and pointing. These effects are discussed for the whole sample of the THROES catalogue in \citet{2017arXiv171105992R}. 
We also confirm the anticorrelation between the line-to-continuum ($F_{\rm CO\ 15-14}$/PACS$_{\rm 170}$) ratio and the IRAS [12]-[25] color (both distance-independent) for the AGB stars, which was noted in Paper I. For the more evolved targets the trend is not so obvious, but the sample size is small. We also see that, in general, the ratio between the molecular emission and the dust emission is higher in less evolved objects than in the most evolved ones, which could be partially attributed to more prominent CO photodissociation as the objects evolves along the AGB-to-PNe track. AFGL 618 and IRAS\,16594-4656 are two clear outliers in this relation since they show a line-to-continuum emission ratio as large as that of the AGB class.  The Red Rectangle (HD\,44179) is well isolated in all the panels due to its comparatively weak CO emission and low CO-to-dust ratio. This surely reflects the different nature of this object with respect to the rest of post-AGB and yPNe in our sample, which is well known from previous works. The Red Rectangle belongs to a special class of post-AGB objects with relatively weak CO emission coming from large ($\sim$1000-2000\,AU) circumbinary rotating disks, with very prominent IR emission by warm/hot dust in the disk, but lacking massive molecular outflows found in many other evolved stars \citep[e.g.,][and references therein]{2016A&A...593A..92B}.

% =========================================================== %
% =========================================================== %
\section{Rotational diagram analysis}
\label{section:RD}

\begin{figure*}[!htp]
%\centering
	\begin{tabular}{ccc} 
		\includegraphics[valign=t]{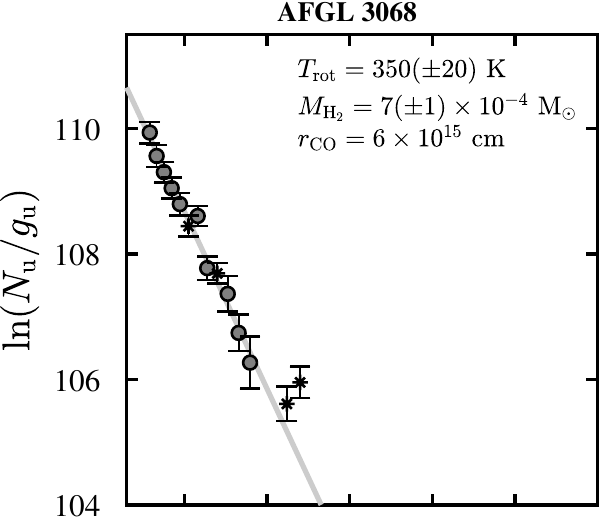} & \hspace{-0.3cm} \includegraphics[valign=t]{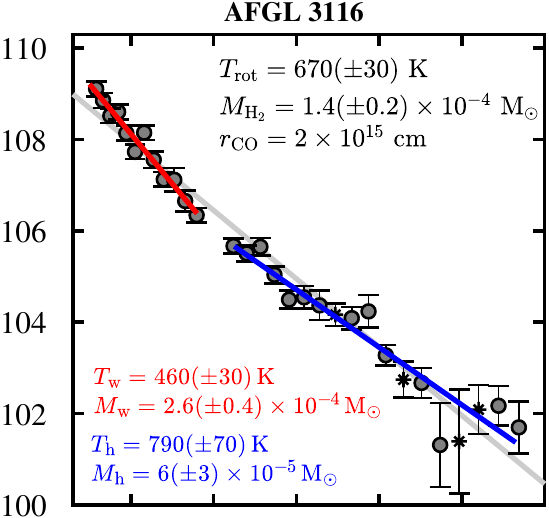} & \hspace{-0.3cm} \includegraphics[valign=t]{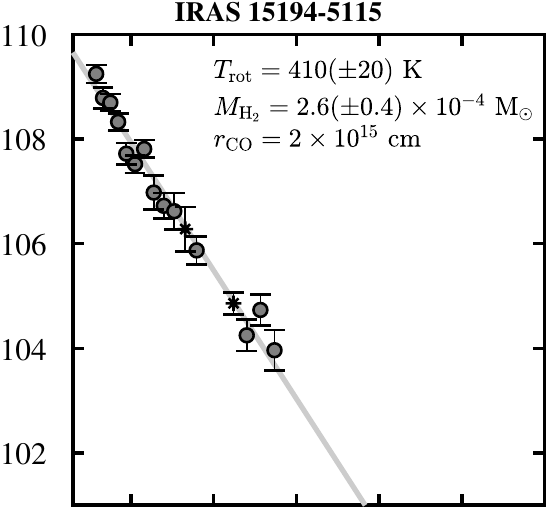}\\
        \includegraphics[valign=t]{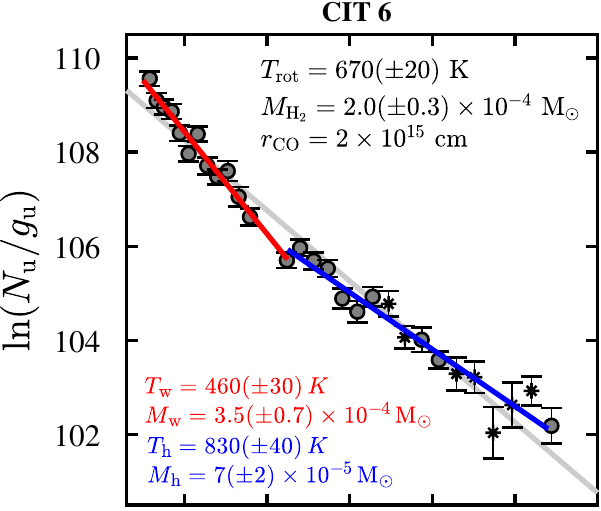} & \hspace{-0.3cm} \includegraphics[valign=t]{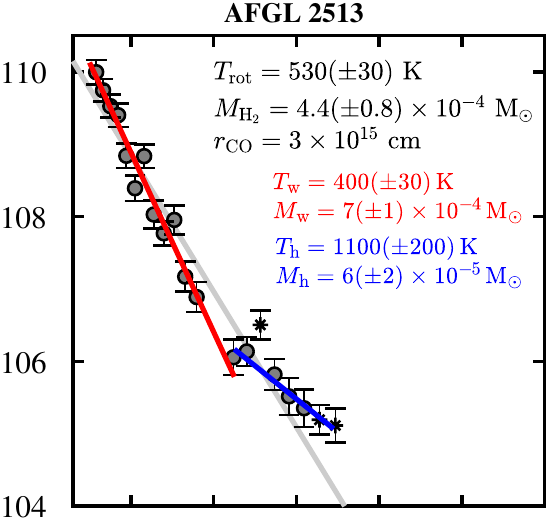} & \hspace{-0.3cm} \includegraphics[valign=t]{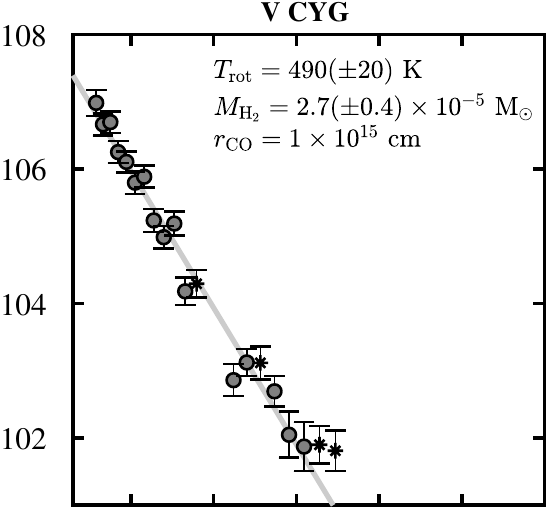}\\ 
        \includegraphics[valign=t]{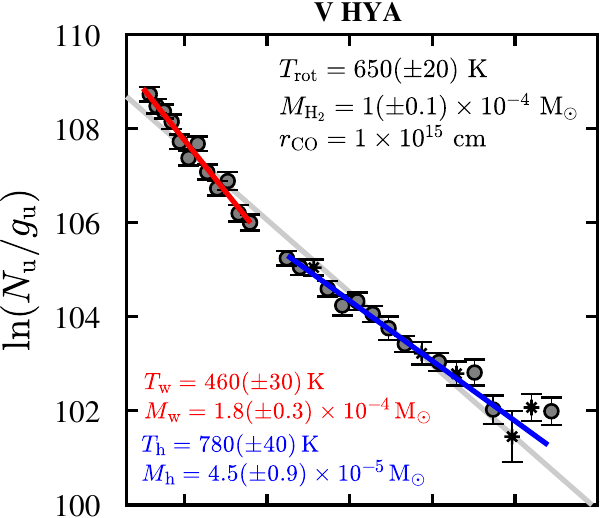} & \hspace{-0.3cm} \includegraphics[valign=t]{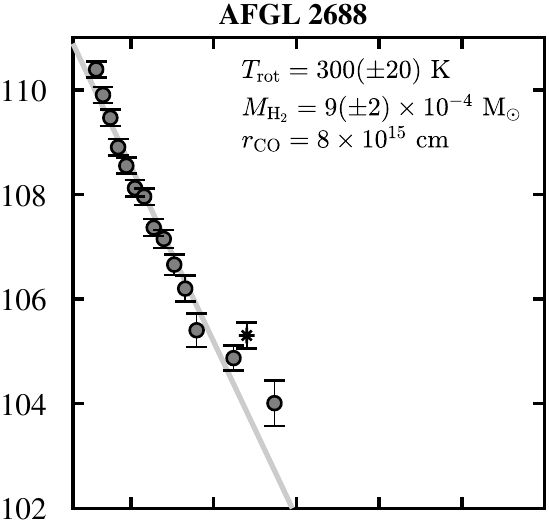} & \hspace{-0.3cm} \includegraphics[valign=t]{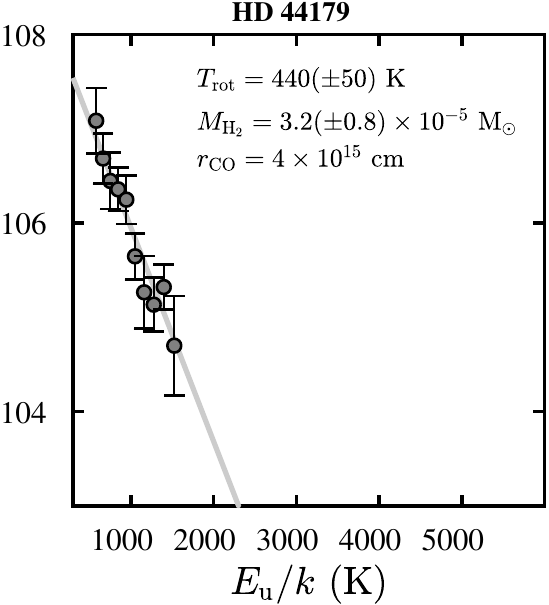}\vspace{-0.8cm} \\ 
        \includegraphics[valign=t]{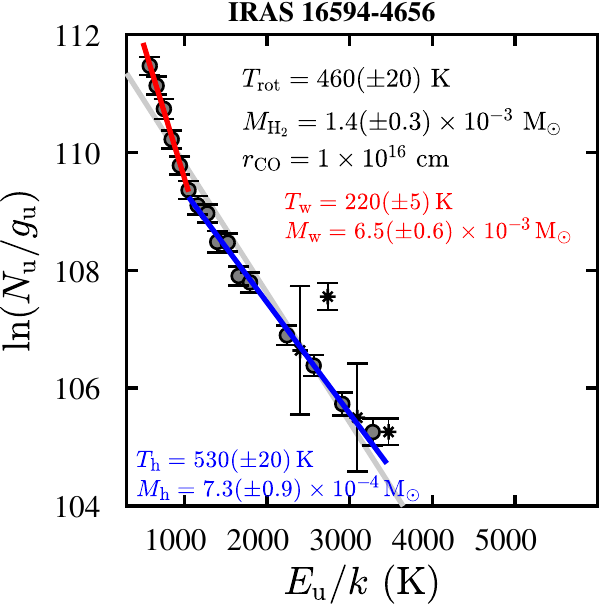} & \hspace{-0.3cm} \includegraphics[valign=t]{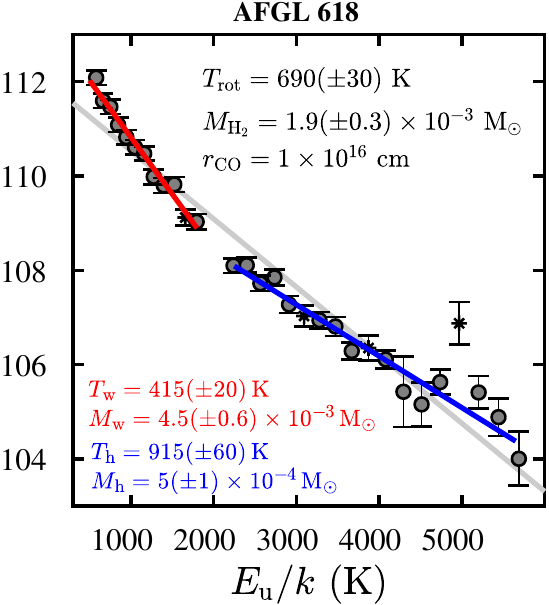} & 
	\end{tabular}
    \caption{Rotational diagrams of the CO molecule. The gray line correspond to a single least-squares fit to the full range of transitions from where a rotational temperature, \trot, and total gas mass, \mass, is computed. The  characteristic radius of the CO-emitting volume ($r_{\rm CO}$) adopted is indicated. The red and blue lines correspond to a two-component model consisting of a "warm" and "hot" region, respectively. Asterisks mark line blends that were excluded in the fit.}  \label{fig:rots}
\end{figure*}
\begin{figure*}[!htp]
\ContinuedFloat
	\begin{tabular}{ccc} 
		\includegraphics[]{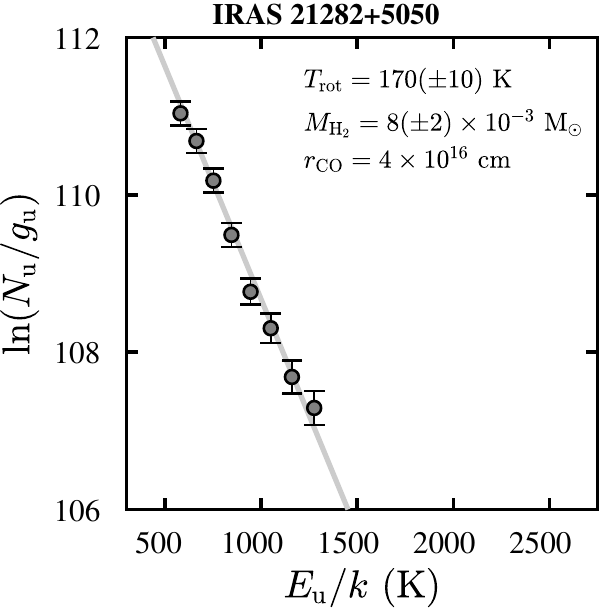} & \hspace{-0.2cm}\includegraphics[]{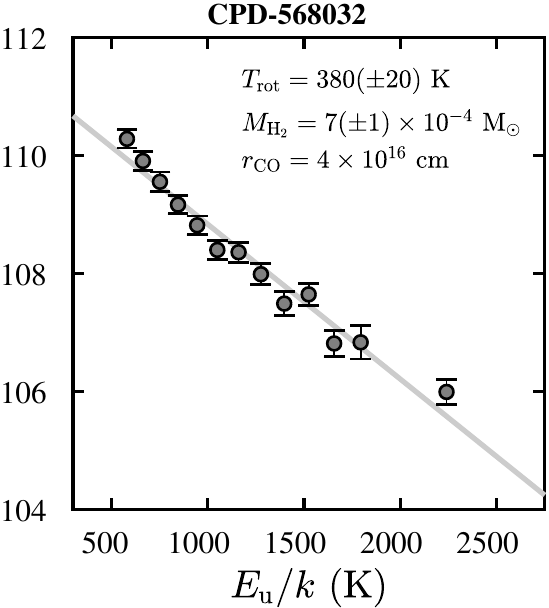} & \hspace{-0.2cm}\includegraphics[]{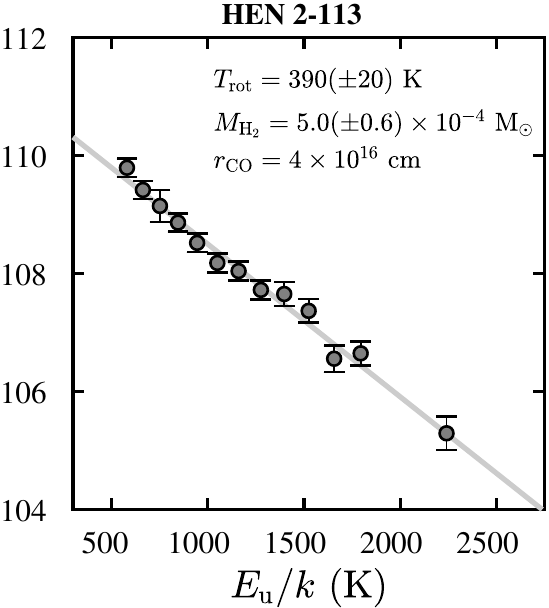}\\
	\end{tabular}
    \caption{Continued.}  %\label{fig:rots2}
\end{figure*}

\begin{figure}
\includegraphics[width=\linewidth]{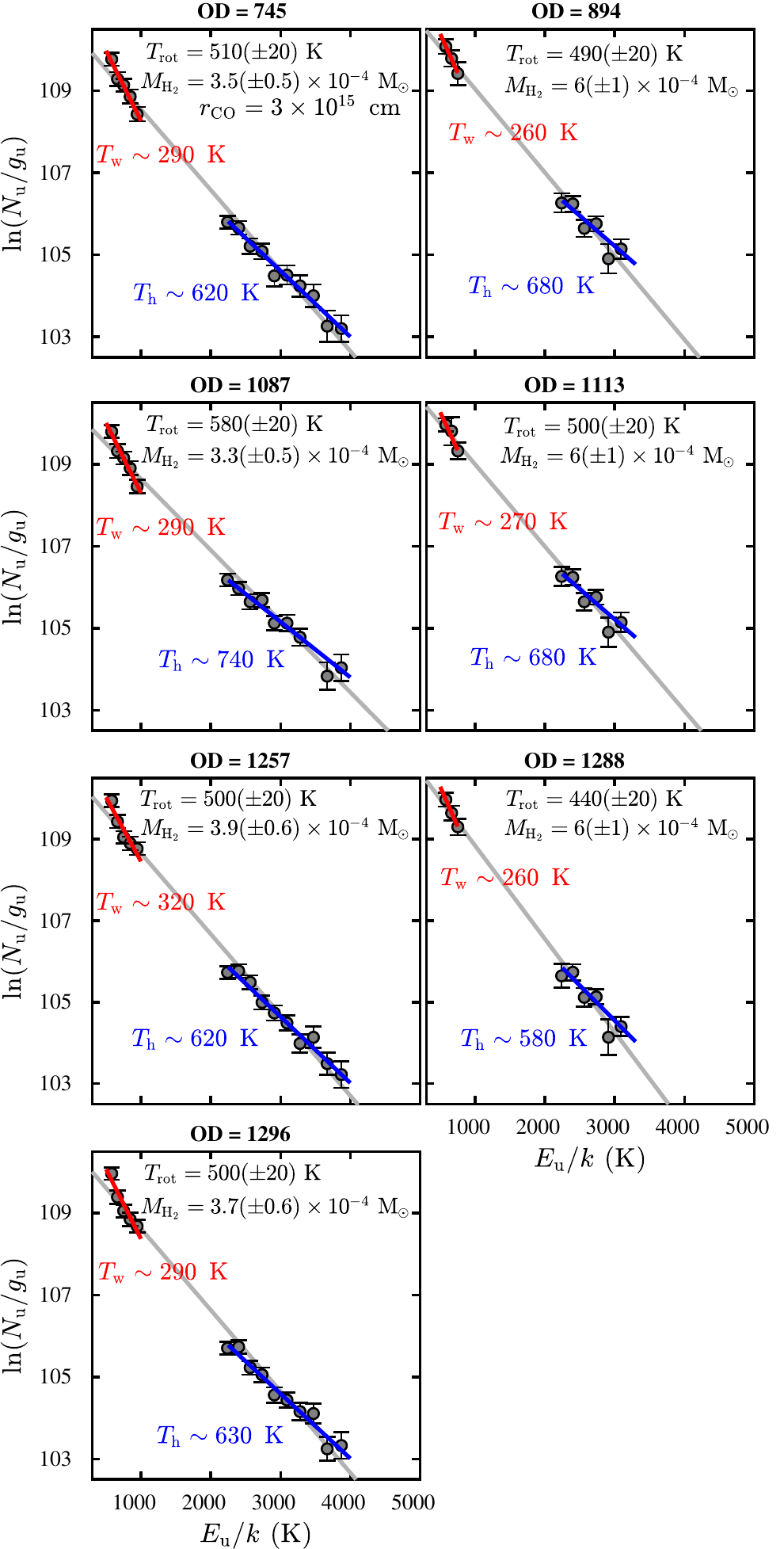} 
\caption{Rotational diagram of the CO molecule in IRC+10216 at different epochs. Analogous to Fig.\,\ref{fig:rots}.} \label{fig:irc10216_rds}
\end{figure}

Following the same approach of Paper I, we have used the well-known and widely employed rotational diagram (RD) technique \citep[e.g.][]{1999ApJ...517..209G} to obtain a first estimate of the (average) excitation temperature (\trot) and total mass (\mass) of the warm inner layers of the molecular envelopes of our sample.  
A canonical opacity correction factor, $C_\tau$, as defined by \citet{1999ApJ...517..209G}, has been included to take into account moderate optical depth effects (\S\,\ref{appendix:opCorrection}). We refer to Paper\,I for a more detailed description of the method.

% ================================================ %
% ================================================ %
\subsection{The characteristic size of the CO-emitting layers}
\label{section:rco}

To compute the optical depth of the line and to make the corresponding \ctau\ correction, we need an estimate of the column density of CO, which is computed in a simplified manner dividing the total number of CO molecules (\nco) by the projected area of the CO-emitting volume on the sky. The {\sl characteristic} size of the envelope regions where the CO PACS emission is produced (\rco) is one of the main sources of uncertainty since, except for a few targets (see below), these high-$J$ CO-emitting layers are unresolved by PACS. Therefore, the value of \rco\ needs to be adopted based on several criteria. These criteria are described in much detail in paper I, Appendix B. In the following, we provide a brief summary of the general method and provide additional arguments particularized to the C-rich targets here under study.

For AGB CSEs, a first estimate of the size of the CO-emitting volume can be derived from the envelope temperature structure, $T(r)$, estimated from detailed non-LTE molecular excitation and radiative transfer (nLTEexRT) calculations in the literature. These have been done for many AGB CSEs using low-$J$ CO transitions (typically \ju$\la$6), and, for a few cases, also using certain CO transitions from higher $J$ levels \citep[see e.g.,][]{2010A&A...518L.143D,2014A&A...561A...5K,2014A&A...569A..76D,2016A&A...591A..44M,2018A&A...609A..63V}. As deduced from these studies, the gas temperature is approximately 1000-2000\,K close to the dust condensation radius ($\sim$5-15 \rs), and decreases gradually towards the outermost layers approximately following a power-law\footnote{See also the temperature profiles for the O-rich AGBs in Fig.\,B.1 of Paper I.} of the type $\sim$1/r$^{\alpha}$, with $\alpha \sim$0.5-1.0. Such models also exist in literature for three targets in our C-rich sample: IRAS\,15194-5115, CIT\,6 and IRC+10216 \citep{1999A&A...345..841R,2002A&A...391..577S,2012A&A...539A.108D}.
According to this, the high-excitation transitions observed with PACS require relative proximity to the central star and, in particular, we anticipate that regions with a few hundred kelvin, as deduced from our RDs (Figs.\,\ref{fig:rots} and \ref{fig:irc10216_rds}), should not extend much farther than a few $10^{15}$\,cm ($\rm \la10^2\,R_{\star}$) in AGB stars.

An additional constraint on the radius can be imposed from the fact that the deepest layer traced by the observed CO emission must be such that $\tau<1$ because of the almost null escape probability from deeper, very optically thick regions (\S\,\ref{appendix:opCorrection}). For all our AGB CSEs, we have explored a range of radii around 1\ex{15}\,cm (Fig.\,\ref{fig:opacity}). We found that values around \rco$\sim$[1-4]\ex{15}\,cm result in line optical depths close to, but smaller than unity (typically $\tau_{J=14\raw13}\sim$0.5-0.9) that yield moderate $C_\tau$ opacity correction factors for lines with \ju$<$19 and negligible for higher-$J$ transitions. For \rco$<$1\ex{15}\,cm, the opacity of the CO\,$J$=14-13 line, which is the optically thickest transition in our sample, becomes larger than 1 in all our targets (also including post-AGBs and yPNes).

We have checked that the range of plausible radii found for the AGB CSEs in our sample, \rco$\sim$[1-4]\ex{15}\,cm, is consistent with the upper limits to the size of their envelopes deduced from the PACS spectral cubes and/or photometric maps by \citet{2016MsT..........1D} and with any other information on the molecular envelope extent from the literature. IRC+10216 is the only AGB CSE that is partially resolved in the PACS cubes, and it is also the closest in our sample. The PACS cubes show a slightly extended source (with a half-intensity size 4$\sigma$ above the instrument PSF in both bands) that implies a deconvolved Gaussian radius of about 2\arcsec, that is $\sim4\times10^{15}$\,cm at $d$=150\,pc. This value is in good agreement with the lower limit to \rco\ needed to satisfy the $\tau_{J=14\raw13}<$1 criteria in this target, \rco$\sim$2\ex{15}\,cm. In this case, we then rather confidently use an intermediate value of \rco=3\ex{15}\,cm.

Contrary to AGBs, for post-AGBs and yPNe there are no model temperature profiles in the literature of the molecular gas in the CSEs (except for the rotating, circumbinary disk of the Red Rectangle, \S\,\ref{Section:individual}).

The range of representative radius adopted for post-AGBs and yPNe is \rco=[0.4-4]\ex{16}\,cm (Table\,\ref{tab:rot}) based on a moderate opacity criteria, the extent of the emission in the PACS cubes and photometric maps, and on additional information on the extent of the intermediate-to-outer molecular envelope from the literature. In particular, in AFGL\,2688, a deconvolved diameter of $\sim$4\arcsec\ in the PACS blue band suggests that \rco$\la1\times10^{16}$\,cm (at $d$=340\,pc). Previous CO\,$J$=2-1 mapping observations identified a compact shell of radius $\sim$2\arcsec\ ($\sim$5\ex{15}cm) around the center of the nebula \citep{2000A&A...353L..25C}. Since the CO\,$J$=14-13 emission is optically thick at \rco$\la$6\ex{15}\,cm (as shown in Fig.\,\ref{fig:opacity}), we adopt as representative radius an intermediate value of \rco=8\ex{15}\,cm. HD\,44179 (The Red Rectangle) is a point-source in the PACS photometric maps, therefore, for a distance of 710 pc, $r_{\rm CO}$ should be of the same order as in AFGL\,2688, which is roughly consistent with interferometric observations \citep[e.g.,][]{2016A&A...593A..92B}.  For IRAS\,16594-4656, only a very loose upper limit to the radius of \rco$<$3\ex{16}\,cm is inferred from optical images and H$_2$ emission maps in this object \citep[e.g.,][]{2008ApJ...688..327H}. We explored the range, $r_{\rm CO}\sim0.8-2\times10^{16}$, similar to AFGL\,2688, and adopted as a reference value the midpoint value where $\tau_{J=14\raw13}\sim$0.7.

All yPNe in our sample are point-sources in the PACS spectral cubes, which means that the upper limit to the radius is of about
\rco$\la$1\arcsec$\sim$[1-2]\ex{16}\,cm for AFGL\,618, and $\sim$[4-6]\ex{16}\,cm for the rest. They are known to have central \ion{H}{ii} regions that have recently formed as the star has become progressively hotter along the PNe evolution. Because the CO envelope surrounds the ionized nebula, a lower limit to \rco\ can be established from the extent of the latter. Taking this into account, we set a representative radius to \rco=1\ex{16}\,cm for AFGL\,618
\citep{2017arXiv170401773S,2013ApJ...777...37L}, and \rco=4\ex{16}\,cm for the rest \citep[see e.g.,][]{2015MNRAS.449L..56D,2010A&A...523A..59C}. 

Given the uncertainty in \rco, we have systematically explored a range of radii around optimal/plausible values of \rco\ to asses the impact of this parameter in our results (Fig.\,\ref{fig:opacity}). The opacity correction increases the smaller \rco~is, therefore the slope and y-intercept of the RD increase as a direct result of the frequency-dependence of C$_\tau$ (\S\ \ref{appendix:opCorrection}), which results in lower values of \trot\ and larger values of $N_{\rm CO}$, thus \mass. This also means that, in practice, only the lowest-frequency-points are affected by C$_\tau$, while the highest-frequency transitions (i.e., highest excitation energies) are unaltered regardless of the radius that we chose within the reasonable constraints that we have put.
% ------------------------------------------------- %
\subsection{Non-LTE effects}

In Paper I we examine and discuss extensively the impact of non-LTE excitation effects (if present) on the values of \trot\ and \mass\ derived from the RDs in a sample of 26 non C-rich evolved stars with mass-loss rates in the range \mloss$\sim$2\ex{-7}-1\ex{-4}\,$\rm M_{\sun}\,yr^{-1}$. The C-rich targets studied here have on average larger mass-loss rates than those in Paper I, therefore, using a similar reasoning, the CO population levels are also most likely close to thermalization in the inner dense regions of the CSEs under study.

This is further supported by nLTEexRT computations of a selection of high-$J$ CO transitions observed with PACS (from \ju=14 to 38) by \cite{2016A&A...588A.124L}. Their sample included all our targets except for AFGL\,2513 and IRC+10216. These authors conclude that over
a broad range of mass-loss rates (\mloss$\sim$10$^{-7}$-2\ex{-5}\,$\rm M_{\sun}\,yr^{-1}$) the CO molecule is predominantly excited through collisions with H$_2$, with a minor effect of FIR radiative pumping due to the dust radiation field. The role of dust-excitation on FIR CO lines was also investigated by \citet{2002A&A...391..577S} and found to be of minor importance for AGBs with typical mass-loss rates of $\sim10^{-5}\rm \,M_{\sun}\,yr^{-1}$. We assess the FIR pumping effect further in Section\,\ref{section:irc10216_var},
where we study multi-epoch RDs of IRC+10216, which is a source with well-known CO line variability.

We stress that even under non-LTE conditions, for a simple diatomic molecule like CO, the RD method provides a reliable measure of the total mass within the emitting volume. Although \trot\ may deviate from the kinetic temperature in regions where the local density is lower than the critical densities of the transitions considered (\nc$\sim$5\ex{5}-3\ex{6}\,\cm3, for \ju=14 and 27, respectively, and \nc$\approx$10$^6$-10$^7$\,\cm3 for \ju$>$27), it does describe quite precisely the molecular excitation, i.e.\, the real level population. Therefore, the total number of emitting molecules (and, thus, the mass) is quite robustly computed by adding up the populations of all levels. This is also supported by the good agreement (within uncertainties) between the mass-loss rates derived from this (and other works) using similar LTE approximations, and those obtained from nLTEexRT models (including PACS lines for a few targets) -- see paper I and Section\,\ref{section:Mloss}.

As discussed in paper I, conceivable LTE deviations would have its largest impact on the excitation temperature of the hot component, since high-$J$ levels have the lowest critical densities (see above). If this is the case, in low mass-loss rate objects, the value of \trot\ deduced for the hot component could deviate from the temperature of the gas and approach to that of the dust within the CO-emitting volume. We note that, in any case, the gas and dust temperatures, although not equal, are not excessively divergent in the  warm envelope regions around $\sim$10$^{15}$\,cm under study \citep[e.g.,][]{2014A&A...569A..76D,2002A&A...391..577S}.

% ===================================================== %
% ===================================================== %
\section{Results}
\label{section:results}

\subsection{Gas temperatures and masses}

\begin{figure}[t]
\centering
	\includegraphics[width=0.88\linewidth]{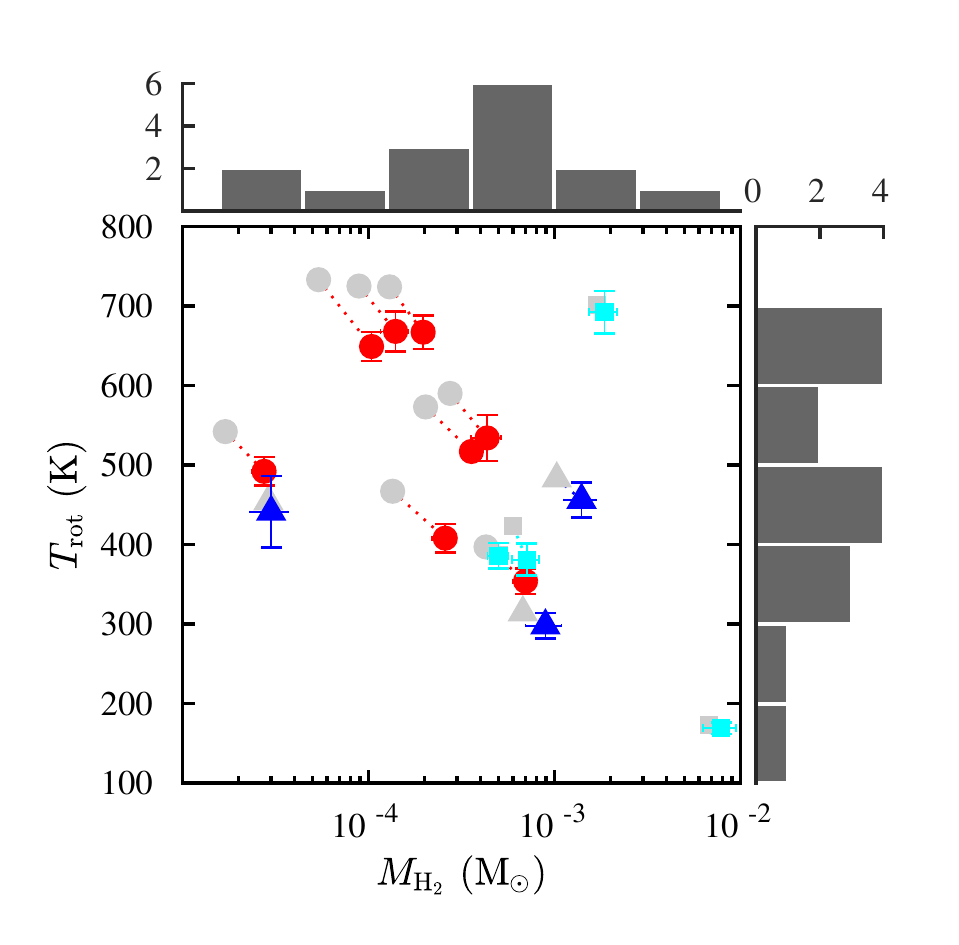} \\ \hspace{-1.15cm}
	\includegraphics[width=0.70\linewidth]{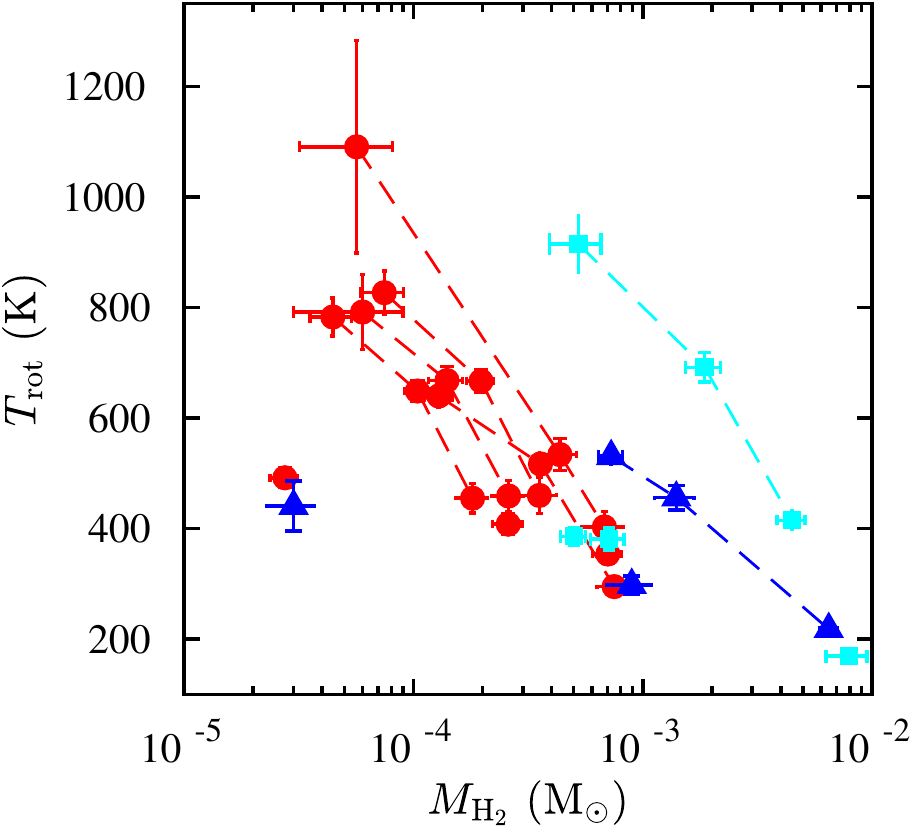}
\caption{Summary of the RD results. Top: The colored symbols correspond to the opacity corrected values (single fit) which are connected by dotted lines to the corresponding uncorrected values in gray; on the sides we show the histograms of temperature and mass of the single fits; in the case of IRC+10216 we show weighted averages of four observations. Bottom: temperature and mass of the warm and hot components connected by dashed lines for the same target (if applicable). } \label{fig:props} 
\end{figure}

The opacity-corrected RDs of the CO molecule are plotted in Fig.\,\ref{fig:rots} and in Fig.\,\ref{fig:irc10216_rds}, where multi-epoch RDs for IRC+10216 are separately shown, together with the best-fit parameters using a single-temperature component (gray line) and double-temperature component (red and blue lines). For 8 out of 14 sources, the RDs include CO transitions with upper-level energies that range from \eu$\sim$580\,K to \eu$\la$2000-3000\,K. For 6 targets, CO transitions with upper-level energies of up to \eu$\sim$5000\,K are also detected. Our temperatures for the group of AGBs are consistent with the ones reported in \citet{2018arXiv180803467N} within the uncertainties, although for many AGBs we find lower values because of the opacity correction term that we have included.

There are cases where it is clear by the analysis of residuals that a single straight line does not fit the entire range of excitation energies. That is the case of V\,Hya, IRAS\,16594-4656, AFGL\,3116, CIT\,6 and AFGL\,618.
Because it is not always clear by eye whether the slope changes and at which point it occurs, we used the Bayesian information criterion (BIC) to help us deciding where to split the diagram and to quantify significance. This is explained in the appendix \ref{appendix:multiFit} and illustrated in the supplementary Fig.\,\ref{fig:stats2}.

We provide the fitting parameters in Table \ref{tab:rot} corresponding to a single-fit and a double-component fit in targets where a single line does not equally fit all data points. As in Paper I, we call these two components "warm" and "hot" with $T_{\rm w}<T_{\rm single}<T_{\rm h}$. Their mean temperatures are $\overline{T}_{\rm w}\sim$ 400\,K and $\overline{T}_{\rm h}\sim$ 820 K respectively. The corresponding masses are $M_{\rm w}$ and $M_{\rm h}$ with the former being 4-10 times larger than the latter.
We find single-fit rotational temperatures in the range \trot$\sim$\,200-700\,K with some of the post-AGBs and yPNe being the targets with the coolest gas, except for the yPNe AFGL\,618 which has the largest rotational temperature in our sample similar to that of the warmest AGB CSEs. 

The total number of CO molecules is in the range \nco$\sim10^{49}-10^{51}$, resulting in column densities of \ncol$\sim10^{16}-10^{19}$\,$\rm cm^{-2}$ for the adopted radii (Table \ref{tab:rot}). To estimate the total gas mass from CO we assumed the same fractional abundance $X_{\rm CO} = 8\times10^{-4}$ \citep[e.g.][]{2006A&A...450..167T} with respect to $\rm H_2$ for all targets. The single-fit values of the total mass of the CO-emitting volume range between \mass$\sim$3$\times$10$^{-5}$\,\msun\ (V\,Cyg and HD\,44179) and $\sim$8$\times$10$^{-3}$\,\msun\ (IRAS\,21282+5050), with a median value of \mass$\sim$4$\times$10$^{-4}$\,\msun.

Figure\,\ref{fig:props} shows the single-fit temperature versus mass for the opacity corrected diagram (colored symbols) and uncorrected (gray symbols). The opacity-correction results in changes in \trot\ of 10-15\% in AGBs, and lower than 5\% in post-AGBs and yPNe. 
In mass this typically corresponds to 60\% in AGBs and lower in the post-AGBs and AFGL\,618 ($<10$\% in $M_{\rm H_2}$), and negligible in the other three yPNes. Figure \ref{fig:opacity} shows that $\tau_{J=14-13}$ is close to unity but it quickly falls off with increasing $J$ meaning that $C_\tau \rightarrow 0$ for $E_{\rm u}/k > 2000$\,K, and that $M_{\rm h}$, in particular, is not underestimated by opacity effects.

As in Paper I, we find an anti-correlation between \mass\ and \trot, especially if we consider only the group of AGBs (Fig. \ref{fig:props}). In general, post-AGBs and yPNe have the the highest masses. One clear exception to trend is The Red Rectangle, which is one of the least massive targets (with only a few $\sim$10$^{-5}$\,\rm \my) in contrast to the rest of post-AGBs and yPNe. This is not surprising given the nature of this object, which is the prototype of a special class of post-AGB objects with hot rotating disks and tenuous winds very different from the massive and fast (high-momentum) outflows of standard pre-PNe \citep[][and references therein]{2016A&A...593A..92B}.

We also investigated the correlation between the CO $J=15-14$ flux and the gas mass and find that the strongest CO emitters have tendentially more massive envelopes, although there is a significant scatter (Fig.\,\ref{fig:orichcomp}).
Also, the targets with the highest temperatures have the highest line-to-continuum (F$_{\rm CO\,15-14}$/PACS$_{\rm 170}$) ratios. IRAS\,21282+5050, which has the most massive warm envelope in our sample, defies this trend since it shows relatively strong CO $J=15-14$ emission, although its \trot$\sim170$\,K is the lowest in the sample. The Red Rectangle appears isolated in a region of rather weak CO emission in spite of relatively high temperatures ($\sim$400-500\,K). We compare these results to Paper I in Section\,\ref{Section:comparison}.

% ===================================================== %
% ===================================================== %
\subsection{Mass-loss rates}

\begin{figure}[]
\centering \hspace{+0.1cm}
\includegraphics[width=0.9\linewidth]{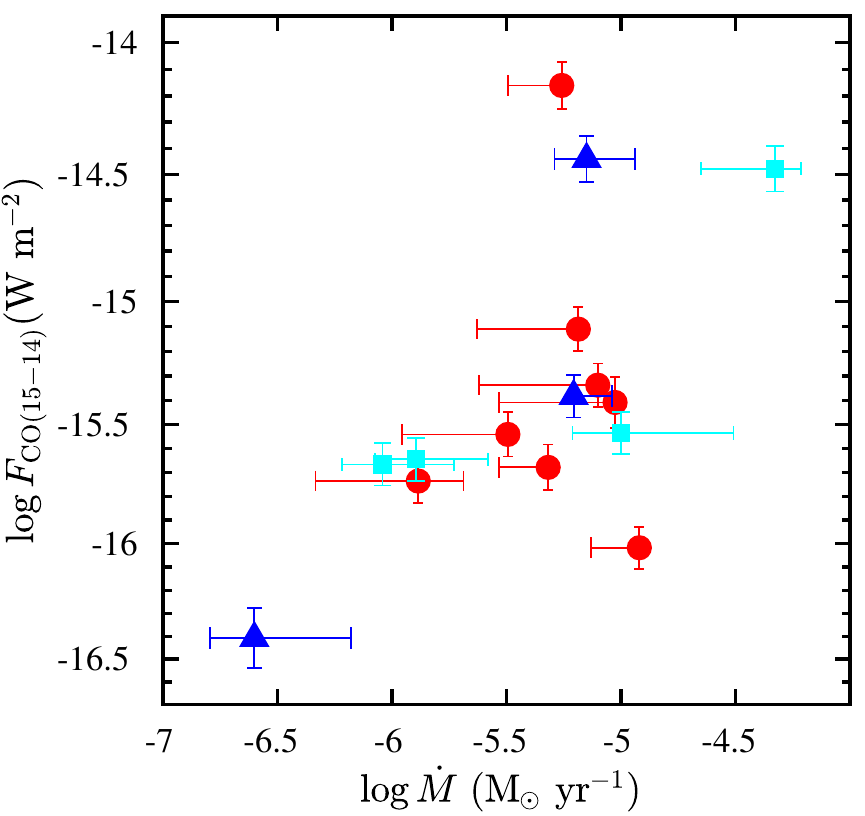}
\caption{Logarithm of the mass-loss rate versus total CO $J=$15-14 flux.} \label{fig:mloss_results}
\end{figure}

The mass-loss rates have been estimated by simply dividing the total mass by the crossing time of the CO-emitting layers, that is:
\begin{equation}\label{MassLossEquation}
 \mloss = \frac{\mtot\,\vexp}{\rco}
\end{equation}
\noindent where \vexp\ is the expansion velocity of the gas, which has been taken from literature (Table\,\ref{tab:2}). For the characteristic radius of the CO-emitting region we use the same value (\rco) adopted for the opacity correction. We note that this estimate represents a mean or "equivalent" mass-loss rate assuming constant-velocity spherically-symmetric mass-loss during the time when the warm-inner envelope layers where the CO PACS lines arise were ejected, that is, during the last $\sim$20-50\,yr and $\la$300\,yr for AGBs and post-AGBs/yPNe, respectively, given the values of \vexp\ and \rco.

The mass-loss rates are listed in Table\,\ref{tab:rot}. We find a range of values of $\dot{M}\sim10^{-7}-10^{-5}\rm~M_{\sun}~yr^{-1}$, with a median value in our AGB stars of \mloss$\sim6\times10^{-6} \rm~M_{\sun}~yr^{-1}$. These values are not to be taken as representative of the whole class of C-rich evolved stars, since our sample is small and not unbiased. This is because the objects in the THROES catalogue were originally selected for \hso\ observations due to various reasons, probably including their strong CO emission.

As in Paper I, we investigated a possible correlation between $\dot{M}$, and \trot, \vexp. We see little evidence of an anti-correlation between \mloss\ and \trot, although the relation is strongly influenced by AFGL\,618, which is a strong outlier in this parameter space (Fig.\,\ref{fig:orichcomp}). In this, and maybe other objects (mainly post-AGB/yPNe), we expect departures from the simple (constant mass-loss rate, spherically symmetric) model adopted to estimate the "equivalent" \mloss.  
We compare the results obtained here and in Paper I in Section \ref{Section:comparison}.

Figure\,\ref{fig:mloss_results} shows the logarithm of the integrated flux of the CO $J=15-14$ line versus the logarithm of the mass-loss rate of the single component fit. The upper and lower limit of the errorbars in $\dot{M}$ correspond to a range of radii around the representative one (see Fig.\,\ref{fig:opacity}). We find a positive trend which is consistent with a power-law relation similar to that found by \citet{2016A&A...588A.124L} in their sample of C-rich AGB CSEs with \water\ FIR emission lines. 

Separate values of $\dot{M}$ for the hot and warm components are computed for completeness, but the difference found ($M_{\rm h}<M_{\rm w}$) should not be overinterpreted as a recent decrease of the mass-loss rate. The hot and warm components most likely trace adjacent layers of the inner-winds of our targets, with the hot component presumably best sampling regions closer to the center. However, for simplicity and since we ignore the true CO excitation structure, we use the same radius to formally compute $\dot{M}$ for both components. We note that due to the $1/r_{\rm CO}$ dependence, the values of $M_{\rm h}$ and $M_{\rm w}$ can be brought closer to the single-fit value if the warm and hot correspond to different $r_{\rm CO}$. We refrain from discussing $M_{\rm h}$ and $M_{\rm w}$ separately since a more sophisticated analysis, including nLTEexRT modeling, is needed in order to assess mass-loss time variability. For this reason we only compare our single-component mass-loss rates to the literature (\S\,\ref{section:Mloss}).
% ===================================================== %
\subsection{The influence of line variability on \trot\ and \mass}
\label{section:irc10216_var}

We have shown in Fig.\,\ref{fig:irc10216specs} the temporal variability of the continuum and line fluxes in the case of the Mira-type variable AGB star IRC+10216. The higher-$J$ CO lines (\eu$>$2000\,K) are the ones that show the strongest variations with time. Here, we are interested in studying how CO line variability affects the values of \mass\ and \trot\ derived using the RD method but using data acquired at different epochs.
We use the same seven available OBSIDS\footnote{Which have been reprocessed and are part of the THROES catalogue.} of IRC+10216 as \citet{2015ASPC..497...43T}, which span a time period of 551 days (Table \ref{tab:1}).

The RDs of IRC+10216 for these seven different observing epochs are shown in Fig.\,\ref{fig:irc10216_rds}, and the results of the RD analysis are tabulated in Table \ref{tab:rot}, together with the remaining targets. Since three of the OBSIDs corresponding to the ODs 894, 1133 and 1288 have a more restricted wavelength coverage, the fits to the RDs have an inherently larger uncertainty because the fit is more sensitive to the low number statistics.  The error-weighted mean (single-fit) rotational temperature and mass are \trot$\sim$520\,K and \mass$\sim$4$\times10^{-4}$\,\msun, respectively.

\begin{figure}[t]
\centering
\includegraphics[width=0.85\linewidth]{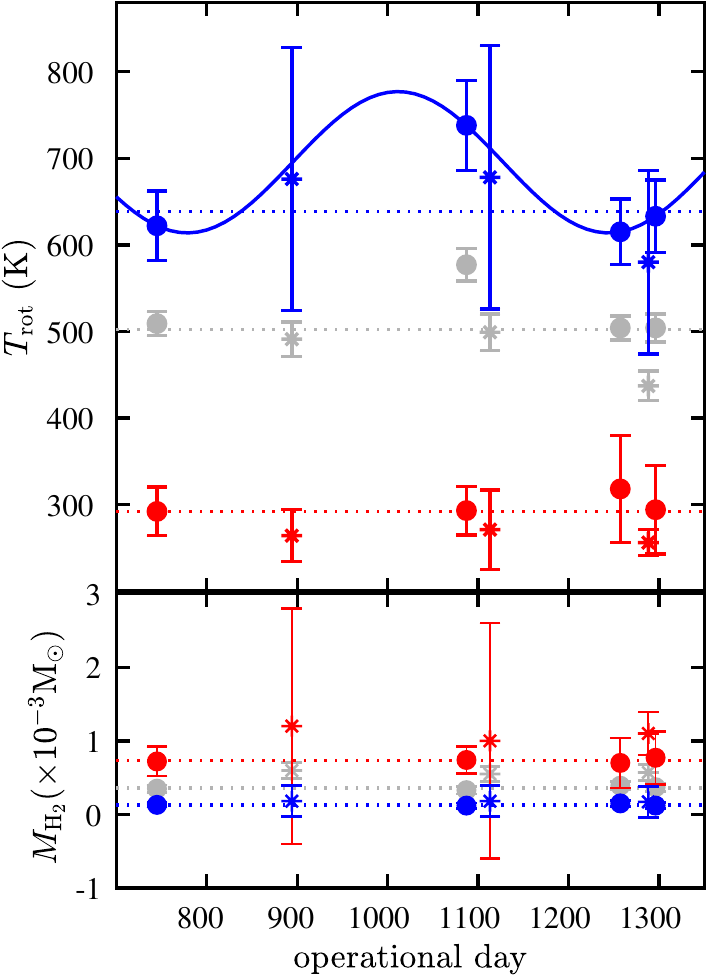}
\caption{Rotational temperature and mass versus operational day in IRC+10216. Top: sinusoidal fit to the \trot variation with fixed period of 630 days (solid line). Bottom: total gas mass over time. In each panel, the dotted lines are the average of each component and the asterisks mark unused data points in the fit (see text). The color code is the same as in Fig.\,\ref{fig:irc10216_rds}.}
\label{fig:irc10216_var}
\end{figure}

In Fig.\,\ref{fig:irc10216_var}, we plot the temperature and
  the mass, for single- and double-temperature components, versus the
  operational day of the observations.  The bottom panel shows that
  the total gas mass deduced from the single-fit of the RD or 
  for the warm and hot components does not reflect the line flux
  variability since it stays essentially constant with time about the
  average value (dotted lines), well within the estimated
  uncertainties.

In a similar manner, the temperature of the warm component does not clearly reflect the CO line flux variations, since all multi-epoch values are in good agreement within uncertainties.  The hot component is the one that shows the largest variations (perhaps periodic) of the temperature, with $T_{\rm h}$ going $\sim$100\,K ($\sim$16\%) above the average ($\sim$\,640\,K) at OD\,1087, and then relaxing back to normal values in the remaining epochs. This variation is echoed in the single-fit value of \trot\, ($\sim$12\%).

Temperature variations are not necessarily expected to be periodic, but if we assume that to be true by fitting a sinusoidal function with the known pulsation period $P=630$ days to the RD parameters, we find that such model accommodates reasonably well the data points corresponding to $T_{\rm h}$ (and also the single-fit \trot). 
It is also possible that IRC+10216 underwent an abrupt change of the physical conditions in its inner wind layers at epoch OD\,1087, since the remaining data points by themselves will not justify/indicate periodic variability.

In summary, probably due to a compensation between the changes in the
y-intercept and the slope of the RD due to CO line flux variations,
which are largest for transitions with the highest $J$, the total gas mass \mass\ appears to be quite robust to FIR pumping (non-LTE) effects. These, however, could have a measurable, yet moderate impact on the rotational temperatures.

\section{Discussion}
\label{section:DIS}

\subsection{Gas temperatures}
\label{section:gasmass}

The detection of high-$J$ CO rotational lines is an indication of a significant amount of molecular gas under relatively high temperature conditions. From our simple RD analysis we inferred that the average gas temperatures of the layers sampled by FIR CO lines are much larger ($T_{\rm rot}\sim200-900$\,K) than those typically derived from mm/sub-mm observations, which are sensitive to $\la$\,100\,K gas from the intermediate-to-outer layers of the envelopes of evolved stars \citep[at $\approx$10$^{16}$-10$^{17}$\,cm, see e.g.][]{2010A&A...523A..18D,2011A&A...530A..83S}.

In a number of targets we identified a double-temperature ("warm" and "hot") component. Deviations from a single straight line fit to the RD have been found also in some of the O-rich objects in Paper I and in other previous works using, for example, ISO and/or \textit{Herschel}/SPIRE CO spectra in a number of AGB and post-AGB CSEs \citep[e.g.,][]{2000A&A...360.1117J,2010A&A...518L.144W,2014MNRAS.437..532M,2015ApJ...806L...3C,2016ApJ...828...51C}. 
AFGL\,618 and IRAS\,16594-4656 are the targets whose RDs show the most obvious departure from linearity. IRAS\,16594-4656 is particularly interesting since we found a much cooler warm-component of just $T_{\rm w}\sim220$\,K, and a breakpoint at lower energies ($E_{\rm u}/k\approx1049.9$\,K) compared to other targets typically with $T_{\rm w}\sim450$\,K up to $E_{\rm u}/k\approx1794.3$\,K.

\begin{figure}[t]
\centering
\includegraphics[width=0.85\linewidth]{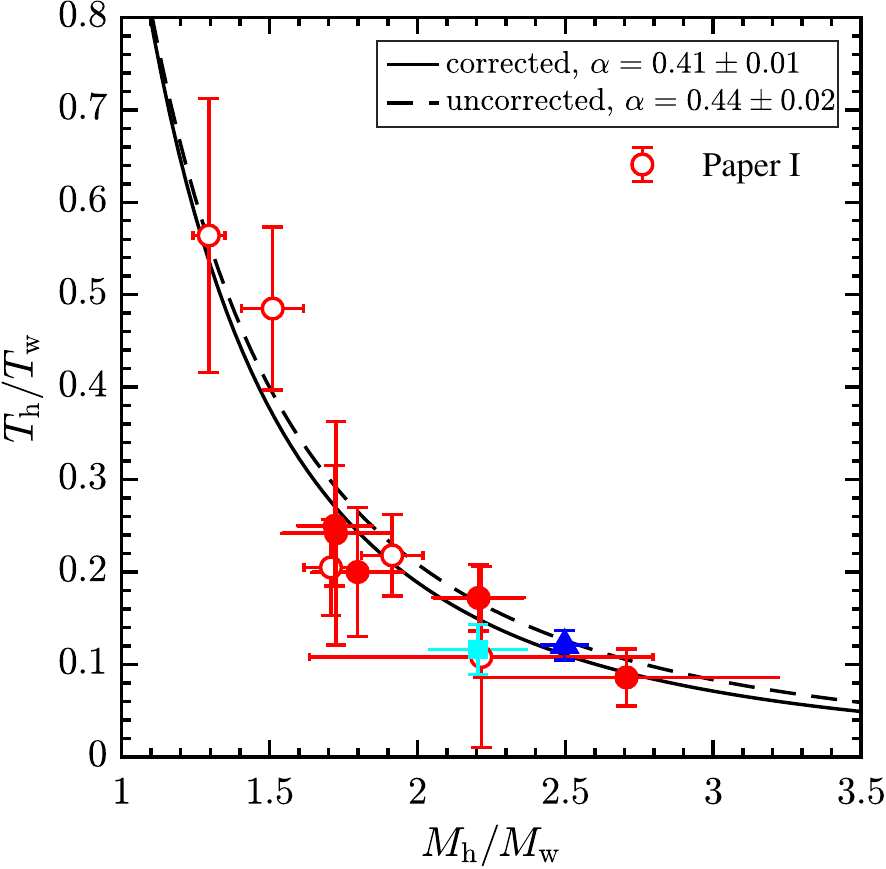}\\
\includegraphics[width=0.85\linewidth]{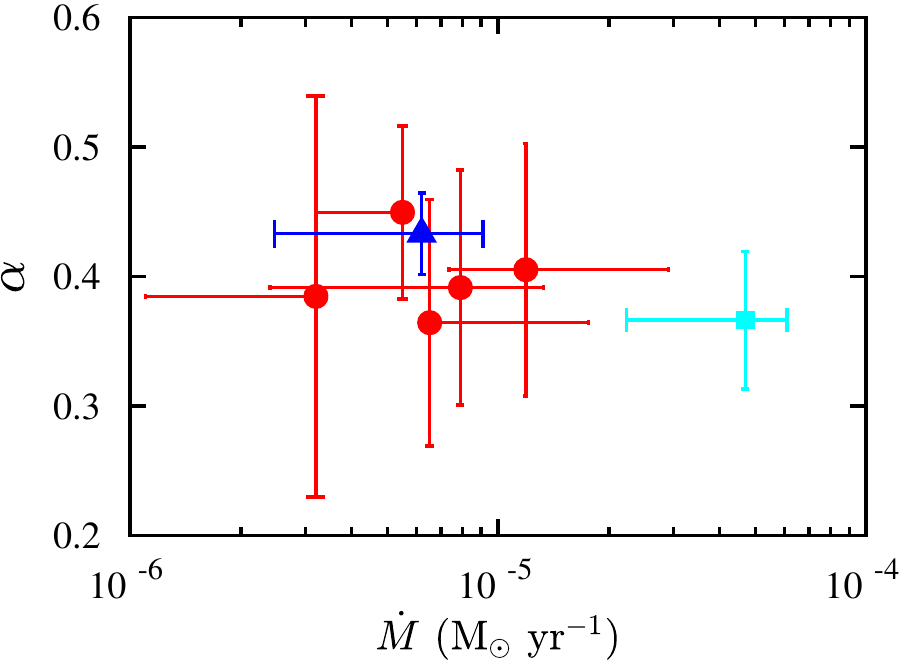}
\caption{Hot/warm temperature ratio versus hot/warm mass ratio for AGBs and post-AGBs and the power-law index. Top: the lines correspond to power-law fits with index $\alpha=0.4$ for the opacity corrected RDs (solid) and $\alpha=0.44$ for the uncorrected RDs (dashed); the open symbols correspond to the O-rich AGB stars in Paper I. Bottom: power law index for each individual target versus mass-loss rate (single fit).} \label{fig:2comps}
\end{figure}

As in Paper I, in order to investigate if the
  double-temperature component is consistent with resulting from the temperature
  stratification within the inner layers of the CSEs, we compared the
  hot-to-warm \mass\ and \trot\ ratios (Fig.\,\ref{fig:2comps}), and
  find that they are correlated. If the temperature profiles in the
  envelope follow a power-law of the type $T(r)\propto r^{-\alpha}$,
  with $\alpha$ being a constant, then the trend in
  Fig.\,\ref{fig:2comps} should also follow approximately a power-law
  function. We find $\alpha\sim0.4$, which is similar to the value
  found for the O-rich targets studied in Paper I. The value of
  $\alpha$ is in agreement with past works that suggested that the
  kinetic temperature distribution is shallower, with values of
  $\alpha$ down to $\sim$0.4-0.5, for the inner
  ($\sim$5\ex{14}-3\ex{15}\,cm) CSE layers
  \cite[][]{2012A&A...539A.108D,2016A&A...588A.124L,2014MNRAS.437..532M} than for the
  outer regions, where the steepest temperature variations ($\alpha
  \sim$1-1.2) are found
  \cite[$\ga$10$^{16}$\,cm;][]{2006A&A...450..167T}.

In addition to the average power-law exponent obtained by fitting all the targets simultaneously, we can also derive a value for each individual case applying:
\begin{equation}
\alpha = -\frac{\log(T_{\rm h}/T_{\rm w})}{\log(M_{\rm h}/M_{\rm w})}
\end{equation}
\noindent as shown in the bottom panel of Fig.\,\ref{fig:2comps}. In the case of IRC+10216, we find $\alpha=0.45\pm0.06$ which is in good agreement with the power-law exponent in the inner CSE between 9 and 65 stellar radii (i.e., up to $\sim3\times10^{15}\rm\,cm$) deduced from detailed non-LTE excitation and radiative transfer models \citep[][]{2012A&A...539A.108D}.

Therefore, the empirical relation found between the
  hot-to-warm ratio of \mass\ and \trot\ is consistent with the
  double-\trot\ component in some of our targets stemming (at least
  partially) from the temperature stratification across the inner
  envelope layers. The two components in the RD do not necessarily
  imply two distinct/detached shells of gas at different temperatures,
  but they most likely reflect the temperature decay laws. As
  explained in Paper I, in case of LTE deviations (not impossible in
  the lowest mass-loss rate objects), the value of $\alpha$ obtained
  from this simple approach would more closely represent the dust
  (rather than the gas) temperature distribution. This needs
  confirmation by detailed nLTEexRT models to the individual targets,
  which will be done in a future publication.

\subsubsection{Individual targets: comparison with previous works} \label{Section:individual}

It is not possible to directly compare most of our results with literature because past studies have been focusing on the cold, outer components of CSEs. 
Prior to \textit{Herschel} there was a study based on ISO LWS data in roughly the same wavelength range by \citet{2000A&A...360.1117J} who also performed RD analysis (without opacity correction). They found $T_{\rm rot}\sim700(\pm 90)$\,K and $T_{\rm rot}\sim380(\pm 30)$\,K for AFGL\,618 and AFGL\,2688 respectively. We obtained similar results without opacity correction, but introducing this effect lowered these values to $T_{\rm rot}\sim690(\pm30)$\,K and $T_{\rm rot}\sim300(\pm20)$\,K. In AFGL\,618 the central star is hot enough ($T_{\rm eff}\sim 33000$\,K) to produce FUV photons that heat the gas. In AGFL\,2688 this is probably not the case since the central star is much cooler ($T_{\rm eff}\sim7250$\,K), so low-velocity shocks are the most likely heating mechanism. 

Also partially based on ISO data, \citet{1999A&A...345..841R}
  used a number of spectral lines of CO between $J=1-0$ and $J=21-20$
  to infer the kinetic temperature profile across the CSE of
  IRAS\,15194-5115. According to their model, $T_{\rm kin}\sim400$\,K
  at $r_{\rm CO}\sim$\,(1-2)$\times10^{15}\rm\,cm$, which is in
  excellent agreement with our opacity-corrected single component
  $T_{\rm rot}\sim410$\,K for $r_{\rm CO}=2\times10^{15}\rm\,cm$. This
  is because, as already pointed out by \citet{1999A&A...345..841R},
  these high-$J$ levels are mainly populated by collisions, therefore
  they are proxies for the kinetic temperature, at least out to a few
  $\sim10^{15}$\,cm. Also using {\sl ISO} data,
  \citet{2002A&A...391..577S} presented a kinetic temperature model
  for CIT\,6 that shows that $T_{\rm kin}\sim400-500$\,K at
  approximately $r_{\rm CO}\sim$\,(1-2)$\times10^{15}\rm\,cm$, which
  is consistent with the warm component ($T_{\rm w}\sim460$\,K) that
  we infer and the representative radius $r_{\rm
    CO}=2\times10^{15}\rm\,cm$ adopted. The hot component ($T_{\rm
    w}\sim830$\,K) found by us would imply that regions closer to the
  star have an important contribution to the emission of the
  highest-$J$ lines.

The only star whose inner/warmer (gaseous) CSE had been studied in detail before using \textit{Herschel}/PACS data is IRC+10216. This was done by \citet{2010A&A...518L.143D} using high-$J$ CO spectral lines to infer the kinetic temperature profile as a function of radial distance. They find $T_{\rm kin}\sim500-600$\,K at $r_{\rm CO}\sim$ (1-2)$\times10^{15}\rm\,cm$.  Here we applied the opacity correction for $r_{\rm CO}=3\times10^{15}$\,cm which seems to be the layer at which $T_{\rm kin}\sim300$\,K in their model. This is also the average value of $T_{\rm w}$ that we found by fitting only the lowest $J$ transitions, wich does not change appreciably with time despite strong line flux variability (Fig.\,\ref{fig:irc10216_var}). The hot component ($\sim$600-700\,K) may correspond to $r_{\rm CO}\sim1\times10^{15}\rm\,cm$ according to their model.

Using \textit{Herschel}/SPIRE data, \citet{2010A&A...518L.144W} also performed RD analysis of CO spectra and derived $T_{\rm rot}\sim$70-230\,K and $T_{\rm rot}\sim$100-200\,K for AFGL\,618 and AFGL\,2688, respectively. The rotational temperatures we find are thus higher than the ones obtained in \citet{2010A&A...518L.144W} as expected. 
Further Large-velocity gradient (LVG) calculations by those authors suggested that hot material at approximately $1000$\,K might exist at the high-velocity wind region of AFGL\,618, which could be the one traced by PACS since we found $T_{\rm h}\sim 900$\,K. 

For HD\,44179 (the Red Rectangle) we obtained $T_{\rm rot}\sim440$\,K, which is about 2-3 times larger than the range of values inferred by \citet{2013A&A...552A.116B} from \textit{Herschel}/HIFI lower-$J$ CO observations. It is not surprising that we find a larger value since the higher $J$ lines probed by PACS are probably formed deeper inside the rotating circum-binary disk at the core of this object. Meanwhile follow up analysis has shown more clearly an outflow with $\sim500$\,K
\citep{2016A&A...593A..92B}, but also seems that such temperature conditions could exist in a region of the inner disk with a radius of about $r_{\rm CO}\sim2\times10^{15}$\,cm, unresolved by PACS. The opacity correction would still be moderate for this value ($\tau_{J=14-13}\sim 0.67$), and it lowers the rotational temperature to $T_{\rm rot}\approx410$\,K which is still within the uncertainties.
For the remaining yPNes (IRAS\,21282+5050, CPD-568032, Hen\,2-113) there are no kinetic temperature models in the literature that we could compare our results to.

% -------------------------------------------------------------- %
% % ========================================================== %

\subsection{Mass-loss rate}
\label{section:Mloss}

As in Paper I, we have compared the values of the mass-loss rates derived from our simple RD analysis with other values found in the literature mostly from low-$J$ observations, paying special attention to a few targets with detailed non-LTE excitation and radiative transfer analysis of CO data including at least some high-$J$ transitions observed with \hso. This is also a way of ascertaining the robustness of the RD method.

\begin{figure}[t]
\centering
\includegraphics[width=0.85\linewidth]{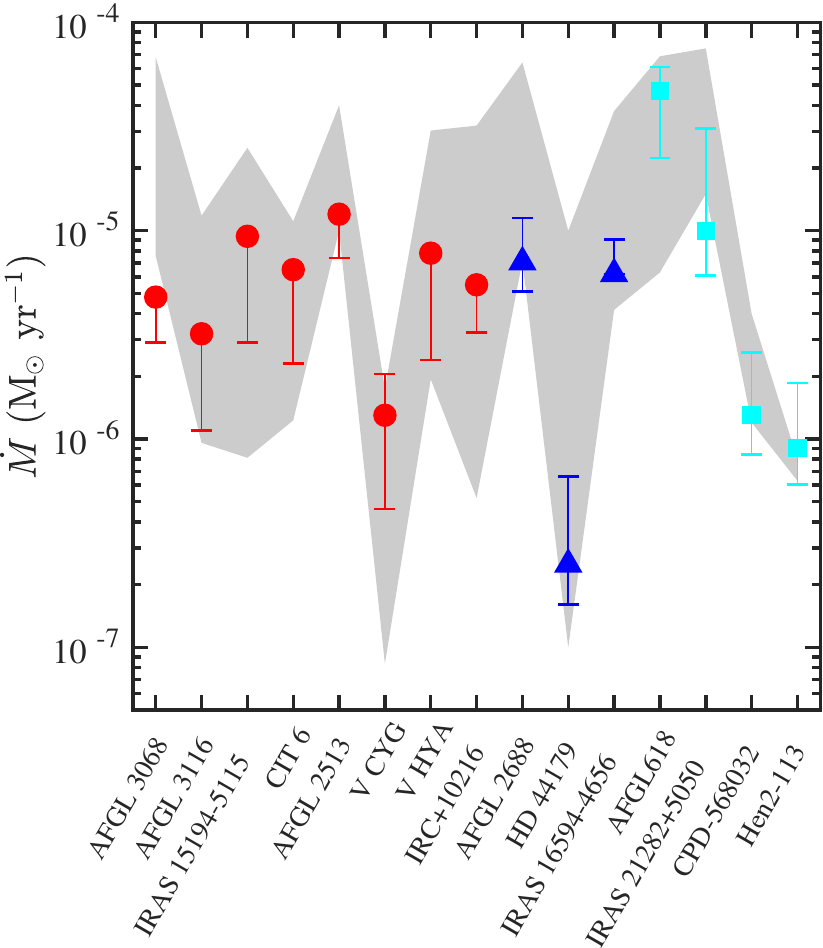}
\caption{Comparison between the mass-loss rate in this work and the literature. The markers correspond to the representative radii listed in Table \ref{tab:rot} for each target and the error bars are mass-loss rates for a given range of radii. The shaded area encloses the range of values found in literature scaled to the same parameters here assumed (see text).} \label{fig:mLoss2}
\end{figure}

Figure\,\ref{fig:mLoss2} shows our estimate of the mass-loss rate versus that found in the literature (see Table\,\ref{tab:2}). For each target the markers correspond to the single temperature component for the radius mentioned in Table\,\ref{tab:rot}, and the error bars represent the uncertainty in the radius for the adopted $v_{\rm exp}$. We show the computed opacities for the considered radii in each target in the supplementary Fig. \ref{fig:opacity}. The range of values found in literature are shown by the gray shaded area whose bounds are set by the maximum and minimum $\dot{M}$ plus uncertainties when reported \citep[factor of $\sim$3 in AGBs, ][]{2008A&A...487..645R,2010A&A...523A..18D}. Those values have been rescaled to the same distance, $v_{\rm exp}$ and $X_{\rm CO}$ here adopted.

Similarly to the non C-rich THROES targets in Paper I, our mass-loss rates are in good agreement with values in the literature within the large uncertainties. In our case, these are dominated by the uncertainty in \rco. We see that the range of radii we explored yields a $\dot{M}$ that fits within the shaded area. In many cases, the error bars are truncated at the upper limit above which the line opacities would be too large to allow reliable estimates of the masses and mass-loss rates (see Fig.\,\ref{fig:opacity}). For example, in the case of AFGL\,3068 this seems to imply smaller radius than what we have adopted to better match the values in literature. 

In the case of IRC+10216, the opacity correction for $r_{\rm CO}=2\times10^{15}$\,cm (instead of \rco=3\ex{15}\,cm) would result in a mass-loss of $\dot{M}\approx1.4\times10^{-5}~\rm M_{\sun}~yr^{-1}$ (not displayed) that would best match the estimate from mm observations. However in these circumstances the expected opacities in the lowest $J$ lines would be too large ($\tau \gg1$). Nonetheless, the derived $T_{\rm w}$ and $T_{\rm h}$ are consistent with kinetic temperature profile models as explained in Section\,\ref{section:gasmass}.
We find an average value of four OBSIDs of $\dot{M}\sim5\times10^{-6}~\rm M_{\sun}~yr^{-1}$, which is lower than the range $\dot{M}\sim$(1-3)$\times10^{-5}~\rm M_{\sun}~yr^{-1}$ scaled from \cite{2006A&A...450..167T, 2010A&A...523A..18D,2012A&A...539A.108D,2017arXiv170904738G}. However the value obtained for the warm component agrees with the lower limit of this range (because of the larger $M_{\rm H_2}$). We obtain $T_{\rm w}\sim300$ K and $\dot{M_{\rm w}}\sim$1$\times10^{-5}~\rm M_{\sun}~yr^{-1}$ for a radius of $r_{\rm CO}=3\times10^{15}$ cm, in good agreement with the results of radiative transfer modeling of the same high $J$ CO lines by \citet{2010A&A...518L.143D}. Scaling their $\dot{M}$ to the same $d$ and $X_{\rm CO}$ gives
$\dot{M}\sim$1.2$\times10^{-5}$\,\my\ with an uncertainty of a factor 2.

For V\,Hya we obtained $\dot{M}\sim8\times10^{-6}~\rm
  M_{\sun}~yr^{-1}$ which is lower than $\dot{M}\sim3\times10^{-5}~\rm
  M_{\sun}~yr^{-1}$ from \citet{thesis-camps} who performed radiative transfer calculations using the same PACS spectrum. In this case, however, the spatio-kinematic structure of the molecular outflow is more complex than assumed here. In particular,  multiple kinematic (fast and slow) components seem to be present \citet{2004ApJ...616L..43H,0004-637X-699-2-1015}, which not only translates into a larger uncertainty in the characteristic value of \vexp\ in the PACS CO-emitting layers, but also implies that the "equivalent" mass-loss rate is particularly questionable. 

As for V\,Hya, for pPNe and yPNe, the assumption of constant-velocity spherically- symmetric mass loss may also not hold. For completeness, the mass-loss rates estimates for these targets are shown Table\,\ref{tab:2}, but they are subject to larger uncertainties and have to be interpreted with caution. We have checked that, even in these cases, our values are in good agreement with previous estimates (making similar simplifying assumptions) in the literature. For example, for the Red Rectangle, we obtained
$\dot{M}\sim2\times10^{-7}$\,\my\ which is within the (scaled) range of $\dot{M}\sim$(0.2-1.4)$\times10^{-7}$\,\my\ reported by \cite{2010A&A...523A..18D}.

In summary, our results are consistent with the literature within the
typical uncertainties, but it is hard to tell what is the exact cause
of the slight discrepancies from case to case. One obvious
  reason is that the simple RD method and our assumption of a
  characteristic value of \rco\ (unknown, but crudely constrained
  from first principles and observations) only provides a rough
  estimate of the mass-loss.  We also note, that the bulk of the CO
  emission under study is produced in the warm inner layers of the
  CSEs of our targets down to a location where $\tau$$\sim$1.  For
  very optically thick CSEs, there may be an additional amount of gas
  that is not fully recovered after the moderate opacity correction
  applied. Another reason for \mloss\ discrepancies is the different
  number of transitions and range of \eu\ covered by different
  studies. Non-LTE excitation and radiative transfer models of the CO
  emission including a wide range of $J$- transitions is needed to
  obtain accurate estimates of the mass-loss rates and, in particular,
  to address \mloss\ time modulations.
  
   %%% new section!! 

% % ============================================================== %
% % ============================================================== %
\subsection{Comparison between the sample of O-rich and C-rich stars}
\label{Section:comparison}

In the Appendix we provide Fig.\,\ref{fig:orichcomp} where we plot together the results presented here to the ones obtained for the sample of O-rich and S-type stars in Paper I. 

We find that the range of \trot\ is approximately the same among C-rich and O-rich stars, but the O-rich AGBs have less massive CSEs by typically one order of magnitude, and lower expansion velocities. This in turn reflects on lower mass-loss rates on average. The scenario is reversed in the groups of PNes with the few O-rich PNes having more warm gas than the carbon counterparts. 
Probably the O-rich AGB stars studied in Paper I are typically low massive stars with very large evolution times, while the O-rich post-AGBs and PNe are very massive objects that have undergone through the Hot Bottom Burning stage. 

We stress that despite these trends being indicative of clear differences in the properties of the CSEs of the targets in the THROES catalogue, they may not be a general property of C-rich versus non C-rich targets, since our samples are not necessarily unbiased and they are definitely not statistically significant.

% % =========================================== %
% % % =========================================== %
 \section{Summary}
 \label{section:conclusion}

In this paper (Paper II), we use \textit{Herschel}/PACS FIR spectra of a sample of 15 C-rich evolved stars, including AGBs, post-AGBs and yPNe, from the THROES catalogue \citep{2017arXiv171105992R}.
These data contain valuable information about the physical-chemical properties of evolved stars as shown, for instance, by the striking differences of spectral features (molecular, atomic/ionized and solid state) as a function of evolutionary stage. In this work, we focus on the rotational spectrum of CO (up to $J=45-44$) which was used as a proxy for the molecular component of the gas in the warm regions of the CSEs. Our findings can be summarized as follows:

\begin{itemize}
\item Due to \textit{Herschel's} higher sensitivity compared to ISO, the range of detected \docem\, transitions has been extended to high rotational levels of up to \ju=45 in low-to-intermediate mass evolved stars. Rotational diagrams using high-excitation CO\,($v$=0) rotational emission lines, with upper-level energies \eu$\sim$580 to 5000 K, have been plotted to estimate rotational temperatures (\trot), total molecular mass in the CO-emitting layers (\mtot) and average mass-loss rates during the ejection of these layers (\mloss).

\item The range of temperatures found in our sample, \trot$\sim$200-700\,K, is larger than what had been deduced from mm/sub-mm observations, and even \hso/HIFI and SPIRE observations, confirming that PACS CO lines probe deeper layers yet poorly studied to date (typically, $\approx$10$^{15}$ cm for AGBs and $\approx$10$^{16}$ cm for post-AGBs and yPNes).)

\item The total gas mass of the warm envelope layers sampled by PACS data are between $M\sim10^{-5}-10^{-3}~\rm M_{\sun}$, with post-AGBs and yPNe being overall more massive.

 \item We find clearly different temperature distributions for the different classes with AGBs having typically hotter gas (up to $T_{\rm rot}\sim1000$\,K) than post-AGBs ($T_{\rm rot}\la500$\,K) and yPNes ($T_{\rm rot}\la400$\,K). The yPN AFGL\,618 is a clear outlier with a very high amount (\mass$\sim$2\ex{-3}\,\msun) of rather hot (up to T$_{\rm h}$$\sim$900\,K) gas, similar to the most massive AGBs in the sample.

\item For AFGL\,3116, CIT\,6, AFGL\,2513, V\,Hya, IRAS\,16594-4656 and AFGL\,618 a double temperature (hot and warm) component is inferred from the RDs. The mean temperatures of the warm and hot components are $\sim$400\,K and $\sim$820\,K, respectively. The mass of the warm component
($\sim$10$^{-5}$-8$\times$10$^{-3}$\msun) is always larger than that of the hot component, by a factor $\sim$4-10. 

\item The warm-to-hot \mass\ and \trot\ ratios in our sample are correlated and are consistent with an average temperature radial profile of $T\propto$\,$r^{-0.4}$, that is, slightly shallower than in the outer envelope layers, in agreement with recent studies.

\item The mass-loss rates estimated are in the range
\mloss$\approx$10$^{-7}$-10$^{-4}$\my, in agreement (within the uncertainties) with values found in the literature for our targets. 

\item We investigated the impact of CO line flux variability on the values of \mass\ and \trot\ derived from the simple RD analysis. We studied in detail the case of the Mira-variable AGB star IRC+10216, for which multi-epoch PACS data exist. In spite of strong line flux variability we find that the total gas mass and the average temperature derived from the RDs at different epochs are minimally affected. Only the hot component does show the sign of line variability ($\delta T/T\sim16$\%), roughly in-phase with the continuum periodicity. 

\item Similarly to Paper I, we find an anti-correlation between \trot\ and \mass, which may result from a combination of CO line cooling and opacity effects, and we find a correlation between \mloss\ and \vexp, which is consistent with the wind acceleration mechanism being more efficient the more luminous/massive the star is. These trends had been reported in previous studies using low-$J$ CO transitions.
\end{itemize}

We show that high-$J$ CO emission lines probed by \textit{Herschel}/PACS are good tracers of the warm gas ($T\sim200-900$\,K) surrounding evolved carbon stars. Using the simple RD technique, we have provided systematic and homogeneous insight into the deepest layers of these CSEs, though it relies on several approximations. Detailed non-LTE excitation and radiative transfer calculations are needed to determine the temperature stratification of the CSEs, to infer mass-loss rates and to address their time-variability.

% % =========================================== %
\renewcommand{\arraystretch}{1.2}
\begin{sidewaystable*}[]
\centering
\caption{Results of the CO rotational diagram analysis with opacity correction and double-component fit. The assumed distance ($d$), expansion velocity ($v_{\rm exp}$) and radius ($r_{\rm CO}$), the derived rotational temperature (\trot), CO column density (\ncol), total mass of $\rm H_2$ within the CO emitting-volume (\mass), mass-loss rate ($\rm \dot{M}=\mass\times v_{\rm exp}/r_{\rm CO}$) and ratio between opacity corrected and uncorrected total masses.}
%\begin{longtab}
%\begin{table*}[]
\label{tab:rot}
\begin{tabular}{lcccccccc}
\hline\hline
\multirow{2}{*}{\begin{tabular}[c]{@{}c@{}}Target\\ \end{tabular}} & $d$ (pc) & $v_{\rm exp}$ $\rm (km~s^{-1})$ & $r_{\rm CO}$ (cm) & \trot (K) & \ncol $\rm (cm^{-2})$ & \mass $\rm (\rm M_{\sun})$ & $\dot{M}$ $\rm (\rm M_{\sun}~yr^{-1})$ & \multirow{2}{*}{\begin{tabular}[c]{@{}c@{}} Correction \mass\end{tabular}}\\ 
& & & & & & & & \\ \hline
AFGL 3068 & 1100 &13 & $6\times 10^{15}$ & 350$\pm$20 & $3\times 10^{18}$ & $7(\pm 1)\times 10^{-4}$ & $5\times 10^{-6}$ &1.6\\ \hline
AFGL 3116 & 630 &14 & $2\times 10^{15}$ & 670$\pm$30 & $5.4\times 10^{18}$ & $1.4(\pm 0.2)\times 10^{-4}$ & $3\times 10^{-6}$ &1.6\\
 &  & &  &   &  &  &  \\
 &  & &  & 460$\pm$30 & $9.8\times 10^{18}$ & $2.6(\pm 0.4)\times 10^{-4}$ & $6\times 10^{-6}$  &1.8\\
 &  & &  & 790$\pm$70 & $2.4\times 10^{18}$ & $6(\pm 3)\times 10^{-5}$ & $1\times 10^{-6}$ &1.3\\ \hline
 IRAS 15194-5115 & 500 & 23 & $2\times 10^{15}$ & 410$\pm$20 & $9.8\times 10^{18}$ & $2.6(\pm 0.4)\times 10^{-4}$ & $9\times 10^{-6}$ &1.9\\ \hline
CIT 6 & 440 &21 & $2\times 10^{15}$ & 670$\pm$20 & $7.4\times 10^{18}$ & $2(\pm 0.3)\times 10^{-4}$ & $7\times 10^{-6}$ & 1.6\\
 &  & &  &   &  &  &  \\
 &  & &  & 460$\pm$30 & $1.3\times 10^{19}$ & $3.5(\pm 0.7)\times 10^{-4}$ & $1\times10^{-5}$ &1.7\\
 &  & &  & 830$\pm$40 & $2.8\times 10^{18}$ & $7(\pm 2)\times 10^{-5}$ & $2\times10^{-6}$ &1.3\\ \hline
AFGL 2513 & 1760 & 26 & $3\times 10^{15}$ & 530$\pm$30 & $7.3\times 10^{18}$ & $4.4(\pm 0.8)\times 10^{-4}$ & $1\times 10^{-5}$ & 1.6 \\ %\hline
 &  & &  &  &   &  &  \\
 &  & &  & 400$\pm$30 & $1.1\times 10^{19}$ & $7(\pm 1)\times 10^{-4}$ & $2\times10^{-5}$ &1.7\\
 &  & &  & 1100$\pm$200 & $9.5\times 10^{17}$ & $6(\pm 2)\times 10^{-5}$ & $2\times10^{-6}$ &1.3\\ \hline
V Cyg & 271 &15 & $1\times 10^{15}$ & 490$\pm$20  & $4.2\times 10^{18}$ & $2.7(\pm 0.4)\times 10^{-5}$ & $1\times 10^{-6}$ & 1.6\\ \hline
V Hya & 380 &24 & $1\times 10^{15}$ & 650$\pm$20  & $1.6\times 10^{19}$ & $1.0(\pm 0.1)\times 10^{-4}$ & $8\times 10^{-6}$ &1.9\\
 &  & &  &  &  &  &  \\
 &  & &  & 460$\pm$30 & $2.7\times 10^{19}$ & $1.8(\pm 0.3)\times 10^{-4}$ & $1\times10^{-5}$ &2.1\\
 &  & &  & 780$\pm$40 & $6.7\times 10^{18}$ & $4.5(\pm 0.9)\times 10^{-5}$ & $3\times10^{-6}$ &1.4\\ \hline\hline
AFGL 2688 & 340 & 20 & $8\times 10^{15}$ & 300$\pm$20  & $2\times 10^{18}$ & $9(\pm 2)\times 10^{-4}$ & $7\times 10^{-6}$ &1.2\\ \hline
HD 44179 & 710 & 10 & $4\times 10^{15}$ & 440$\pm$50  & $3\times 10^{17}$ & $3.2(\pm 0.8)\times 10^{-5}$ & $2\times 10^{-7}$ &1.1\\ \hline
IRAS 16594-4656 & 1800 & 14 & $1\times 10^{16}$ & 460$\pm$20 & $2\times 10^{18}$ & $1.4(\pm 0.3)\times 10^{-3}$ & $6\times 10^{-6}$ &1.4\\
 &  & &  &  &  &  &  \\
 &  & &  & 220$\pm$5 & $1\times 10^{19}$ & $6.5(\pm 0.6)\times 10^{-3}$ & $3\times 10^{-5}$ &1.5\\
 &  & &  & 530$\pm$20 & $1\times 10^{18}$ & $7.3(\pm 0.9)\times 10^{-4}$ & $3\times 10^{-6}$ &1.3\\ \hline\hline
AFGL 618 & 900 & 80 & 690$\pm$30 & $1\times 10^{16}$ & $2.8\times 10^{18}$ & $1.9(\pm 0.3)\times 10^{-3}$ & $5\times 10^{-5}$ &1.1\\
 &  & &  &  &  &   &  \\
 &  & &  & 415$\pm$20  & $6.8\times 10^{18}$ & $4.5(\pm 0.6)\times 10^{-3}$ & $1\times 10^{-4}$ &1.1\\
 &  & &  & 915$\pm$60  & $7.9\times 10^{17}$ & $5(\pm 1)\times 10^{-4}$ & $1\times 10^{-5}$ &1\\ \hline
IRAS 21282+5050 & 2000 & 14 & $4\times 10^{16}$ & 170$\pm$10  & $7.5\times 10^{17}$ & $8(\pm 2)\times 10^{-3}$ & $1\times 10^{-5}$ &1.2\\ \hline
CPD-568032 & 1530 & 22.6 & $4\times 10^{16}$ & 380$\pm$20  & $6.8\times 10^{16}$ & $7(\pm 1)\times 10^{-4}$ & $1\times 10^{-6}$ &1\\ \hline
Hen 2-113 & 1230 & 23 & $4\times 10^{16}$ & 390$\pm$20 & $4.7\times 10^{16}$ & $5.0(\pm 0.6)\times 10^{-4}$ & $9\times 10^{-7}$ &1\\ \hline
\end{tabular}
%\end{table*}
%\end{longtab}
\end{sidewaystable*}

\renewcommand{\arraystretch}{1.2}
\begin{sidewaystable*}[]
%\begin{longtab}
%\begin{table*}[]
\ContinuedFloat
\centering
\caption{Continued.}
\begin{tabular}{lcccccccc}
\hline\hline
\multirow{2}{*}{\begin{tabular}[c]{@{}c@{}}Target\\ \end{tabular}} & $d$ (pc) & $v_{\rm exp}$ $\rm (km~s^{-1})$ & $r_{\rm CO}$ (cm) & \trot (K) & \ncol $\rm (cm^{-2})$ & \mass $\rm (\rm M_{\sun})$ & $\dot{M}$ $\rm (\rm M_{\sun}~yr^{-1})$ & \multirow{2}{*}{\begin{tabular}[c]{@{}c@{}} Correction \mass\end{tabular}}\\ 
& & & & & & &  \\ \hline
\multirow{2}*{IRC+10216} \\ \footnotesize{(OD 745)} & 150 & 14.5 & $3\times 10^{15}$ & 510$\pm$20 & $5.8\times 10^{18}$ & $3.5(\pm 0.5)\times 10^{-4}$ & $5\times 10^{-6}$ &1.8\\
 &  & &  &  &  &  &  \\
&  & &  & 290$\pm$30 & $1.2\times 10^{19}$ & $7(\pm 2)\times 10^{-4}$ & $1\times 10^{-5}$ &1.9\\
& & &  & 620$\pm$40 & $2.3\times 10^{18}$ & $1.3(\pm 0.3)\times 10^{-4}$ & $2\times 10^{-6}$ &1.3\\ \hline
\footnotesize{(OD 894)}* & 150 & 14.5 & $3\times 10^{15}$ & 490$\pm$20 & $1\times 10^{19}$ & $6(\pm 1)\times 10^{-4}$ & $9\times 10^{-6}$ &2.2\\
 & & &  &  &  &  &  \\
&  & &  & 260$\pm$30  & $2\times 10^{19}$ & \textbf{$1(\pm 1)\times 10^{-3}$} & $2\times 10^{-5}$ &2.1\\
&  & &  & 680$\pm$150 & $3\times 10^{18}$ & $2(\pm 2)\times 10^{-4}$ & $3\times 10^{-6}$ &1.5\\ \hline
\footnotesize{(OD 1087)} & 150 & 14.5 & $3\times 10^{15}$ & 580$\pm$20 & $5.5\times 10^{18}$ & $3.3(\pm 0.5)\times 10^{-4}$ & $5\times 10^{-6}$ &1.7\\
 &  &  & &  &  &  &  \\
&  &  & & 290$\pm$30 & $1.2\times 10^{19}$ & $7(\pm 2)\times 10^{-4}$ & $1\times 10^{-5}$ &1.7\\
&  &  & & 740$\pm$50 & $2.1\times 10^{18}$ & $1.2(\pm 0.4)\times 10^{-4}$ & $2\times 10^{-6}$ &1.3\\ \hline
\footnotesize{(OD 1113)}* & 150 & 14.5 & $3\times 10^{15}$ & 500$\pm$20 & $9\times 10^{18}$ & $6(\pm 1)\times 10^{-4}$ & $8\times 10^{-6}$ &2.1\\
  &  & &  &  &  &  &  \\
&  & & & 270$\pm$50 & $1.7\times 10^{19}$ & $1(\pm 3)\times 10^{-3}$ & $2\times 10^{-5}$ &2\\
&  & & & 680$\pm$150 & $3\times 10^{18}$ & $2(\pm 2)\times 10^{-4}$ & $3\times 10^{-6}$ &1.5\\ \hline
\footnotesize{(OD 1257)} & 150 & 14.5 & $3\times 10^{15}$ & 500$\pm$20 & $6.6\times 10^{18}$ & $3.9(\pm 0.6)\times 10^{-4}$ & $6\times 10^{-6}$ &1.8\\
&  &  & &  &  &  &  \\
&  &  & & 320$\pm$60  & $1.2\times 10^{19}$ & $7(\pm 3)\times 10^{-4}$ & $1\times 10^{-5}$ &1.9\\
&  &  & & 620$\pm$40  & $2.5\times 10^{18}$ & $1.5(\pm 0.4)\times 10^{-4}$ & $2\times 10^{-6}$ &1.3\\ \hline
\footnotesize{(OD 1288)}* & 150 & 14.5 & $3\times 10^{15}$ & 440$\pm$20 & $1\times 10^{19}$ & $6(\pm 1)\times 10^{-4}$ & $9\times 10^{-6}$ &2.2\\
 &  & & &  &  &  &  \\
&  & & & 260$\pm$15  & $1.9\times 10^{19}$ & $1.1(\pm 0.3)\times 10^{-3}$ & $2\times 10^{-5}$ &2.1\\
&  & & & 580$\pm$120  & $2.9\times 10^{18}$ & $2(\pm 2)\times 10^{-4}$ & $3\times 10^{-6}$ &1.5\\ \hline
\footnotesize{(OD 1296)} & 150 & 14.5 & $3\times 10^{15}$ & 500$\pm$20 & $6.3\times 10^{18}$ & $3.7(\pm 0.6)\times 10^{-4}$ & $6\times 10^{-6}$ &1.8\\
 &  & & &  &  &  &  \\
&  & & & 290$\pm$50 & $1.3\times 10^{19}$ & $8(\pm 4)\times 10^{-4}$ & $1\times 10^{-5}$ &1.9\\
&  & & & 630$\pm$40 & $2.1\times 10^{18}$ & $1.2(\pm 0.4)\times 10^{-4}$ & $2\times 10^{-6}$ &1.3\\ 
\hline
\end{tabular}
\tablefoot{In the case of IRC+10216 the fit was performed using 15 spectral lines except for the OBSIDS marked with an asterisk (OD 894, 1113, 1296) for which we were only able to use 9 lines which then reflects on the accuracy of the best-fit parameters.}
%\end{table*}
%\end{longtab}
\end{sidewaystable*}

% % =========================================== %
\begin{acknowledgements}
We thank the referee for the useful comments and remarks. PACS has been developed by a consortium of institutes led by MPE (Germany) and including UVIE (Austria); KU Leuven, CSL, IMEC (Belgium); CEA, LAM (France); MPIA (Germany); INAF-IFSI/OAA/OAP/OAT, LENS, SISSA (Italy); IAC (Spain). This development has been supported by the funding agencies BMVIT (Austria), ESA-PRODEX (Belgium), CEA/CNES (France), DLR (Germany), ASI/INAF (Italy), and CICYT/MCYT (Spain). This publication makes use of data products from the THROES catalog, which is a project of the Centro de Astrobiología (CAB-CSIC) with the collaboration of the Spanish Virtual Observatory (SVO), funded by the European Space Agency (ESA). 
J.M.S.S.\, acknowledges financial support from the ESAC Faculty and the ESA Education Office under the ESAC trainee program. The Institute for Solar Physics is supported by a grant for research infrastructures of national importance from the Swedish Research Council (registration number 2017-00625). C.S.C.\,acknowledges financial support by the Spanish MINECO through grants AYA2016-75066-C2-1-P and by the European Research Council through ERC grant 610256: NANOCOSMOS.   
\end{acknowledgements}

% % =========================================== %
\bibliographystyle{aa} 
% \bibliography{bib.bib} 

% % =========================================== %
\begin{appendix}

\section{Methods}
\label{section:methods}

% % =========================================== %
\subsection{Opacity correction}
\label{appendix:opCorrection}

\begin{figure*}[!htp]
\includegraphics[width=\linewidth]{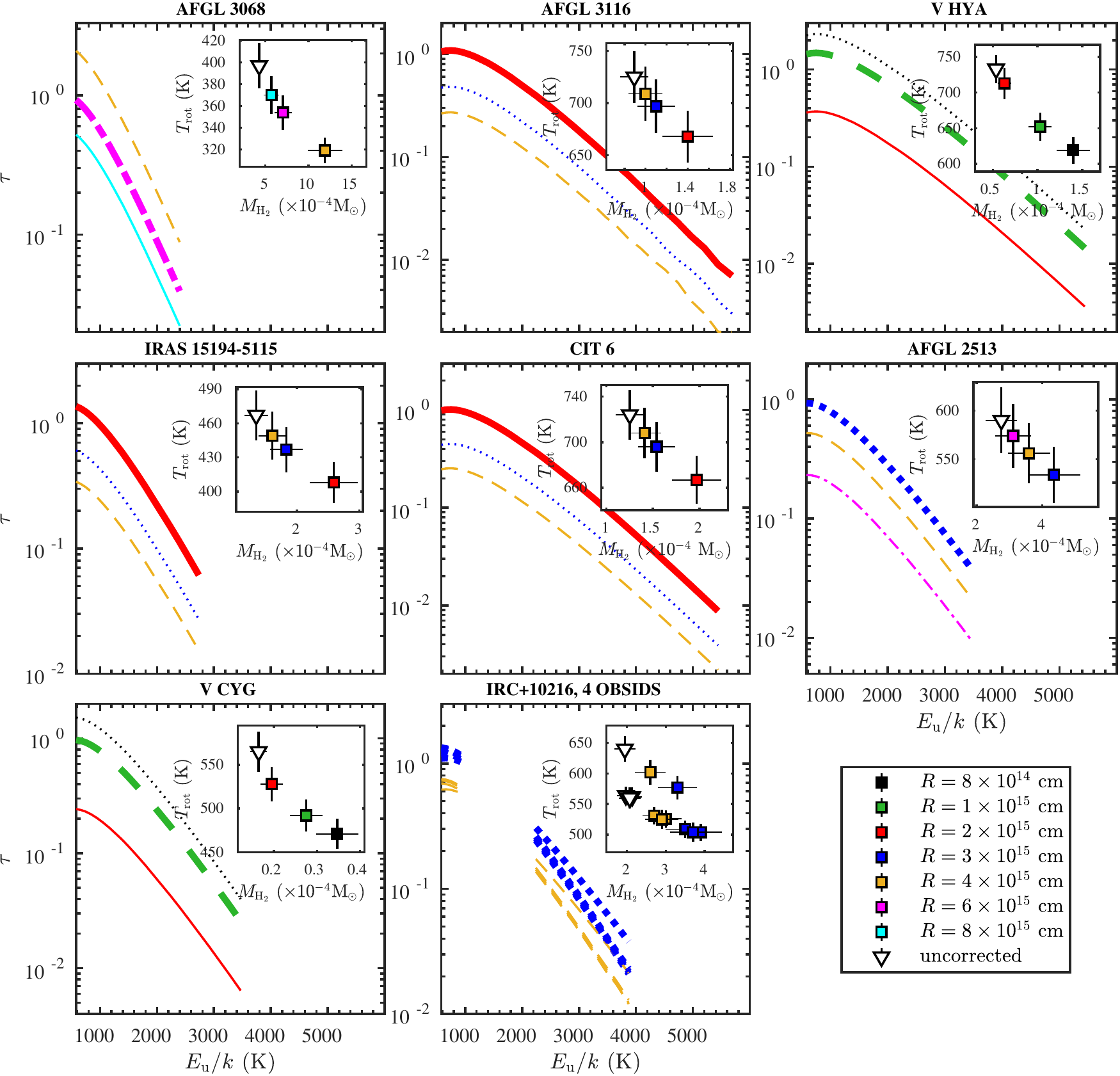}
\caption{Optical depth at line center ($\tau$) as a function of the energy of the upper level ($E_{\rm u}$). $\tau$ was computed for a range of radii; the characteristic radius adopted ($r_{\rm CO}$, Table \ref{tab:rot}) is marked with a thick line. In each panel, the inset shows the values of \trot\ and \mass\ after the $C_\tau$ correction has been applied for the corresponding radii, along with the opacity-uncorrected results (open triangle).} \label{fig:opacity}
\end{figure*}
\begin{figure*}[!htp]
\ContinuedFloat
\includegraphics[]{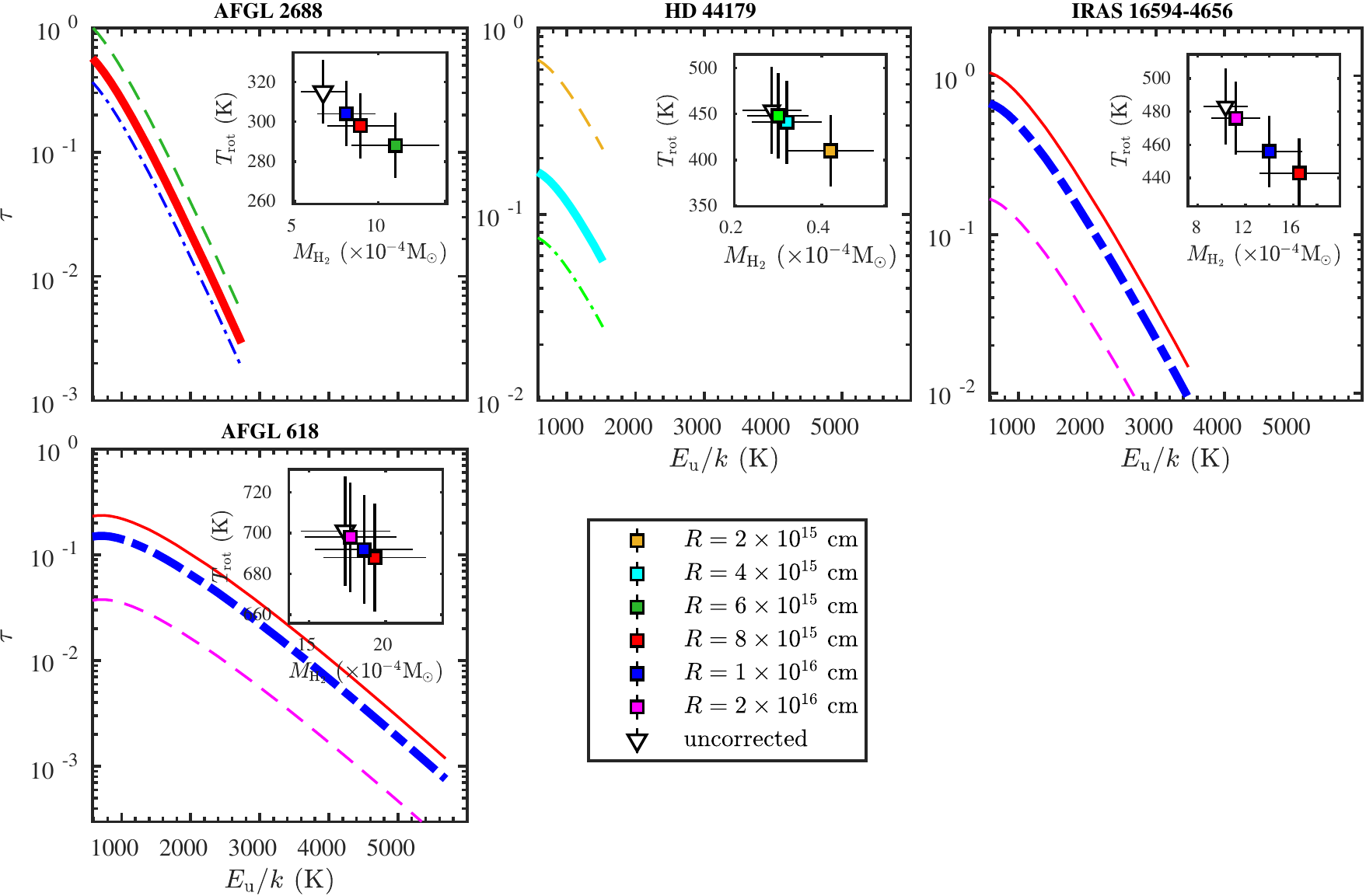} 
%\caption{Continued.} 
\end{figure*}
\begin{figure*}[!htp]
\ContinuedFloat
\includegraphics[]{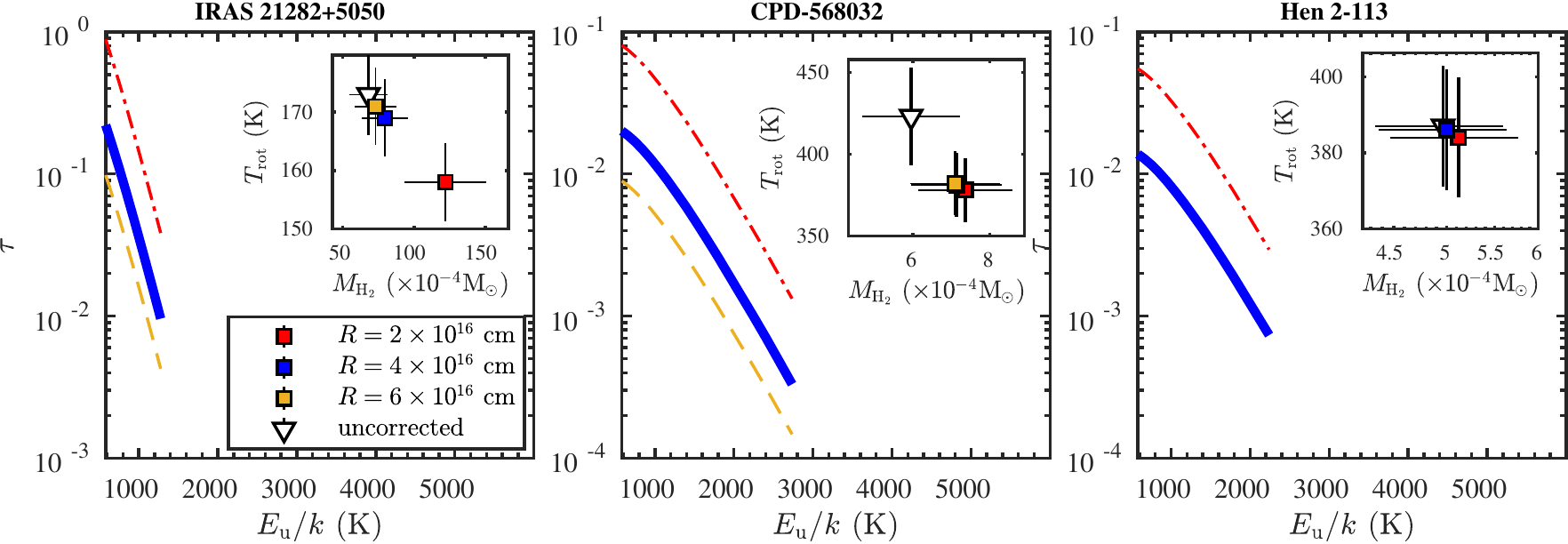} 
\caption{Continued.} 
\end{figure*}
In case of optically thick emission, an optical depth correction factor $\rm C_{\tau}$ should be added to the RD to compensate for an underestimation of the mass within the CO-emitting volume.
%% optical depth effects also result in an overestimate of \trot. 
The optical depth at the line center  ($\tau$) of a given CO\,(\ju-\jl) transition
is:
\begin{equation}
\tau = \frac{A_{\rm ij}\lambda^3_0 N^{\rm col}_{\mathrm{u}}}{8\pi V}\times[e^{(h\nu/kT)}-1] \label{eq:Tau}
\end{equation}
\noindent where $V = v_{\rm exp} \sqrt{\pi}/(2\sqrt{\log 2})~[\rm km~s^{-1}]$ with $v_{\rm exp}$ being the expansion velocity of the gas, $N^{\rm col}_{\mathrm{u}}=N_{\mathrm{u}}/\pi r_{\rm CO}^2$ is the column density of the upper level, $A_{\rm ij}$ is the Einstein coefficient for spontaneous emission and $\lambda_0$ is the peak wavelength. It follows that the correction factor is written as:
\begin{equation}
C_{\tau} = \frac{\tau}{1-e^{-\tau}}
\end{equation}

To compute \ctau, we perform a first fit of the RD datapoints, $\ln\left(\frac{N_{\rm u}}{g_{\rm u}}\right)_0$, starting with null opacity correction ($\ln C_\tau = 0$) which yields initial (or opacity-uncorrected) values of \trot\ and \nco. We use these values to calculate $\tau$ and \ctau\ and to apply the opacity correction, that is, $\ln\left(\frac{N_{\rm u}}{g_{\rm u}}\right)_1$ = $\ln\left(\frac{N_{\rm u}}{g_{\rm u}}\right)_0 + \ln C_\tau$.
%Since \ctau\ depends on the wavelength of the transition,
%the opacity correction results in a larger value of the mass and a lower value of \trot. 
A second fit to $\ln\left(\frac{N_{\rm u}}{g_{\rm u}}\right)_1$ is then performed, which renders the so-called opacity-corrected values of \trot\ and
\nco\ (Table\,\ref{tab:rot}).

We want to highlight that, as explained in \cite{1999ApJ...517..209G}, the opacity correction is only reliable for moderate values of the optical depth. For this reason, for all objects in our sample, the minimum acceptable value of \rco\ used to compute the CO column density is always chosen so that it results in values of $\tau$ close to, but smaller than, unity (Fig.\,\ref{fig:opacity}). Indeed, the envelope layers where $\tau\sim$1 are the deepest regions observationally accessible, because of the almost null escape probability (1/\ctau\,$\raw$0) from deeper, optically thicker regions. As seen in Fig.\,\ref{fig:opacity}, for \rco$\la$1\ex{15}\,cm the opacity of the CO\,$J$=14-13 line (\eu=580\,K), which is the  optically thickest transition in our sample, becomes larger than 1 in all our targets. The expected size of the CO-emitting volume
in our sample is discussed in detail in Section.\,\ref{section:RD}.

The optical depth of the line is also sensitive to the expansion velocity of the gas (Eq.\,\ref{eq:Tau}), which is unconstrained from PACS data and has been assumed to be the terminal expansion velocity of the AGB CSE from the literature (values and references are given in Table\,\ref{tab:1}). In the case of pAGBs and yPNe, we adopt the average expansion velocity of the bulk of the envelope from different previous works. The uncertainty of \vexp\ is normally less than 10\%.
% % ============================================= %
\begin{figure}
\centering
\includegraphics[width=\linewidth]{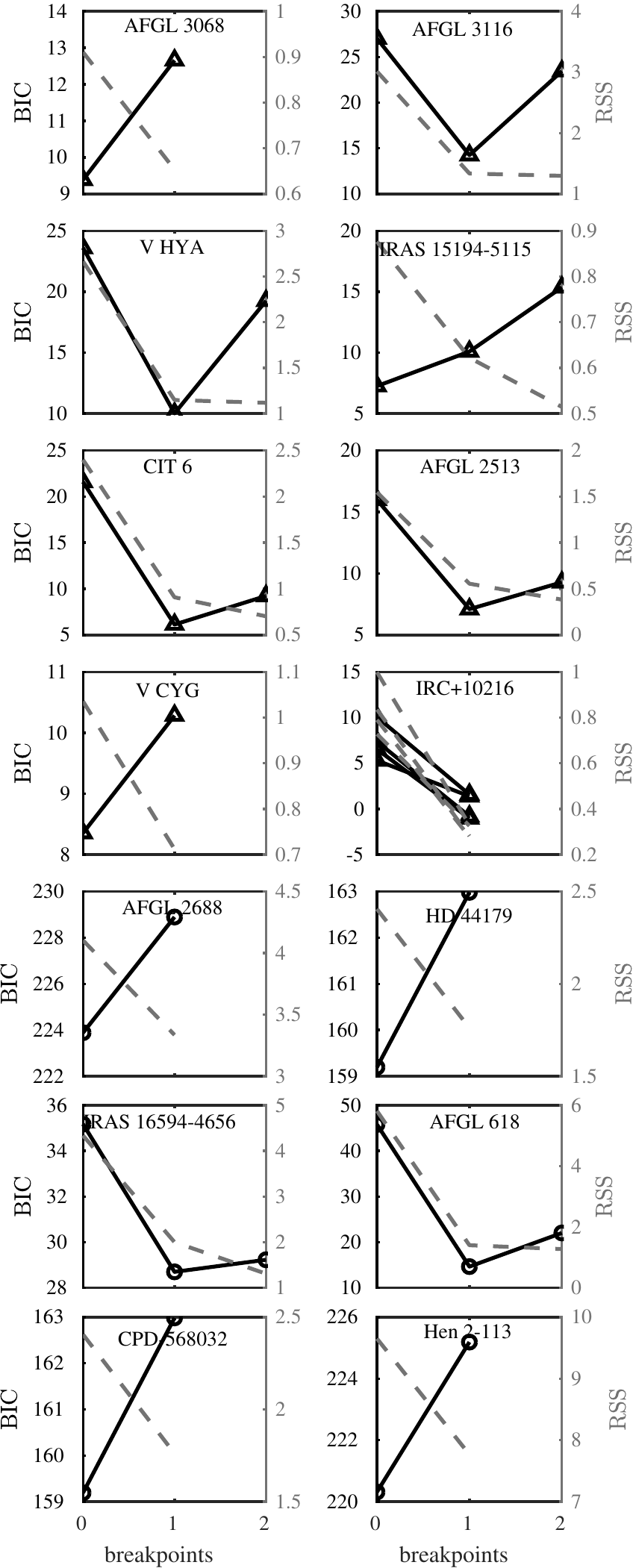} 
\caption{Bayesian information criterion and residual sum of squares as a function of number of breakpoints. The BIC is given by the solid line and the RSS is given by the dashed line. The statistic was not computed for IRAS\,21282+5050 due to insufficient number of points.} 
\label{fig:stats2}
\end{figure}
% % ============================================= %
\subsection{Model selection with BIC}
\label{appendix:multiFit}

Here we describe the method used to automatically find a break point in the linear relation in the rotational diagram using the Bayesian information criterion (BIC)\footnote{Implemented in R using the package \textit{strucchange} \citep{JSSv007i02}.}.

Our approach concerns the minimization of the RSS (residual sum of squares) by computing the BIC for a range of possible locations of break points. The BIC can be regarded as a rough estimate of the Bayes factor \citep{schwarz1978}, and it is given by:
\begin{equation}
	BIC = n \ln(RSS/n) + k \ln(n)
\end{equation}
\noindent where a penalty term, $ k \ln(n)$, penalizes model complexity  depending on the number of parameters $k$ and data points $n$ \citep[see also][]{Maindonald:2010:DAG:1875091}. We also tested the Akaike's information criterion (AIC) \citep{1974ITAC...19..716A}, but we concluded that for this particular application it tends to overfit essentially because the penalty term is smaller. In fact, due to the small number of data points, we are not interested in splitting the rotational diagram too much in order to obtain robust results. It can happen that BIC suggests more than one breakpoint due to the smaller accuracy of the higher $J$ line fluxes, and in that case only the first break point is taken into account. Obvious line blends were excluded upon the computation of both statistics. 

Figure \ref{fig:stats2} shows the BIC test in a graphical way where we see that the BIC curve is usually minimal at 0 or 1 breakpoints. For example for AFGL\,2513 it is shown that the residuals of the fit keep decreasing with 2 breakpoints, although this is strongly penalized by the BIC, so a 2-component fit is favored. On the contrary, in AFGL\,3068, IRAS\,15194-5115 and IRC+10216 a single temperature component suffices to reproduce the data. 

We note that these methods do not prove that a double temperature component is physically true, neither they provide an alternative explanation for the trends in the data. They simply highlight hidden patterns in the residuals which can be caused by many effects, being line blend the most obvious among them, or heteroscedasticity. 

% ........................................ %
\section{Tables}
\label{appendix:tables}

\begin{table*}
\centering
\caption{Characterization of PACS observations.}
\label{tab:1}
\begin{tabular}{llllllll}
\hline\hline
\multicolumn{2}{l}{\multirow{2}{*}{Target name}} & \multirow{2}{*}{Alt. name} & \multirow{2}{*}{Class} & \multirow{2}{*}{R.A. (J2000)} & \multirow{2}{*}{DEC (J2000)} & \multirow{2}{*}{OBSID} & \multirow{2}{*}{Obs. date} \\
\multicolumn{2}{l}{} &  &  &  &  &  &   \\ \hline
\multicolumn{2}{l}{ \object{AFGL 3068}} & LL Peg & C-rich AGB & $23^{\rm h} 19^{\rm m} 12^{\rm s}.39$ & $17º 11' 35\arcsec.4$ & \begin{tabular}[c]{@{}c@{}}1342199417\\ 1342199418\end{tabular} & 2010-06-30  \\
\hline
\multicolumn{2}{l}{\object{AFGL 3116}} & LP And & C-rich AGB & $23^{\rm h} 34^{\rm m} 27^{\rm s}.67$ & $43º 33' 2\arcsec.52$ & \begin{tabular}[c]{@{}c@{}}1342212512\\ 1342212513\end{tabular} & 2011-01-11  \\ 
\hline
\multicolumn{2}{l}{\object{IRC+10216}} & CW Leo & \multicolumn{1}{l}{C-rich AGB} & \multicolumn{1}{l}{$9^{\rm h} 47^{\rm m} 57^{\rm s}.41$} & \multicolumn{1}{l}{$13º 16' 43\arcsec.60$} & \begin{tabular}[c]{@{}c@{}}1342221889\\1342241328\\ 1342245395\\1342246381\\1342253754\\1342255741\\1342256262\end{tabular}  & \begin{tabular}[c]{@{}c@{}}2011-05-29\\2011-10-25\\ 2012-05-05\\2012-05-30\\2012-10-21\\2012-11-21\\2012-11-30\end{tabular}  \\
\hline
\multicolumn{2}{l}{\object{IRAS 15194-5115}} & II Lup & C-rich AGB & $15^{\rm h} 23^{\rm m} 4^{\rm s}.91$ & $-51º 25' 59\arcsec.0$ & \begin{tabular}[c]{@{}c@{}}1342215685\\ 1342215686\end{tabular} & 2010-03-10  \\
\hline
\multicolumn{2}{l}{\object{CIT 6}} & RW Lmi & C-rich AGB & $10^{\rm h} 16^{\rm m} 2^{\rm s}.28$ & $30º 34' 18\arcsec.48$ & \begin{tabular}[c]{@{}c@{}}1342197799\\ 1342197800\end{tabular} & 2010-06-05 \\
\hline
\multicolumn{2}{l}{\object{AFGL 2513}} & V1969 Cyg & C-rich AGB & $20^{\rm h} 09^{\rm m} 14^{\rm s}.25$ & $31º 25' 44\arcsec.9$ & \begin{tabular}[c]{@{}c@{}}1342270010\\ 1342269936\end{tabular} & \begin{tabular}[c]{@{}c@{}}2013-04-14\\ 2013-04-12\end{tabular}  \\
\hline
\multicolumn{2}{l}{\object{V Cyg}} &  & C-rich AGB & $20^{\rm h} 41^{\rm m} 18^{\rm s}.27$ & $48º 08' 28\arcsec.8$ & \begin{tabular}[c]{@{}c@{}}1342208939\\ 1342208940\end{tabular} & 2010-11-15  \\ \hline
\multicolumn{2}{l}{\object{V Hya}} &  & C-rich AGB & $10^{\rm h} 51^{\rm m} 37^{\rm s}.25$ & $21º 15' 00\arcsec.3$ & \begin{tabular}[c]{@{}c@{}}1342197790\\ 1342197791\end{tabular} & 2010-06-05  \\
\hline
\multicolumn{2}{l}{\object{AFGL 2688}} & Egg Nebula & C-rich post-AGB & $21^{\rm h} 2^{\rm m} 18^{\rm s}.74$ & $36º 41' 37\arcsec.68$ & \begin{tabular}[c]{@{}c@{}}1342199233\\ 1342199234\end{tabular} & 2010-06-26  \\ \hline
\multicolumn{2}{l}{HD 44179} & Red Rectangle & mixed post-AGB & $6^{\rm h} 19^{\rm m} 58^{\rm s}.22$ & $-10º 38' 14\arcsec.7$ & \begin{tabular}[c]{@{}c@{}}1342220928\\ 1342220929\end{tabular} & 2011-04-30  \\
\hline
\multicolumn{2}{l}{\object{IRAS 16594-4656}} & Water Lily Nebula & mixed post-AGB & $17^{\rm h} 03^{\rm m} 10^{\rm s}.03$ & $-47º 00' 27\arcsec.7$ & \begin{tabular}[c]{@{}c@{}}1342228414\\ 1342228415\end{tabular} & 2011-09-10  \\ \hline
\multicolumn{2}{l}{\object{AFGL 618}} & CRL 618 & C-rich yPNe & $4^{\rm h} 42^{\rm m} 53^{\rm s}.66$ & $36º 6' 53\arcsec.28$ & \begin{tabular}[c]{@{}c@{}}1342225838\\ 1342225839\end{tabular} & 2011-08-07  \\ \hline
\multicolumn{2}{l}{\object{IRAS 21282+5050}} &  & C-rich yPNe & $21^{\rm h} 29^{\rm m} 58^{\rm s}.42$ & $51º 3' 59\arcsec.76$ & \begin{tabular}[c]{@{}c@{}}1342220741\\ 1342223375\end{tabular} & \begin{tabular}[c]{@{}c@{}}2011-05-12\\ 2011-06-30\end{tabular}  \\ \hline
\multicolumn{2}{l}{\object{CPD-568032}} & Hen 3-1333 & mixed yPNe & $17^{\rm h} 7^{\rm m} 0^{\rm s}.87$ & $-56º 54' 48\arcsec.0$ & \begin{tabular}[c]{@{}c@{}}1342228201\\ 1342228202\end{tabular} & 2011-09-06  \\
\hline
\multicolumn{2}{l}{\object{Hen 2-113}} &  & mixed yPNe & $14^{\rm h} 59^{\rm m} 53^{\rm s}.52$ & $-54º 18' 07\arcsec.20$ & \begin{tabular}[c]{@{}c@{}}1342225142\\ 1342225143\end{tabular} & 2011-08-02  \\ \hline
\end{tabular}
\tablefoot{Target name, alternative name, evolutionary classification, coordinates given by right ascension (R.A) and declination (DEC), observation identifier (OBSID) and observation date. The spectra correspond to bands B2B and R1 in IRC+10216 and B2A, B2B and R1 for the remaining targets.}
\end{table*}

%reference for HD 44179 being mixed chemistry Waters et al 1998
%reference for IRAS 16594 being mixed chemistry Woods et al A&A 429, 977–992 (2005), and references therein

\begin{table*}[tbp]
\centering
\caption{Properties of stars and their CSEs taken from the bibliography.}
\label{tab:2}
\begin{tabular}{lccccccc}
\hline\hline
Target name & Var. & $T_{\rm eff}$ (K) & $d$ (pc) & $v_{\rm exp}~\rm (km~s^{-1})$ & \multicolumn{2}{c}{$\dot{M} \rm~(M_{\sun}~yr^{-1})$} & references \\ \hline
AFGL 3068 & Mira & 2000 & 1100 & 13 & (0.9, 6) $\times10^{-5}$ & (0.8, 6.8) $\times10^{-5}$ & 2, 10, 23, 29 \\
AFGL 3116 & Mira & 2000 & 630 & 14 & (4.6, 12) $\times10^{-6}$ & (1, 11.8) $\times10^{-6}$ & 2, 10, 23 \\
IRC+10216 & Mira & 2330 & 150 & 14.5 & (0.2, 4) $\times10^{-5}$ & (0.05, 3.2) $\times10^{-5}$ & 2, 12, 23, 24, 31 \\
IRAS 15194-5115 & Mira & 2400 & 500 & 23 & (0.4, 1.5) $\times10^{-5}$ & (0.08, 2.5) $\times10^{-5}$ & 2, 10, 13 \\
CIT 6 & SRa & 2450 & 440 & 21 & (5, 6) $\times10^{-6}$ & (1.2, 11) $\times10^{-6}$ & 2, 10, 23 \\
AFGL 2513 & Mira & 2500 & 1760 & 26 & 2 $\times10^{-5}$ & (1, 4) $\times10^{-5}$ & 6, 14 \\
V Cyg & Mira & 2580 & 271 & 15 & (0.4, 6.3) $\times10^{-6}$ & (0.08, 1.7) $\times10^{-6}$ & 2, 10, 15 \\
V Hya & SR/Mira & 2650 & 380 & 24 & (0.25, 6) $\times10^{-5}$ & (0.03, 3) $\times10^{-5}$ & 1, 10, 17, 25 \\ \hline
AFGL 2688 &  & 7250 & 340 & 20 & (0.7, 2) $\times10^{-4}$ & (0.06, 0.6) $\times10^{-4}$ & 3, 8, 19, 27 \\
HD 44179 &  & 7750 & 710 & 8.3 & $10^{-7}$-$10^{-5}$ & $10^{-7}$-$10^{-5}$ & 5, 7, 10, 32 \\
IRAS 16594-4656 &  & 10000 & 1800 & 14 & 1 $\times10^{-5}$ & (0.4, 3.7) $\times10^{-5}$ & 9, 18 \\ \hline
AFGL 618 &  & 33000 & 900 & 80 & (0.3, 2) $\times10^{-4}$ & (0.06, 0.7) $\times10^{-4}$ & 4, 16, 26, 27, 30 \\
IRAS 21282+5050 &  & 30000 & 2000 & 14 & (2, 10) $\times10^{-5}$ & (1.5, 7.5) $\times10^{-5}$ & 11, 21, 26 \\
CPD-568032 &  & 30000 & 1530 & 22.6 & (1.2-4) $\times10^{-6}$ & (1.2-4) $\times10^{-6}$ & 20, 28 \\
Hen 2-113 &  & 30900 & 1230 & 23 & (6.3-8) $\times10^{-7}$ & (6.3-8) $\times10^{-7}$ & 20, 28 \\ \hline
\end{tabular}
\tablefoot{Type of variability, effective temperature ($T_{\rm eff}$), distance ($d$), expansion velocity ($v_{\rm exp}$) and range of gas mass-loss rate ($\dot{M}$) including uncertainty, with and without scaling to the same $X_{\rm CO}$, $d$ and $v_{\rm exp}$ used in this paper on the right and left columns, respectively. In case there is only one $\dot{M}$ value  in literature with no estimated uncertainty we assumed an error factor of 3 which is the typical uncertainty in mass-loss rates of AGB stars \citep[e.g.][]{2010A&A...523A..18D,2008A&A...487..645R}.}
\tablebib{$~{}^{(1)}$\citet{2005A&A...429..235B}, $~{}^{(2)}$\citet{2014A&A...566A.145R}, $~{}^{(3)}$\citet{2012ApJ...745..188B}, $~{}^{(4)}$\citet{2004ApJ...617.1142S}, $~{}^{(5)}$\citet{2016A&A...593A..92B} $~{}^{(6)}$\citet{2006A&A...445.1069G}, $~{}^{(7)}$\citet{2002A&A...393..867M}, $~{}^{(8)}$\citet{2012MNRAS.425..997I}, $~{}^{(9)}$\citet{2015ApJ...802...39M} $~{}^{(10)}$\citet{2010A&A...523A..18D}, ${}^{(11)}$\citet{2003ApJ...585..475H}, $~{}^{(12)}$\citet{2015A&A...575A..91C},$~{}^{(13)}$\citet{1999A&A...345..841R}, $~{}^{(14)}$\citet{2002A&A...390..511G}, $~{}^{(15)}$\citet{2010A&A...521L...5N}, $~{}^{(16)}$\citet{2016ApJ...820..134H}, $~{}^{(17)}$\citet{1997A&A...326..318K}, $~{}^{(18)}$\citet{2005A&A...429..977W}, $~{}^{(19)}$\citet{2009ApJ...690..837M}, $~{}^{(20)}$\citet{1998MNRAS.296..419D}, $~{}^{(21)}$\citet{1988A&A...198L...1L}, $~{}^{(22)}$\citet{2016A&A...593A..92B}, $~{}^{(23)}$\citet{2006A&A...450..167T}, $~{}^{(24)}$\citet{2012A&A...539A.108D},  $~{}^{(25)}$\citet{1999A&A...351...97K},  $~{}^{(26)}$\citet{0004-637X-509-1-392}, $~{}^{(27)}$\citet{2001A&A...377..868B}, $~{}^{(28)}$\citet{1996A&A...312..167L}, $~{}^{(29)}$\citet{2012ApJS..203...16S}, $~{}^{(30)}$\citet{2010A&A...518L.144W},
$~{}^{(31)}$\citet{2018A&A...610A...4G},
$~{}^{(32)}$\citet{2013A&A...552A.116B}}
\end{table*}

\begin{table*}[]
\caption{Carriers of some of the spectral lines detected towards our sample of C-rich evolved stars.}
\label{tab:lines}  
\begin{tabular}{l|cccccc|cccc|c}
\hline\hline
\multicolumn{1}{c|}{\multirow{2}{*}{Target Name}} & \multicolumn{6}{c|}{Molecular} & \multicolumn{4}{c|}{Atomic/ionized} & Dust \\
\multicolumn{1}{c|}{} & CO & HCN & CS & OH & $\rm H_{2}O$ & CH+ & \begin{tabular}[c]{@{}c@{}}{[}OI{]}\\ $63.18\rm\mu m$\end{tabular} & \begin{tabular}[c]{@{}c@{}}{[}NII{]}\\ $121.89\rm\mu m$\end{tabular} & \begin{tabular}[c]{@{}c@{}}{[}OI{]}\\ $145.52\rm\mu m$\end{tabular} & \multicolumn{1}{c|}{\begin{tabular}[c]{@{}c@{}}{[}CII{]}\\ $157.74\rm\mu m$\end{tabular}} & \multicolumn{1}{c}{\begin{tabular}[c]{@{}c@{}}forsterite\\ $69\rm~\mu m$\end{tabular}} \\ \hline
AFGL 3068 & \checkmark & \checkmark & \checkmark &  & \checkmark &  &  &  &  &  &  \\
AFGL 3116 & \checkmark & \checkmark & \checkmark &  & \checkmark &  &  &  &  &  &  \\
IRC+10216 & \checkmark & \checkmark & \checkmark &  & \checkmark &  &  &  &  &  &  \\
IRAS 15194-5115 & \checkmark & \checkmark & \checkmark &  & \checkmark &  &  &  &  &  &  \\
CIT 6 & \checkmark & \checkmark & \checkmark &  & \checkmark &  &  &  &  &  &  \\
AFGL 2513 & \checkmark & \checkmark & \checkmark &  & \checkmark &  &  &  &  &  &  \\
V Cyg & \checkmark & \checkmark & \checkmark &  & \checkmark &  &  &  &  &  &  \\
V Hya & \checkmark & \checkmark & \checkmark &  & \checkmark &  &  &  &  &  &  \\ \hline
AFGL 2688 & \checkmark & \checkmark &  &  & \checkmark &  &  &  &  &  &  \\
HD 44179 & \checkmark &  &  &  &  &  & \checkmark &  & \checkmark & \checkmark & \checkmark \\
IRAS 16594-4665 & \checkmark &  &  & \checkmark &  & \checkmark & \checkmark &  & \checkmark & \checkmark &  \\ \hline
AFGL 618 & \checkmark & \checkmark &  & \checkmark & \checkmark &  & \checkmark &  & \checkmark & \checkmark &  \\
IRAS 21282+5050 & \checkmark &  &  &  &  & \checkmark & \checkmark & \checkmark & \checkmark & \checkmark &  \\ 
CPD-568032 & \checkmark &  &  & \checkmark &  & \checkmark & \checkmark &  & \checkmark & \checkmark & \checkmark \\
Hen 2-113 & \checkmark &  &  & \checkmark &  & \checkmark & \checkmark & \checkmark & \checkmark & \checkmark & \checkmark \\
\cline{1-12}
\end{tabular}
\end{table*}

\begin{sidewaystable*}[]
\centering
\caption{Line fluxes of CO rotational transitions in a sample of C-rich evolved stars.} %The line fluxes measured in a PACS spectrum can be converted from [Jy $\rm \mu m$] to [W $\rm m^{-2}$] via $F=2.99\times 10^{-12}F\lambda^{-2}_{0}$, where $\lambda_{0}$ is the line central wavelength.}
\label{tab:fluxes}
\begin{tabular}{lccccrrrrr}
\hline\hline
Transition & \multicolumn{1}{c}{$\rm \nu_{\rm rest}~(MHz)$} & \multicolumn{1}{c}{$ \rm \lambda_{\rm rest}~(\mu m)$} & \multicolumn{1}{c}{$\rm E_{\rm up}/k~(K)$} & \multicolumn{1}{c|}{$\rm A_{ij}~(s^{-1})$} &  & &$F~(\rm \times 10^{-16} ~W~m^{-2})$ &  &  \\ \hline
$\rm {}^{12}CO~(\nu =0)$ & & & & \multicolumn{1}{r|}{} & CIT 6& V Hya& IRAS 15194-5115& AFGL 2513&V Cyg\\
$J = 14 \rightarrow$ 13 & 1611794 & 185.999 & 580.497 & \multicolumn{1}{r|}{0.00027} & 8.9$\pm$0.3 & 4.3$\pm$0.2 & 4.3$\pm$0.4 & 0.88$\pm$0.07 & 1.8$\pm$0.2 \\
$J = 15 \rightarrow$ 14 & 1726603 & 173.631 & 663.361 & \multicolumn{1}{r|}{0.00034} & 7.7$\pm$0.3 & 4.6$\pm$0.2 & 3.9$\pm$0.5 & 0.96$\pm$0.05 & 1.8$\pm$0.1 \\
$J = 16 \rightarrow$ 15 & 1841346 & 162.812 & 751.733 & \multicolumn{1}{r|}{0.00041} & 9.2$\pm$0.5 & 5.6$\pm$0.2 & 5.0$\pm$0.3 & 1.07$\pm$0.06 & 2.6$\pm$0.1 \\
$J = 17 \rightarrow$ 16 & 1956018 & 153.267 & 845.608 & \multicolumn{1}{r|}{0.00048} & 11.3$\pm$0.6 & 6.1$\pm$0.2 &4.9$\pm$0.3 & 1.29$\pm$0.08 & 2.3$\pm$0.1 \\
$J = 18 \rightarrow$ 17 & 2070616 & 144.784 & 944.983 & \multicolumn{1}{r|}{0.00057} & 9.6$\pm$0.7 & 5.3$\pm$0.2 &3.7$\pm$0.5 & 0.99$\pm$0.07 & 2.7$\pm$0.1 \\
$J = 19 \rightarrow$ 18 & 2185135 & 137.196 & 1049.854 & \multicolumn{1}{r|}{0.00067} & 8.1$\pm$0.6 & 5.0$\pm$0.2 &4.1$\pm$0.3 & 0.84$\pm$0.08 & 2.7$\pm$0.2 \\
$J = 20 \rightarrow$ 19 & 2299570 & 130.369 & 1160.217 & \multicolumn{1}{r|}{0.00077} & 16.2$\pm$0.9 & 8.9$\pm$0.3 &7.4$\pm$0.6 & 1.7$\pm$0.1 & 3.9$\pm$0.3 \\
$J = 21 \rightarrow$ 20 & 2413917 & 124.193 & 1276.068 & \multicolumn{1}{r|}{0.00088} & 11$\pm$1.0 & 6.4$\pm$0.3 &4$\pm$1 & 1.0$\pm$0.1 & 2.6$\pm$0.2 \\
$J = 22 \rightarrow$ 21 & 2528172 & 118.581 & 1397.402 & \multicolumn{1}{r|}{0.00101} & 10.9$\pm$0.6 & 5.9$\pm$0.2 &4.3$\pm$0.8 & 1.0$\pm$0.07 & 2.7$\pm$0.2 \\
$J = 23 \rightarrow$ 22 & 2642330 & 113.458 & 1524.215 & \multicolumn{1}{r|}{0.00114} & 16$\pm$2 & 9$\pm$1 &5$\pm$2 & 1.5$\pm$0.2 & 4.1$\pm$0.4 \\
$J = 24 \rightarrow$ 23 & 2756388 & 108.763 & 1656.502 & \multicolumn{1}{r|}{0.00128} & 11$\pm$2 & 5.6$\pm$0.5 &5$\pm$2$^{*}$ & 0.9$\pm$0.1 & 1.9$\pm$0.3 \\
$J = 25 \rightarrow$ 24 & 2870339 & 104.445 & 1794.258 & \multicolumn{1}{r|}{0.00143} & 9$\pm$1 & 5.9$\pm$0.5 &3.8$\pm$0.8 & 0.8$\pm$0.1 & 2.7$\pm$0.4$^{*}$ \\
$J = 28 \rightarrow$ 27 & 3211519 & 93.349 & 2240.286 & \multicolumn{1}{r|}{0.00194} & 6.8$\pm$0.4 & 5.3$\pm$0.3 &2.5$\pm$0.4$^{*}$ & 0.7$\pm$0.1 & 1.2$\pm$0.2 \\
$J = 29 \rightarrow$ 28 & 3325005 & 90.163 & 2399.862 & \multicolumn{1}{r|}{0.00213} & 11$\pm$1.0 & 5.3$\pm$0.4 &1.6$\pm$0.4 & 0.8$\pm$0.1 & 1.8$\pm$0.2 \\
$J = 30 \rightarrow$ 29 & 3438365 & 87.19 & 2564.879 & \multicolumn{1}{r|}{0.00232} & 9.7$\pm$0.9 & 6.4$\pm$0.5$^{*}$ &3.2$\pm$0.8 & 1.5$\pm$0.2$^{*}$ & 2.1$\pm$0.4$^{*}$ \\
$J = 31 \rightarrow$ 30 & 3551592 & 84.411 & 2735.331 & \multicolumn{1}{r|}{0.00252} & 9.8$\pm$0.9 & 4.9$\pm$0.3 &1.7$\pm$0.6 & 0.9$\pm$0.1 & 1.6$\pm$0.3 \\
$J = 32 \rightarrow$ 31 & 3664684 & 81.806 & 2911.209 & \multicolumn{1}{r|}{0.00274} & 6.1$\pm$0.9 & 4.1$\pm$0.6 & … & 0.7$\pm$0.2 & 1.0$\pm$0.3 \\
$J = 33 \rightarrow$ 32 & 3777636 & 79.359 & 3092.509 & \multicolumn{1}{r|}{0.00295} & 5.4$\pm$0.9 & 5.2$\pm$0.6 & … & 0.7$\pm$0.2 & 0.96$\pm$0.3 \\
$J = 34 \rightarrow$ 33 & 3890443 & 77.059 & 3279.223 & \multicolumn{1}{r|}{0.00318} & 9$\pm$1 & 4.6$\pm$0.4 & … & 0.7$\pm$0.1$^{*}$ & 1.1$\pm$0.3$^{*}$ \\
$J = 35 \rightarrow$ 34 & 4003101 & 74.89 & 3471.344 & \multicolumn{1}{r|}{0.0034}  & 8$\pm$2$^{*}$ & 3.9$\pm$0.8 & … & 0.8$\pm$0.1$^{*}$ & 1.2$\pm$0.3$^{*}$ \\
$J = 36 \rightarrow$ 35 & 4115606 & 72.843 & 3668.864 & \multicolumn{1}{r|}{0.00364} & 4.7$\pm$0.9$^{*}$ & 3.2$\pm$0.3 & … & … & … \\
$J = 37 \rightarrow$ 36 & 4227953 & 70.907 & 3871.775 & \multicolumn{1}{r|}{0.00388} & 5$\pm$1 & 3.0$\pm$0.6$^{*}$ & … & … & … \\
$J = 38 \rightarrow$ 37 & 4340138 & 69.074 & 4080.071 & \multicolumn{1}{r|}{0.00412} & 3.7$\pm$0.4 & 2.8$\pm$0.3 & … & … & … \\
$J = 39 \rightarrow$ 38 & 4452157 & 67.336 & 4293.743 & \multicolumn{1}{r|}{0.00437} & 3$\pm$1$^{*}$ & 2.5$\pm$0.5$^{*}$ & … & … & … \\
$J = 40 \rightarrow$ 39 & 4564006 & 65.686 & 4512.783 & \multicolumn{1}{r|}{0.00461} & 3.2$\pm$1.0$^{*}$ & 2.9$\pm$0.7 & … & … & … \\
$J = 41 \rightarrow$ 40 & 4675679 & 64.117 & 4737.183 & \multicolumn{1}{r|}{0.00486} & 1.1$\pm$0.6$^{*}$ & 1.4$\pm$0.4 & … & … & … \\
$J = 42 \rightarrow$ 41 & 4787174 & 62.624 & 4966.933 & \multicolumn{1}{r|}{0.00511} & 2$\pm$1$^{*}$ & 0.9$\pm$0.5$^{*}$ & … & … & … \\
$J = 43 \rightarrow$ 42 & 4898485 & 61.201 & 5202.026 & \multicolumn{1}{r|}{0.00536} & 3.3$\pm$0.9$^{*}$ & 1.8$\pm$0.4$^{*}$ & … & … & … \\
$J = 44 \rightarrow$ 43 & 5009608 & 59.843 & 5442.452 & \multicolumn{1}{r|}{0.00561} & 1.7$\pm$0.6 & 1.9$\pm$0.5 & … & … & … \\
$J = 45 \rightarrow$ 44 & 5120539 & 58.547 & 5688.201 & \multicolumn{1}{r|}{0.00585} & … & … & … & … & … \\ \hline
\end{tabular}
\tablefoot{The ellipsis mark absent or noisy (SNR<3) lines, and asterisks (*) flag line blends which may have caused overestimated fluxes.}
\end{sidewaystable*}

\begin{sidewaystable*}
\ContinuedFloat
\centering
\caption{Continued. }
\label{my-label2}
\begin{tabular}{rrrrrrrrr}
\hline
AFGL 3068 & AFGL 3116 & AFGL 618 & HD 44179 & IRAS 16594-4656 & AFGL 2688 & Hen 2-113& CPD-568032 & IRAS 21282+5050 \\ 
2.1$\pm$0.2 & 2.7$\pm$0.2 & 39$\pm$1 & 0.4$\pm$0.1 & 4.1$\pm$0.1 & 41$\pm$1 & 2.24$\pm$0.09& 2.4$\pm$0.1 & 2.96$\pm$0.06 \\
2.1$\pm$0.2 & 2.9$\pm$0.2 & 33$\pm$1 & 0.39$\pm$0.09 & 4.1$\pm$0.1 & 36$\pm$1 & 2.16$\pm$0.07 & 2.3$\pm$0.1 & 2.92$\pm$0.05 \\
2.3$\pm$0.1 & 2.8$\pm$0.1 & 40$\pm$1 & 0.4$\pm$0.1 &  3.9$\pm$0.3 & 33$\pm$1 & 2.3$\pm$ 0.5& 2.2$\pm$0.1 & 2.42$\pm$0.06 \\
2.5$\pm$0.2 & 4.1$\pm$0.3 & 37$\pm$1 & 0.53$\pm$0.09 &  3.2$\pm$0.1 & 26$\pm$1 & 2.3$\pm$ 0.1& 2.0$\pm$0.07 & 1.63$\pm$0.04 \\
2.7$\pm$0.3 & 3.5$\pm$0.2 & 38$\pm$1 & 0.6$\pm$0.1 & 2.76$\pm$0.06 & 25$\pm$1 & 2.1$\pm$ 0.1& 1.9$\pm$0.1 & 1.04$\pm$0.07 \\
2.5$\pm$0.2 & 3.0$\pm$0.2 & 39$\pm$2 & 0.45$\pm$0.09 & 2.42$\pm$0.01 & 22$\pm$1 & 2.0$\pm$ 0.1& 1.6$\pm$0.1 & 0.9$\pm$0.1 \\
3.9$\pm$0.2 & 6.1$\pm$0.4 & 44$\pm$1 & 0.4$\pm$0.1 & 2.5$\pm$0.1 & 24$\pm$1 & 2.2$\pm$ 0.1& 2$\pm$0.2 & 0.59$\pm$0.09 \\
2.3$\pm$0.3 & 4.3$\pm$0.5 & 34$\pm$2 & 0.4$\pm$0.1 & 2.8$\pm$0.1 & 17$\pm$1 & 2.0$\pm$ 0.1& 1.7$\pm$0.2 & 0.50$\pm$0.08 \\
2.7$\pm$0.2 & 3.6$\pm$0.2 & 36$\pm$2 & 0.7$\pm$0.1 & 2.2$\pm$0.2 & 18$\pm$1 & 2.4$\pm$ 0.3& 1.3$\pm$0.2 & … \\
2.5$\pm$0.6 & 4.6$\pm$0.9 & 46$\pm$2 & 0.4$\pm$0.2 & 2.7$\pm$0.1 & 14$\pm$2 & 2.2$\pm$0.3& 1.9$\pm$0.2 & … \\
1.6$\pm$0.4 & 3.7$\pm$0.6 & 28$\pm$2$^{*}$ & … & 1.9$\pm$0.1 & 11$\pm$2 & 1.2$\pm$0.2& 1.0$\pm$0.2 & … \\
1.3$\pm$0.5 & 3.4$\pm$0.2 & 31$\pm$2 & … & 2.1$\pm$0.2 & 6$\pm$2 & 1.6$\pm$0.2& 1.2$\pm$0.3 & … \\
1.2$\pm$0.3 & 3.2$\pm$0.3 & 21$\pm$1 & … & 1.6$\pm$0.1 & 6$\pm$1 & 0.7$\pm$0.2& 0.9$\pm$0.1 & … \\
1.9$\pm$0.4$^{*}$& 3.3$\pm$0.3 & 25$\pm$1 & … & 1$\pm$2$^{*}$ & 11$\pm$2$^{*}$ & …& bad fit & … \\
… & 4.5$\pm$0.5$^{*}$ & 20$\pm$1 & … & 1.3$\pm$0.1 & bad fit$^{*}$ & …& … & … \\
… & 2.9$\pm$0.3 & 26$\pm$2 & … & 4.9$\pm$0.8$^{*}$ & 4$\pm$2 & …& … & … \\
… & 2.0$\pm$0.2 & 17$\pm$2 & … & 0.9$\pm$0.1 & … & …& … & … \\
… & 2.4$\pm$0.5 & 15$\pm$2$^{*}$ & … & 0.8$\pm$0.8$^{*}$ & … &… & … & … \\
… & 2.4$\pm$0.7$^{*}$ & 16$\pm$2 & … & 0.7$\pm$0.1 & … & …& … & … \\
… & 2.2$\pm$0.4 & 16$\pm$2 & … & 0.8$\pm$0.1$^{*}$ & … & …& … & … \\
… & 2.3$\pm$0.5 & 11$\pm$1 & … &  … & … &… & … & … \\
… & 3$\pm$1$^{*}$ & 13$\pm$3$^{*}$ & … & … & … & …& … & … \\
… & 1.3$\pm$0.2$^{*}$ & 11$\pm$1 & … &  … & … & …& … & … \\
… & 0.9$\pm$0.3 & 6$\pm$5 & … & … & … & …& … & … \\
… & 0.9$\pm$0.3 & 5$\pm$2 & … & … & … & …& … & … \\
… & … & 10$\pm$2 & … & … & … & …& … & … \\
… & … & 37$\pm$20$^{*}$ & … & … & … & … &…& … \\
… & … & 9$\pm$3 & … & … & … &  … & …&… \\
… & … & 6$\pm$2 & … &  … & … & … & …& … \\
… & … & 3$\pm$2 & … & … & … & … & … & … \\ \hline
\end{tabular}
\end{sidewaystable*}

\begin{sidewaystable*}[]
\centering
\caption{CO line fluxes of seven observations of IRC+10216. Asterisks flag known line blends.}
\label{tab:fluxes_10216}
\begin{tabular}{lccccccccccc}
\hline\hline
\multicolumn{1}{c}{Transition} & \multicolumn{1}{c}{$\rm \nu_{\rm rest}~(MHz)$} & \multicolumn{1}{c}{$ \rm \lambda_{\rm rest}~(\mu m)$} & \multicolumn{1}{c}{$\rm E_{\rm up}/k~(K)$} & \multicolumn{1}{c|}{$\rm A_{\rm ij}~(s^{-1})$} &  &  & & $F~\rm(\rm \times 10^{-16} ~W~m^{-2})$ &  & \\ \hline
$\rm {}^{12}CO~(\nu =0)$ &  &  &  & \multicolumn{1}{l|}{} & OD 745 & OD 894 & OD 1087 & OD 1113& OD 1257 & OD 1288 & OD 1296 \\
$J = 14 \rightarrow$ 13 & 1611794 & 185.99 & 580.5 & \multicolumn{1}{l|}{0.00027} & 85$\pm$3 & 98$\pm$10 & 92$\pm$4 & 90$\pm$5 & 97$\pm$3 & 86$\pm$8 & 100$\pm$3 \\
$J = 15 \rightarrow$ 14 & 1726603 & 173.63 & 663.36 & \multicolumn{1}{l|}{0.00034} & 74$\pm$5 & 103$\pm$13 & 80$\pm$6 & 108$\pm$30 & 82$\pm$5 & 88$\pm$8 &80$\pm$5 \\
$J = 16 \rightarrow$ 15 & 1841346 & 162.81 & 751.73 & \multicolumn{1}{l|}{0.00041} & 89$\pm$4 & 100$\pm$24 & 93$\pm$3 & 93$\pm$12 & 78$\pm$2 & 89$\pm$12 &79$\pm$3 \\
$J = 17 \rightarrow$ 16 & 1956018 & 153.27 & 845.61 & \multicolumn{1}{l|}{0.00048} & 92$\pm$7 & & 99$\pm$7 & & 92$\pm$7 & &89$\pm$6 \\
$J = 18 \rightarrow$ 17 & 2070616 & 144.78 & 944.98 & \multicolumn{1}{l|}{0.00057} & 82$\pm$7 & & 85$\pm$6 & & 110$\pm$5 & &102$\pm$5 \\
 &  &  &  & \multicolumn{1}{l|}{} &  &  &  &  \\
$J = 28 \rightarrow$ 27 & 3211519 & 93.35 & 2240.29 & \multicolumn{1}{l|}{0.00194} & 69$\pm$2 & 104$\pm$18 & 97$\pm$3 & 104$\pm$18 & 64$\pm$2 & 59$\pm$15 & 62$\pm$2 \\
$J = 29 \rightarrow$ 28 & 3325005 & 90.16 & 2399.86 & \multicolumn{1}{l|}{0.00213} & 72$\pm$4 & 124$\pm$15 & 95$\pm$6 & 124$\pm$15 & 80$\pm$5 & 78$\pm$10 & 77$\pm$5 \\
$J = 30 \rightarrow$ 29 & 3438365 & 87.19 & 2564.88 & \multicolumn{1}{l|}{0.00232} & 55$\pm$7 & 82$\pm$12 & 82$\pm$9 & 82$\pm$12 & 72$\pm$6 & 50$\pm$9 & 55$\pm$4 \\
$J = 31 \rightarrow$ 30 & 3551592 & 84.41 & 2735.33 & \multicolumn{1}{l|}{0.00252} & 57$\pm$7 & 108$\pm$11 & 101$\pm$9 & 108$\pm$11 & 52$\pm$5 & 60$\pm$6 & 55$\pm$5 \\
$J = 32 \rightarrow$ 31 & 3664684 & 81.81 & 2911.21 & \multicolumn{1}{l|}{0.00274} & 37$\pm$8 & 54$\pm$18 & 68$\pm$6 & 54$\pm$18 & 47$\pm$5 & 26$\pm$11 & 39$\pm$4 \\
$J = 33 \rightarrow$ 32 & 3777636 & 79.36 & 3092.51 & \multicolumn{1}{l|}{0.00295} & 43$\pm$8 & 80$\pm$15& 79$\pm$10 & 80$\pm$15 & 42$\pm$5 & 39$\pm$7 &40$\pm$4 \\
$J = 34 \rightarrow$ 33 & 3890443 & 77.06 & 3279.22 & \multicolumn{1}{l|}{0.00318} & 38$\pm$8 & & 64$\pm$9 & & 30$\pm$5 & &35$\pm$5 \\
$J = 35 \rightarrow$ 34 & 4003101 & 74.89 & 3471.34 & \multicolumn{1}{l|}{0.0034} & 34$\pm$8$^{*}$ & & 64$\pm$9$^{*}$ & & 39$\pm$8$^{*}$ & &38$\pm$7$^{*}$ \\
$J = 36 \rightarrow$ 35 & 4115606 & 72.84 & 3668.86 & \multicolumn{1}{l|}{0.00364} & 19$\pm$7 & & 32$\pm$9 & & 23$\pm$5 & &18$\pm$4 \\
$J = 37 \rightarrow$ 36 & 4227953 & 70.91 & 3871.78 & \multicolumn{1}{l|}{0.00388} & 20$\pm$6 & & 45$\pm$10 & & 20$\pm$6 & &22$\pm$6 \\ \hline
\end{tabular}
\end{sidewaystable*}

%\newpage
\FloatBarrier

\section{Supplement figures}
\begin{figure*}[!htp]
\centering
	\begin{tabular}{cc}
\includegraphics[valign=t, width=0.44\linewidth]{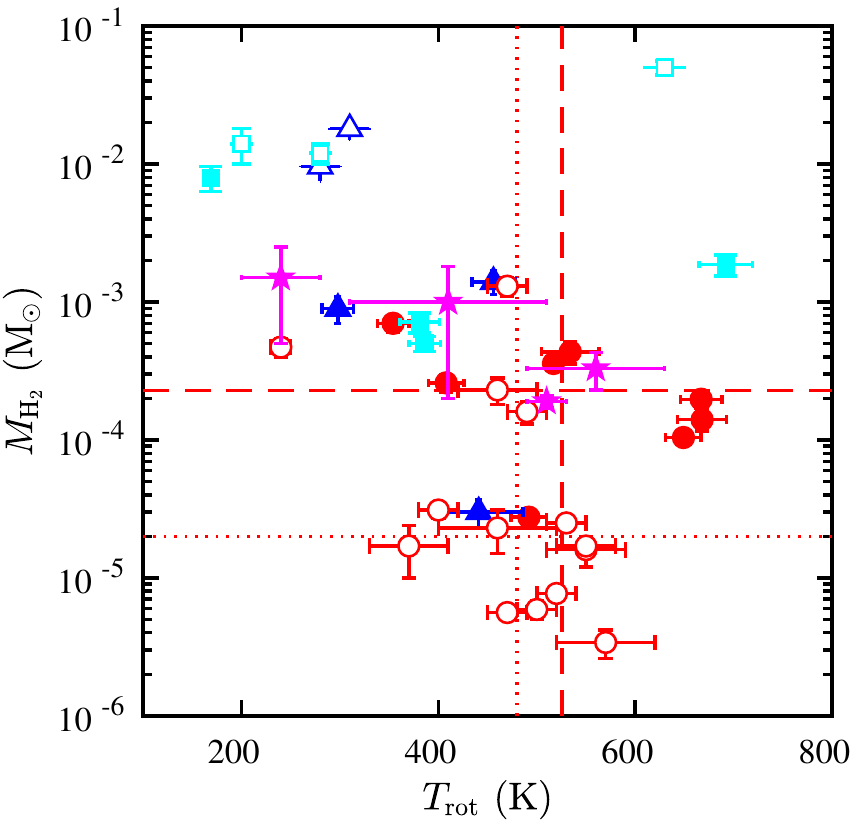} & \hspace{-0.3cm} 
        \includegraphics[valign=t, width=0.44\linewidth]{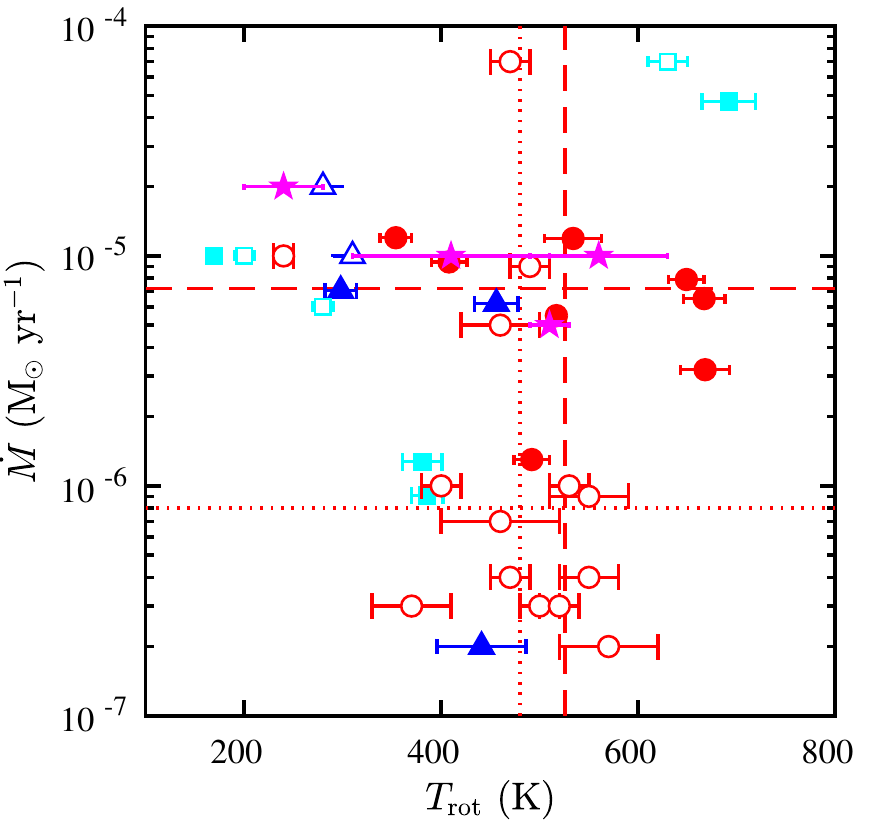} \\ 
\includegraphics[valign=t, width=0.44\linewidth]{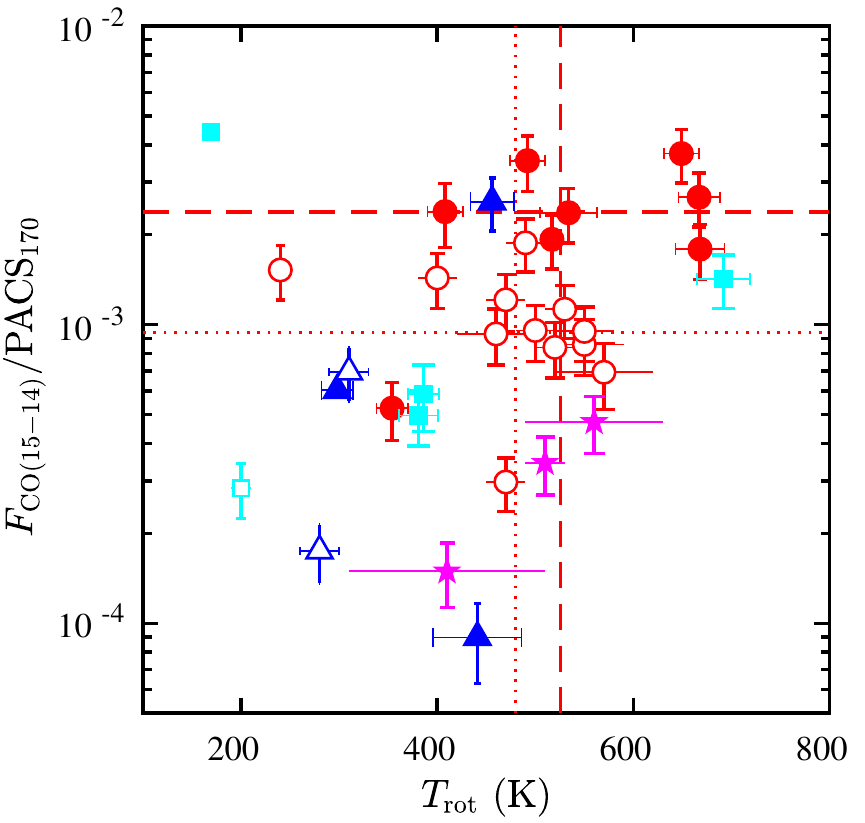} & \hspace{-0.3cm} 
        \includegraphics[valign=t, width=0.44\linewidth]{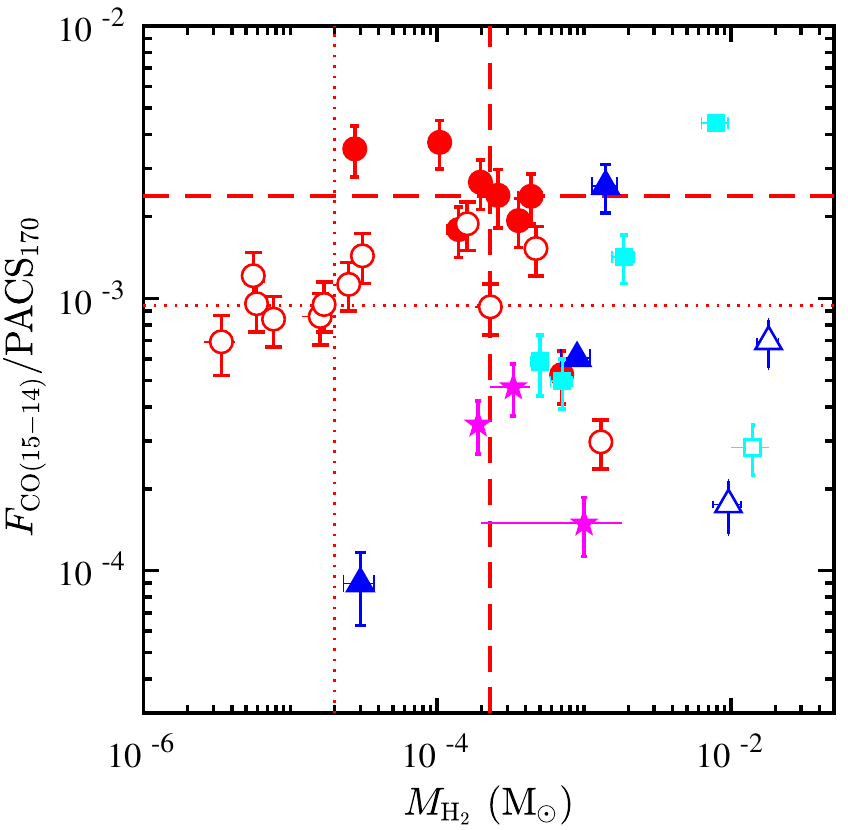} \\ 
        \includegraphics[valign=t, width=0.44\linewidth]{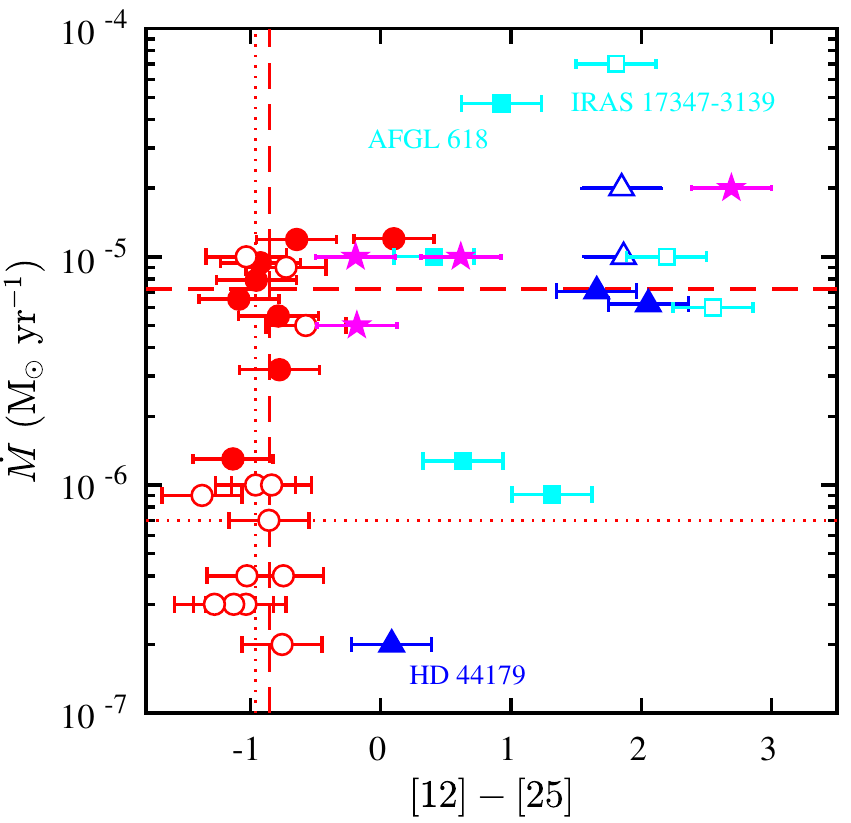} & \hspace{-0.3cm} \includegraphics[valign=t, width=0.44\linewidth]{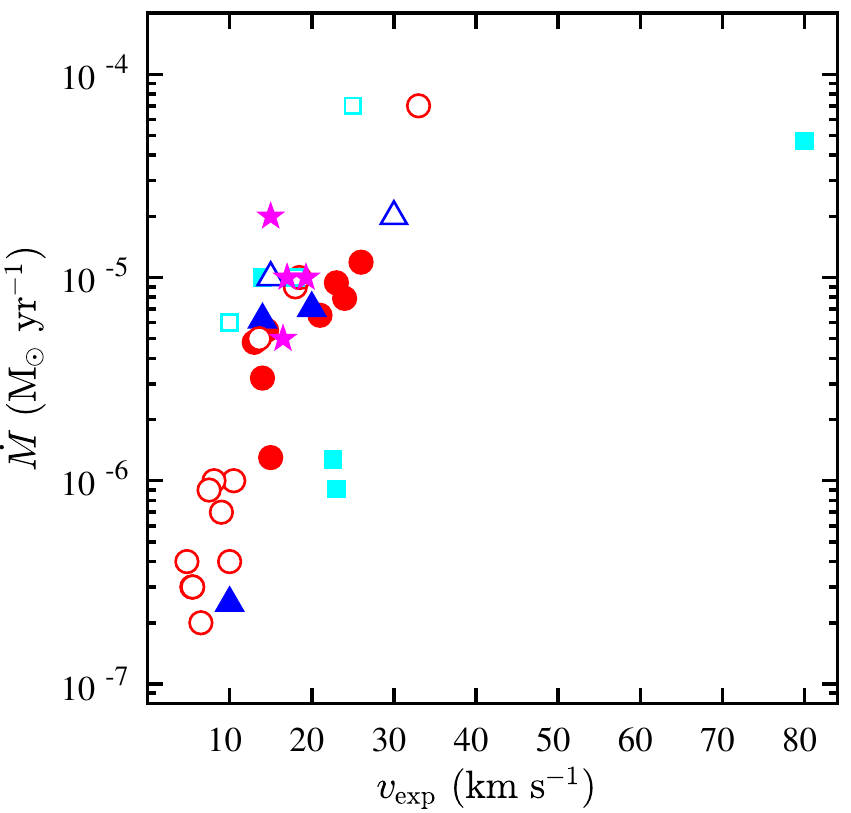}\\
	\end{tabular}
\caption{Supplementary plots. Temperature, mass, mass-loss-rate, line to continuum ratio and IRAS color of our sample of C-rich (filled symbols) and O-rich stars in Paper I (open symbols and magenta pentagrams). The color code is the same as in Fig. \ref{fig:IRAS}.} \label{fig:orichcomp}
\end{figure*}

\begin{figure*}
\centering
\includegraphics[width=\linewidth]{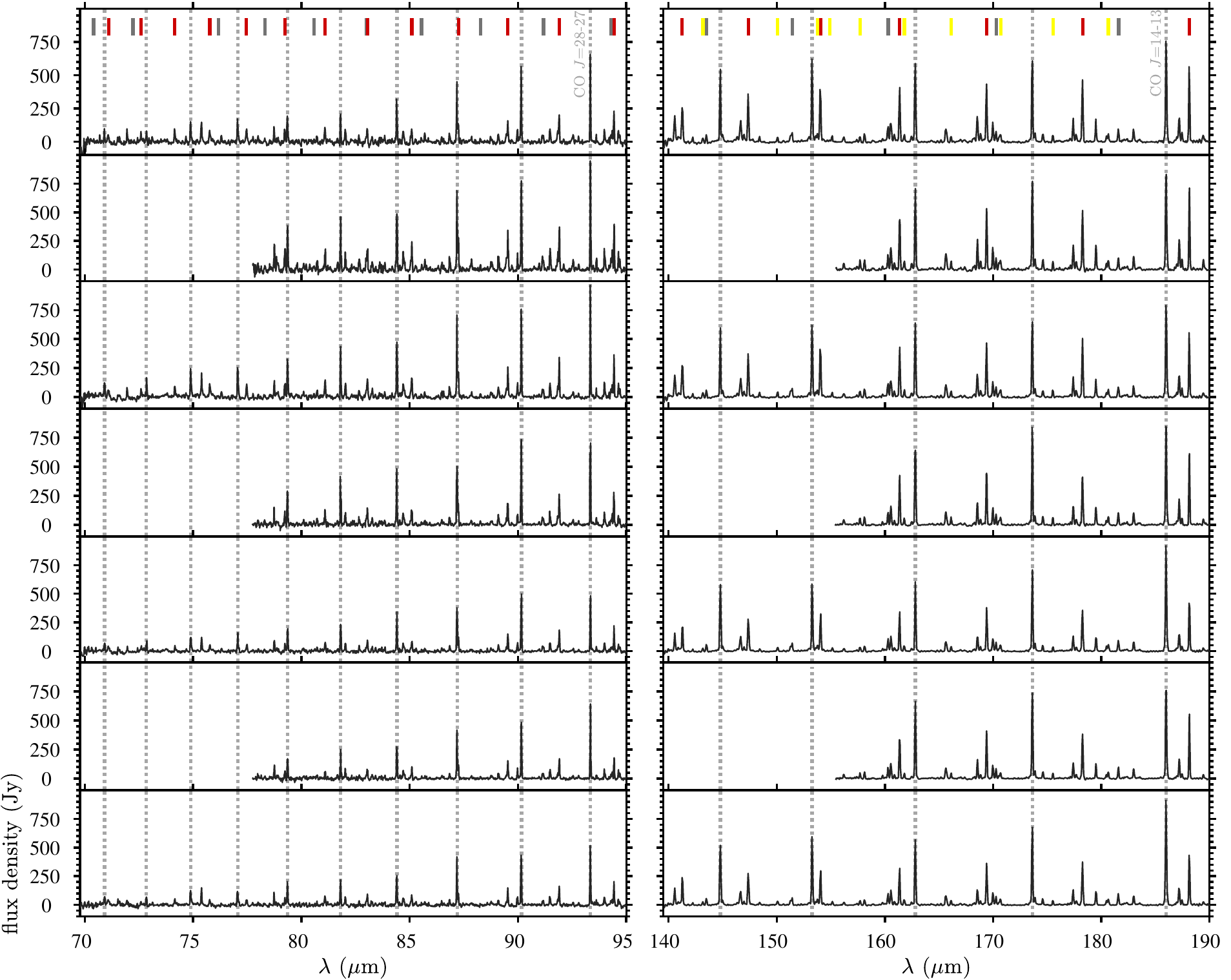}
\caption{Spectra of IRC+10216 at different epochs. Each row corresponds to a different observing day (from top to bottom: OD = 745, 894, 1087, 1113, 1257, 1288, 1296). Same color code for the top tick marks as in Fig.\ref{fig:specs}.} \label{fig:irc10216_split}
\end{figure*}

\end{appendix}

\end{document}